\documentstyle[12pt]{article}

\newcommand{\be}{\begin{equation}}
\newcommand{\ee}{\end{equation}}
\newcommand{\ba}{\begin{eqnarray}}
\newcommand{\ea}{\end{eqnarray}}
\newcommand{\la}{\lambda}
\newcommand{\La}{\Lambda}

\newcommand{\tr}{\rm tr}

\begin{document}
\hoffset=-.4truein\voffset=-0.5truein
\setlength{\textheight}{8.5 in}
\begin{titlepage}
\begin{center}
%\hfill April 7, 2006\\
\hfill LPTENS 06-17\\
\vskip 0.6 in
{\large  WKB-expansion of the HarishChandra-Itzykson-Zuber integral for  
arbitrary $\beta$}
\vskip
.6 in
\end{center}
\begin{center}
{\bf S. Hikami$^{a)}$}{\it and} {\bf E. Br\'ezin$^{b)}$}
\end{center}
\vskip 5mm
\begin{center}

{$^{a)}$ Department of Basic Sciences,
} {University of Tokyo,
Meguro-ku, Komaba, Tokyo 153, Japan.  
e-mail:hikami@rishon.c.u-tokyo.ac.jp}\\
{$^{b)}$ Laboratoire de Physique
Th\'eorique, Ecole Normale Sup\'erieure}\\ {24 rue Lhomond 75231, Paris
Cedex
05, France. e-mail: brezin@lpt.ens.fr{\footnote{\it
Unit\'e Mixte de Recherche 8549 du Centre National de la
Recherche Scientifique et de l'\'Ecole Normale Sup\'erieure.
} }}\\
\end{center}
\vskip 3mm

\begin{center}
{\bf Abstract}
\end{center}
\vskip 3mm

%*******************************

This article is devoted to the asymptotic expansion of
the generalized Harish Chandra-Itzykson-Zuber
matrix integral for  non-unitary symmetries characterized by a   
parameter $\beta$ (as
usual
$\beta= 1,2$ and $4$ correspond to the orthogonal, unitary and
symplectic group integrals).

The results are of the   form
$\sum_{perm.}\exp ( \sum x_i \lambda_i) f(\tau_{ij})/[\prod_{i<j}  
\tau_{ij}]^{\beta-1}$,
in which $x_i$ and
$\lambda_i$ are the eigenvalues of the  two $k \times k$ matrices, and
  $\tau_{ij} = (x_i - x_j)(\lambda_i - \lambda_j)$ .
  A WKB-expansion for $f$ is
derived from the heat kernel differential equation, for general values  
of $k$ and $\beta$.
 From  an expansion in terms of zonal polynomials, one obtains an  
expansion in powers of the
$\tau$'s for
$\beta=1$, and  generalizations are considered for general $\beta$.
A duality
relation,  and  a transformation of products of pairs
of symmetric  functions into $\tau$ polynomials, is used to obtain
the expression for $f(\tau_{ij})$ for  general $\beta$.
\end{titlepage}

\section{  Introduction }

We study the  HIZ-integral $I$
\be\label{I}
I(X,\Lambda) = \int  \exp (  {\rm \tr} {\hskip 2mm}g X g^{-1} \Lambda )  
{\rm dg}
\ee
where $g$ is one of  the Lie groups $O(k)$, $U(k)$ or $Sp(k)$, and   
$dg$ is the
corresponding Haar measure.

  The $k \times k$ matrices $X$ and $\Lambda$
are Hermitian (real, complex and quaternion).
This integral appears at several key points in  random matrix theory.  
When $g$
varies over the unitary group, the result is well-known : due to
Harish-Chandra \cite{Harish-Chandra}, it has been rederived in the  
context
of random matrix applications by Itzykson and Zuber
\cite{Itzykson-Zuber}. The HIZ result turns out to be "WKB exact" in  
the unitary case, i.e. to
be equal to  the sum of Gaussian integrals over the $k!$ saddle-points  
of the
integrand.

   For  non-unitary cases, this semi-classical property is no longer  
true.  However the
full WKB expansion is needed in several problems arising in  random
matrix theory, for instance in the study of universal short-distance  
correlations of
the eigenvalues. There are several studies of those non-unitary cases.  
For instance,
in the orthogonal case the result is known as an expansion in terms of  
zonal
polynomials $Z_p(x_1,...,x_k)$
\cite{James}, which are symmetric polynomials in the $x_i$. However, the
expressions in terms of zonal polynomials do not provide  the desired
WKB expansion around the saddle-points, which involves combinations of  
the form
$\prod_{i<j}(x_i-x_j)$.  Furthermore there is no explicit expression  
for zonal polynomials
of arbitrary order.

For general $\beta$, zonal polynomials are generalized, and known under  
the name of Jack
polynomials
\cite{Jack}
but the situation is not improved.
  In this article we extend the HIZ result to non-unitary groups and  
obtain
expressions of  the  form  $$e^{\sum x_i \lambda_i}
\chi(\tau_{ij})/
\prod\tau_{ij}^{\beta/2}$$ or $$e^{ \sum x_i  
\lambda_i}f(\tau_{ij})/\prod
\tau_{ij}^{\beta - 1},$$ where $\tau_{ij} = (x_i - x_j)(\lambda_i -
\lambda_j)$,
with the parameter $\beta$ corresponding to
the  $O(k)$,$U(k)$ and $Sp(k)$ group integrals for $\beta=1,2$ and 4,
respectively. Although the group integrals correspond to those three  
values, the
expressions may be continued to arbitrary $\beta$
through the differential equation for $I(X,\Lambda)$. The fact that,  
once
  the exponential prefactor $e^{\sum x_i \lambda_i}$ is extracted, the  
result
may be expressed in terms of the variables $\tau$'s is far from  
trivial. Although
it follows from this paper, through a first representation in terms of  
extended
zonal polynomials, and further identities on products of symmetric
  functions in the variables $x_i$ and $\lambda_j$,  an a priori proof  
would be welcome.

Recently, a similar expansion, using the formalism of Baker-Akhiezer  
functions,
has been considered for $I(X,\Lambda)$ by Berest \cite{Berest}.

However this is still not in terms of
the  $\tau_{ij}$, and it involves differential operators which make the  
calculation
of $I(X,\Lambda)$ difficult, except in simple cases.

In the case of the
symplectic group ($\beta=4$), we have found earlier that the WKB  
expansion
for $I(X,\La)$ terminates after a finite number of terms\cite{BHb,BHc}.  
We have
obtained these explicit finite expansions for k=2,3 and 4. In this  
article we
study the coefficients of this expansion for general k ; however the  
structure
for arbitrary k is still only partially known. Again a proof of the fact
that the WKB expansion terminates after a finite number of terms would  
be welcome.
Recently the problem has been considered by Ben Said and {\O}rsted {\cite  
{BO}} but they have simply
verified like ourselves
this amazing property.

For
other even integer values of $\beta$,
($\beta=6,8,10,...)$,  similarly the WKB series stops after a finite  
number of
terms. In this article we discuss also the coefficients of these  
expansions, but again proofs
of the finiteness are lacking.

  The expressions that we have obtained are of the form
\be\label{eq2}
I_{\beta}(x,\lambda) =\sum_{perm. of \lambda_i} \frac{1}{[\prod_{i<j}^k
(x_i - x_j)(\lambda_i - \lambda_j)]^{\beta - 1}}
e^{ \sum_{i=1}^k x_i \lambda_i} f(\tau_{ij})
\ee

where $f$ is given as  a series in terms  of the $\tau_{ij}= (x_i -  
x_j)(\lambda_i-
\lambda_j)$,
\be
f(\tau_{ij}) = 1 + c_1 \sum_{i<j} \tau_{ij}+
O(\tau^2).
\ee

and the successive coefficients $c_1, c_2,\cdots$ are functions of  
$\beta$ and $k$. The "perm."
in (\ref{eq2})
means a sum over the $k!$ permutations of the $\lambda_i$'s.

In the case $k=3,\beta=4$, we have
\ba\label{k3b3}
f(\tau_{ij}) &=& 1 - \frac{1}{3}(\tau_{12} + \tau_{23} + \tau_{13})
+ \frac{1}{6}(\tau_{12}\tau_{23} + \tau_{23}\tau_{13} +  
\tau_{13}\tau_{12})\nonumber\\
&-& \frac{1}{12}\tau_{12}\tau_{23}\tau_{13}.
\ea
In the case $k=4,\beta=4$, the result is again a similar finite  
polynomial which is given in
\cite{BHb}. Our aim is to find similar compact expressions for  general  
values of $k$ and
$\beta$.

Let us note that in the cases $\beta = 2 m (m=2,3,...)$, since  the  
function $f(\tau_{ij})$ is a
polynomial,
  it means that the HIZ integrals may be written as
\be
I_{\beta} = \sum_{perm.} e^{\sum x_i \lambda_i}
\frac{1}{[\Delta(x)\Delta(\lambda)]^{\frac{\beta}{2}}}
[ 1 + a_1 \sum \frac{1}{\tau_{ij}} + O(\frac{1}{\tau^2})]
\ee
which is a WKB expansion corrected by a {\it{finite}} number of terms.

The integral $I$ is also known to be expressible as an infinite sum  
over extended zonal
polynomials
$Z_p(x)$, with a parameter $\alpha = \frac{2}{\beta}$
\cite{James,Jack,Macdonald}
\be
I_{\beta} = \sum_{m=0}^\infty \frac{1}{m!}\frac{1}{\prod_{q=0}^{m-1}(1  
+ q \alpha)}
\sum_p \chi_p(1) \frac{Z_p( X)Z_p( \Lambda)}{Z_p(\rm I)}
\ee
where $\chi_p(1)$ is a character, and $Z_p(\rm I)$ is a dimensional  
constant which
depends upon $k$. (Explicit values of $Z_p(x)$, $\chi_p(1)$, $Z_p(\rm  
I)$ are given
in Appendix A for the lower orders).

To extract the WKB-exponential $e^{\sum x_i \lambda_i}$ factor in  
(\ref{eq2}), we  shift
$X \rightarrow \tilde X = X - \frac{1}{k}{\rm tr} X$. Then, the above  
expression becomes
\be\label{zonalI}
I_{\beta} = e^{\sum x_i \lambda_i -\frac{1}{k}\sum_{i<j}\tau_{ij}}
\sum_{m=0}^\infty \frac{1}{m!}\frac{1}{\prod_{q=0}^{m-1}(1 + q \alpha)}
\sum_p \chi_p(1) \frac{Z_p( \tilde X)Z_p(\tilde  \Lambda)}{Z_p(\rm I)}
\ee

In the case $\beta=1$, the power of the Vandermonde factor in the  
denominator of
(\ref{eq2}) vanishes. Therefore one may compare directly the expression  
of
(\ref{zonalI})
with
(\ref{eq2}) ; consequently the expression for  $f(\tau_{ij})$ is  
obtained by rewriting the
products of symmetric functions in terms of  polynomials in the  
$\tau$'s.

In the other cases, $\beta \ne 1$, one needs to extract the Vandermonde  
factor.  For instance if
$\beta$ is an even integer, this factor
  is  antisymmetric  under permutations of  the $x_i$    and it
makes the problem difficult.

In this article  a dual representation is derived , which solves the  
problem
of the Vandermonde factor.
The duality comes from  another expression of  the integral $I_{\beta}$
in terms of  extended zonal polynomials. Namely, we can write
the series (\ref{zonalI}) as

\ba\label{DD2}
I_{\beta} &=& \sum_{perm.} [\frac{e^{\sum x_i \lambda_i}}{\prod_{i<j}
[(x_i-x_j)(\lambda_i - \lambda_j)]^{\beta-1}}
\nonumber\\
&\times&e^{-\frac{1}{k}\sum_{i<j} \tau_{ij}}\sum_{m=0}^\infty
\frac{1}{m!}\frac{1}{\prod_{q=0}^{m-1}(1 + q \alpha)}
\sum_p \chi_p(1) \frac{Z_p(\tilde X)Z_p(\tilde \Lambda)}{Z_p(\rm I)}]
\ea
where the parameter $\alpha$ is  now $\frac{2}{2 - \beta}$ instead of
$ \frac{2}{\beta}$. This is a duality transformation of $\beta  
\rightarrow 2 - \beta$.
Note that it is the dual invariant  product
$\beta(\frac{\beta}{2}-1)$, which appears as a coefficient of the  
inverse square potential
in the Calogero-Moser
model.
The expression  (\ref{DD2}) is  close to the desired expression  
(\ref{eq2}).
The WKB-exponential term $e^{\sum x_i \lambda_i}$ and the correct   
power of the Vandermonde
term
are already present in (\ref{DD2}).
The two expressions (\ref{zonalI}) and (\ref{DD2}), are dual under the  
transformation
  $\beta \rightarrow 2-\beta$.
Then if one writes the subsequent series in terms of the variables   
$\tau$,
  one obtains the desired WKB expansion.

This article is organized as follows:

In section two, we discuss the differential equation
satified by  the integral $I(X,\Lambda)$.

In section three, the series expansion in terms of $\tau_{ij} = (x_i-  
x_j)(\lambda_i -
\lambda_j)$ is obtained  by  consideration of residues (residual  
equations).

In section four, the case  $\beta=4$ is investigated separately;  the   
$\tau$-expansion of $f$
is discussed. (The result up to  fourth order  is given in Table A).

In section five, we extend the $\tau$ expansion to arbitrary $\beta$.  
The residual
equations are also used at this effect. (The coefficients   
$C_{[graph]}$ are given in Table
B). We find that the residual equations have no $\beta$ dependence.

In section six, we derive the $\tau$ expansion directly for $\beta=1$
from the zonal polynomial expansion
of the integral $I$, based on the expression (\ref{zonalI}).
  ( These results are expressed in Table C).

In section seven, we extend the result of the section six to  general  
$\beta$, using the
extended zonal polynomials (Jack polynomial). We find a duality  
relation of the type (\ref{DD2}).
By the duality trasformation $\alpha=\frac{2}{\beta} \rightarrow  
\frac{2}{2 - \beta}$,
the expressions for $C_{[graph]}$ given in section five (Table B) are  
rederived.  The residual
equation of section five, is  shown to be also valid for the  
coefficients of
$C_{[graph]}$, which is determined from the extended zonal polynomials.

In section eight, we discuss the case $\beta=4$ by taking $\alpha=-1$  
in (\ref{DD2}).
In the denominator of (\ref{DD2}), one has to deal with the divergent  
factor $\frac{1}{1 +
\alpha}$. This divergence is cancelled by the sum over the partition of  
the zonal
polynomials. We explicitly evaluate (\ref{DD2}) up to  order six in the  
limit
$\alpha\rightarrow -1$.  We find the large k behavior of this expansion.

In section nine, the case $\beta=2$, where the duality transformation  
becomes
singular, is discussed.

In section ten, the large k limit and large $\beta$ limits are  
discussed briefly.

Finally, a summary and  discussions are given.
  The application of this $\tau_{ij}$ expansion to the random matrix
theory is briefly mentioned \cite{BHg}.

In  Appendix A, the extended zonal polynomials
in terms of the classical symmetric functions $s_n$, a sum  of powers,
are given up to order six for  general  $\alpha$.
The characters $\chi_p(1)$ and the dimensional constants $Z_p(\rm I)$  
are explicitly
given.

In Appendix B, the expression of  paired products of classical symmetric
functions,
$s_n(\tilde x)s_n(\tilde \lambda)$, as polynomials in $\tau$ , is  
investigated
with the help of  differential operators
$D_{j_1,j_2,...}^{i_1,i_2,...}$, which act on  $\tau$ series. This  
technique is used to
prove various identities which appear in the $\tau$ transformations.

In Appendix C, we discuss some cubic and quartic identities for the  
variables $\tau_{ij}$.
In view of these identities, the $\tau$ expansion has some freedom  
which is used to fix  some
of the coefficients in the expansion arbitrarily.

In Appendix D, we characterize each $\tau$ terms by the unique factors  
$x_{i_1}^{p_1}x_{i_2}^{p_2}
\cdots \lambda_{j_1}^{q_1}\lambda_{j_2}^{q_2} \cdots$, which are picked  
up by the differentials
$D_{j_1 j_2 ...}^{i_1 i_2 ...}$. Using these unique characteristic  
differentials, we obtain the
coefficients of $\tau$ expansion for $f$.

%%%%%%%%%%%%%%%%%%%%%%%%

\section{Differential equations for the HIZ-integrals}

The Laplacian operator with respect to the matrix elements

\be
L =  \sum_{i,j} \frac{\partial^2}{\partial X_{ij}^2}.
\ee
(a short-hand notation for the  appropriate group invariant Laplacian)
acts on the integrand of
$I(x,\la)$ in (\ref{I}) by
\be \label{Laplace} L e^{  \tr \Lambda X} = (\tr \Lambda ^2) e^{  \tr  
\Lambda X}.
\ee

  Given that $I(x,\la)$ is a function of the eigenvalues $x_i$ of the  
matrix $X$, the
equation (\ref{Laplace}) reads
\cite{BHa},
\be \label {7}
[ \sum_{i=1}^k \frac{\partial^2}{\partial x_i^2} + \beta \sum_{i=1,(i\ne
j)}^k
\frac{1}{x_i - x_j} \frac{\partial}{\partial x_i}] I = \epsilon I\ ,
\ee
with the eigenvalue  $\epsilon$
\be
\epsilon =  \sum_{i=1}^k \lambda_i^2
\ee

Note that the integral $I$ is manifestly symmetric under interchange of
the matrices $\Lambda$ and $X$, but the procedure is dissymetric. The
solutions will of course   restore this  property, which is far from  
obvious if
one
considers the equation  (\ref{7}) alone.
The
$x$-dependent eigenfunctions of this Schr\"{o}dinger-like operator have  
a
scalar
product given by the measure
\be \langle \varphi_1\vert \varphi_2\rangle = \int dx_1\cdots dx_k
\vert\Delta(x_1\cdots x_k)\vert^{\beta} \ \varphi_1^{*}(x_1\cdots x_k)
\varphi_2(x_1\cdots x_k)
\ee
The measure becomes trivial if one
multiplies the wave function by
$\vert\Delta\vert^{\beta/2}$ . Thus  if, for some given ordering of the
$x_i$'s,
one changes
$I(x)$ to
\be\label{psi}
  \psi(x_1\cdots x_k) =
\vert\Delta(x_1\cdots x_k)\vert^{\beta/2} I(x_1\cdots x_k),  \ee one
obtains the quantum
Hamiltonian,
\be\label{r2}
[ \sum_{i=1}^k \frac{\partial^2}{\partial x_i^2} - \beta  
(\frac{\beta}{2}
  -1) \sum_{i<j}
\frac{1}{(x_i - x_j)^2}]\psi = \epsilon \psi.
\ee
This Schr\"{o}dinger equation is  a simple
Calogero-Moser model.

  For a given ordering of the $x_i$'s, pulling out of (\ref{r2}) the    
exponential factor which is
the value of the integrand at its saddle-point, one obtains
\be\label{chi}
  \psi_0 = e^{ \lambda_1 x_1 + \cdots + \lambda_k x_k} \chi
\ee
where $\chi$ satisfies
\be \label{diff}
[ \sum_{i=1}^k \frac{\partial^2}{\partial x_i^2} + 2  \sum_{i=1}^k
\lambda_i
\frac{\partial}{\partial x_i} - y\sum_{i<j} \frac{1}{(x_i - x_j)^2}]  
\chi =
0
\ee
in which $y$
stands for
\be \label{y} y = \beta (\frac{\beta}{2} -
1).\ee

In   \cite{BHc} an expansion of $\chi$ for large $\tau$'s
\be
\chi = 1 -\frac{y}{2} ( \sum_{i<j}^k \frac{1}{\tau_{ij}})
+ \frac{y}{2}(-1 + \frac{y}{4})(\sum_{i<j}^K \frac{1}{\tau_{ij}^2}) +  
\cdots
\ee
has been introduced.

For even integer values of $\beta$, this expansion terminates after a  
finite number of terms. The
term of highest degree is   $\displaystyle{[\prod_{i<j}^k
\tau_{ij}]^{-\beta/2 + 1}}$. Therefore, one may write the whole  
expression as
\be
\chi = \frac{f_\beta}{[\prod_{i<j}^k \tau_{ij}]^{\beta/2 - 1}}
\ee
 From the definitions of $\psi$ and $\chi$ in (\ref{psi},\ref{chi}),
the integral $I$ is  then expressed as
\be\label{expf}
I_{\beta}(x,\lambda) = \sum_{{\rm perm.  of  
}\lambda_i}\frac{1}{[\prod_{i<j}
\tau_{ij}]^{\beta - 1}}
  e^{ \sum_{i=1}^k x_i \lambda_i} f_\beta
\ee
in which we have  normalized  $f_\beta$  such that $f_\beta = 1 +
O(\tau_{ij})$.

Continuing from even integer $\beta$'s to arbitrary
real numbers, we consider below  the expression of $f_\beta$
for arbitrary $\beta $ and arbitrary $k$.

%*******************

\section{Expansion in powers of $\tau$}

We now perform an expansion in powers af the variables $\tau$. Let us  
begin with the simple case $\beta=4$, and $k=3$ .
The differential equation for $\psi$, defined by (\ref{r2}), is
\be
\hat P \psi = 0
\ee
where the operator $\hat P$ is
\be
\hat P =  \sum_{i=1}^k \frac{\partial^2}{\partial x_i^2} -
   \sum_{i<j}
\frac{4}{(x_i - x_j)^2} - \sum_{i=1}^k \lambda_i^2 .
\ee
The expression of  $\psi$ in terms of $\tau_{ij}$ consists,
in this simple k=3 case, of the sum of four terms of increasing degrees:

\be
\psi = \psi_0 + \psi_1 + \psi_2  + \psi_3
\ee
\be
\psi_0 = \frac{1}{\Delta(x)\Delta(\lambda)} e^{\sum x_i \lambda_i}
\ee
\be
\psi_1 = C_1 e^{\sum x_i  
\lambda_i}\frac{1}{\Delta(x)\Delta(\lambda)}[\tau_{12} + \tau_{23}
+ \tau_{13}]
\ee
\be
\psi_2 = C_2 e^{\sum x_i \lambda_i}\frac{1}{\Delta(x)\Delta(\lambda)}[  
\tau_{12}\tau_{23} +
\tau_{23}\tau_{31} + \tau_{31}\tau_{12}]
\ee
\ba
\psi_3 &=&  C_3e^{\sum x_i  
\lambda_i}\frac{1}{\Delta(x)\Delta(\lambda)}[ \tau_{12}\tau_{23}
\tau_{13}]
\nonumber\\
&=&C_3 e^{\sum x_i \lambda_i}
\ea

where $\Delta(x) = (x_1-x_2)(x_1 - x_3)(x_2 - x_3)$ and $\tau_{12} =  
(x_1-x_2)(\lambda_1 -
\lambda_2)$. The unknown coefficients $C_n$ may be obtained from the  
differential equation.

Applying $\hat P$ on $\psi_0$ and $\psi_1$, we have
\be
   \hat P \psi_0 = - 2 \frac{1}{\Delta(x)\Delta(\lambda)} e^{\sum x_i  
\lambda_i}
[\frac{\lambda_1 - \lambda_2}{x_1 - x_2}+\frac{\lambda_1 -  
\lambda_3}{x_1 - x_3}+
\frac{\lambda_2 - \lambda_3}{x_2 - x_3}].
\ee
\ba
\hat P \psi_1 &=& C_1 e^{\sum x_i \lambda_i}[
-\frac{2}{(x_3-x_1)(x_2-x_3)\tau_{13}\tau_{23}}-\frac{2}{(x_1-x_2)(x_2- 
x_3)\tau_{12}\tau_{23}}
\nonumber\\
&-&
\frac{2}{(x_1-x_2)(x_3-x_1)\tau_{12}\tau_{13}}-  \frac{4}{(x_1 -  
x_2)^2\tau_{13}\tau_{23}}
- \frac{4}{(x_1 - x_3)^2\tau_{12}\tau_{23}}]\nonumber\\
&-&\frac{4}{(x_2 - x_3)^2\tau_{12}\tau_{13}} -  
\frac{2}{(x_1-x_3)^2\tau_{23}}-
\frac{2}{(x_2-x_3)^2\tau_{13}}-\frac{2}{(x_1-x_2)^2\tau_{23}}\nonumber\\
&-&\frac{2}{(x_2-x_3)^2\tau_{12}}-\frac{2}{(x_1-x_2)^2\tau_{13}}-
\frac{2}{(x_1-x_3)^2\tau_{12}}]
\ea
where the first three terms are generated by $\displaystyle{\frac{d^2}{d x_i^2}}$,
and the next three ones from the factor
$\displaystyle{- 4 \sum \frac{1}{(x_i-x_j)^2}}$ in $\hat P$.
The last six terms are obtained by the differentiations both of  
$\exp[\sum \lambda_i x_i]$
and $\displaystyle{\frac{1}{\tau_{ij}}}$. Note that $\hat P \psi_1$ consists of terms  
of order  $O(1/x^4)$ and
$O(1/x^3)$.

The unknown coefficients
$C_n$ are determined by the condition  $\hat P \psi =0$.
Since $x_i$ and $\lambda_i$ are arbitrary variables, we have the  
freedom of taking any particular
limit ; for instance
  $x_1\rightarrow x_2$.
 From the condition $\hat P\psi=0$, the pole term of $1/(x_1 - x_2)$  
present in $\hat P \psi_0$
should
be cancelled by the term coming from $\hat P \psi_1$. This requirement  
is sufficient  to fix the
coefficient $C_1$.
Indeed In  $\hat P \psi _0$, the residue of this pole is proportional  
to $\lambda_1$,
$\displaystyle{\hat P \psi_0 \sim -2\frac{\lambda_1}{(x_1 -  
x_2)\Delta(x)\Delta(\lambda)}}$.
 From the second and the fourth terms in $\hat P \psi_1$, we have
$\displaystyle{-6 C_1 \frac{\lambda_1}{(x_1 - x_2)\Delta(x)\Delta(\lambda)}}$ term.
Therefore, the required cancellation fixes the coefficient $C_1$ to be  
$-1/3$.

The term of $\displaystyle{-2 \frac{C_1}{(x_1-x_2)^2 \tau_{23}}}$ in $\hat P \psi_1$ is
cancelled by the term in $\hat P \psi_2$.
Indeed we have
\ba
\hat P \psi_2 &=& C_2 e^{\sum x_i  
\lambda_i}[-\frac{4}{\tau_{12}}(\frac{1}{(x_1-x_3)^2}
  + \frac{1}{(x_2-x_3)^2})\nonumber\\
&-&\frac{4}{\tau_{23}}(\frac{1}{(x_1-x_3)^2}
  + \frac{1}{(x_1-x_2)^2})
-\frac{4}{\tau_{13}}(\frac{1}{(x_1-x_2)^2}+\frac{1}{(x_2-x_3)^2})\nonumber\\
&-&\frac{2}{(x_1-x_2)^2}-\frac{2}{(x_1-x_3)^2}-\frac{2}{(x_2- 
x_3)^2}]
\ea
The pole  term $\displaystyle{-2 \frac{C_1}{(x_1-x_2)^2 \tau_{23}}}$
should be cancelled, and thus
\be
2C_1+ 4 C_2 = 0.
\ee

Thus one obtains  $C_2 = 1/6$ . The last term $\psi_3$ is
\be
\psi_3 = C_3 e^{\sum x_i \lambda_i}
\ee
\be
\hat P \psi_3 = C_3 e^{\sum x_i \lambda_i}(- \frac{4}{(x_1-x_2)^2}-  
\frac{4}{(x_2-x_3)^2}-
\frac{4}{(x_1-x_3)^2})
\ee

The cancellation of the double pole $1/(x_1-x_2)^2$ requires
\be
2 C_2 + 4 C_3=0
\ee

i.e. $C_3 = -1/12$.

We have thus determined the coefficients of the various   terms in the  
expansion for
$\beta=4$ , $k=3$. The results obtained in this manner coincide  with  
our earlier results \cite{BHb}
; this justifies the expression   (\ref{k3b3}) given in the  
introduction.

The remarkable point is that the cancellation conditions  connect  
linearly the  term
$C_{n-1}$ and the n-th order term $C_{n}$
as
\be
  A C_{n-1} + B C_{n} = 0 .
\ee
It turns out that this structure holds  for  arbitary k and $\beta$,  
and it turns out
that the coefficients $A$ and
$B$ hereabove are  functions of
$k$ and  not of $\beta$.

%*****************************************************************

\section{The $\tau$-expansion for the symplectic case $\beta=4$.}
The group parameter $\beta$ remains equal to 4 in this section. (In the  
next section, we will
consider  general $\beta$).
Generalizing the simple $k=3$ example of the previous section,   we  
examine the
cancellation
of the pole $1/(x_1 - x_2)$ when $x_1 \rightarrow x_2$ for arbitrary k.
Let us focus on a specific  pole term in the
various pieces  of $\hat P \psi$.

(i) In $\hat P\psi_{n-1}$, the action of $\displaystyle{\frac{\partial^2}{\partial  
x_1^2} +
\frac{\partial^2}{\partial x_2^2}}$
on $e^{\sum \lambda_i x_i}$ and $1/(x_1 - x_2)$ yields $\displaystyle{-2  
\frac{\lambda_1 - \lambda_2}{x_1 -
x_2}[e^{\sum \lambda_i x_i}
\frac{1}{\Delta(x)\Delta(\lambda)}\prod \tau_{lm}]}$.

(ii) In $\hat P\psi_n$, the multiplication by $\displaystyle{-4 \frac{1}{(x_1 -  
x_2)^2}}$ gives \\
$\displaystyle{-4
\frac{\lambda_1 - \lambda_2}{x_1 - x_2}[e^{\sum \lambda_i x_i}
\frac{1}{\Delta(x)\Delta(\lambda)}\prod \tau_{lm}]}$.

(iii)The derivative $\displaystyle{\frac{d^2}{d x_2^2}}$ of $\psi_n$ yields
$\displaystyle{-2 \frac{1}{(x_1 - x_2)(x_2 - x_m)}}$ (where $m\ne 1,2$). This term may  
be written  as \\
$\displaystyle{e^{\sum
\lambda_i x_i}\frac{1}{\Delta(x)\Delta(\lambda)}[ - 2 \frac{\tau_{1m}}  
{(x_1 - x_2)(x_2 -
x_m)}\prod' \tau_{ij}]}$. Since $\tau_{1m}/(x_2 - x_m)$ reduces to
$\lambda_1 - \lambda_m$ in the limit $x_1\rightarrow x_2$, we are left  
with the
  pole term  \\
$\displaystyle{ - 2 \frac{\lambda_1 -\lambda_m}{x_1 - x_2} e^{\sum  
\lambda_i x_i}
\frac{1}{\Delta(x)\Delta(\lambda)}\prod \tau_{ij}}$ in this limit.

(iv)Leaving aside the overall factor $\displaystyle{e^{\sum \lambda_i  
x_i}\frac{1}{\Delta(x)\Delta(\lambda)}}$,
the pole terms in (i),(ii) and (iii) must cancel. The residue of  
$1/(x_1 -
x_2)$ is the combination of $\lambda_1 \prod \tau_{ij}$ with unknown  
coefficients $C$ for each
monomial in
$\psi_n$.

For instance, in the case of  k=3, n=1, from (i),(ii) and (iii), we  
would find
\be
- 2 C_0 \frac{\lambda_1}{x_1 - x_2} - 4C_1 \frac{\lambda_1 }{x_1 - x_2}
- 2 C_1 \frac{\lambda_1 }{x_1 - x_2} =0
\ee

and the cancellation imposes
$C_0 + 3 C_1=0$ $(C_0=1)$.
The pole term in $\hat P \psi$ is always a single pole, once we factor  
out the Vandermonde
factor in the denominator.

This method of selecting  pole terms is simple but
at increasing orders there are many terms and many coefficients.
  A graphical characterization of those terms is thus necessary, in  
order to
extract the various contributions to the pole terms of
$\hat P \psi_n$.
For the general problem of order $k$ one has
\ba
\psi &=& e^{\sum \lambda_i x_i} \frac{1}{\Delta(x)\Delta(\lambda)}[  
C_{[0]} +
C_{[\rm I]}(\tau_{12} + \tau_{13} + \cdots) +  
C_{[\Lambda]}(\tau_{12}\tau_{13} +
\cdots)\nonumber\\
& +&
C_{[\rm I,\rm I]}(\tau _{12}\tau_{34} + \cdots) + O(\tau^3)]
\ea
We represent the terms which are products of $\tau_{ij}$ by a graph in  
which one draws
  a line between the points (i) and (j) whenever a factor $\tau_{ij}$ is  
present.
For a given graph,  a graph-dependent  coefficient characterizes its  
weight in
the function
$f_{\beta}$.
For instance the graph $[graph]$= [I,I], has a weight $C_{[I,I]}$,  
which is the coefficient of
the
sum  $\Sigma\tau_{n_1 n_2} \tau_{n_3 n_4}$, where the  $n_i$ are all  
different.
The graph $[\wedge]$ consists of two lines meeting  at one point, and  
it comes with a coefficient
$C_{[\wedge]}$ in front of the sum $\Sigma\tau_{n_1 n_2} \tau_{n_1  
n_3}$ in which the three $n_i$
are different.

One may now apply the rules (i)-(iii).\\
{\it  at first order:}
  We consider  the pole term of the form $\displaystyle{\frac{\lambda_1}{x_1 - x_2}}$.  
 From $\psi_{[0]}$
the operation (i) gives $-2C_{[0]} = -2$.
The rule (ii) gives the same pole with the coefficient $\displaystyle{- 4 C_{[\rm  
I]}}$.
The rule (iii), from the terms $-2 (\lambda_1 - \lambda_m)/(x_1 - x_2)$  
, where
m is not equal to 1 or 2 we obtain a  term $\displaystyle{-2 (k-2)C_{[\rm I]}
\frac{\lambda_1}{x_1 - x_2}}$.
Therefore the cancellation yields the relation
\be
-2 C_{[0]} - 4 C_{[\rm I]} - 2 (k-2) C_{[\rm I]} = 0
\ee
which gives $C_{[\rm I]} = - \frac{1}{k}$.\\
{\it at order two:}
There are two coefficients corresponding to the two graphs of that  
order, namely
$C_{[\Lambda]}$ and $C_{[\rm I,\rm I]}$.

To determine
$C_{[\Lambda]}$, we compare the pole term
  $\displaystyle{\lambda_1 \tau_{13}\frac{1}{x_1 - x_2}}$ or equivalently
$\displaystyle{-\lambda_1 \lambda_3 x_1\frac{1}{x_1 - x_2}}$ term in
$\psi_1$ and $\psi_2$.
  Applying (i) for $\psi_1$ yields $-2 (\lambda_1 - \lambda_2)\tau_{13}$.
Applying  (ii) for $\psi_{[\Lambda]}$, we have $-4 (\lambda_1 -  
\lambda_2)\tau_{13}$.
Applying (iii) for $\psi_{[\Lambda]}$ term, we get
$-2 (\lambda_1 - \lambda_m) \tau_{13}$, where m is not equal to
1,2 or 3. Thus wefind the relation
\be\label{40}
-2 C_{[\rm I]} - 4 C_{[\Lambda]} - 2 (k-3) C_{[\Lambda]} =0
\ee
which leads to
$C_{[\Lambda]}= \frac{1}{k(k-1)}$.

To determine $C_{[\rm I,\rm I]}$, we collect the pole terms of  the form
$\displaystyle{-2 \frac{\lambda_1 - \lambda_2}{x_1 - x_2} \tau_{34}}$.
In $\tau_{34}= (\lambda_3 - \lambda_4)(x_3 - x_4)$, we look at the
term $\lambda_3 x_4$, and thus select the pole term $\lambda_1  
\lambda_3 x_4/
(x_1 - x_2)$.
 From (i), one obtains a coefficient $2 C_{[\rm I]}$  for this pole  
(fig.2 (i)).
Since we are looking at terms containing $\lambda_1 \lambda_3 x_4$,
they arise from $\psi_{[\Lambda]}$ as $-2 (\lambda_1 - \lambda_4)  
\tau_{34}$
with the coefficient $C_{[\Lambda]}$ (fig.2(iii)).
Finally from $\psi_{[\rm I,\rm I]}$, we have three type of  
contributions;
the first one is Fig.2(ii),it gives  
$-4(\lambda_1-\lambda_2)\tau_{34}C_{[\rm i,\rm i]}$.
the second one is $2 (\lambda_2 - \lambda_3)\tau_{14}C_{[\rm I,\rm I]}$  
(fig.2 (iv)),
and the last one  is $-2(\lambda_1 - \lambda_m)\tau_{34}C_{[\rm I,\rm  
I]}$ (Fig.2 (v)).
In this term $m$ should be larger than 4, and a factor $(k-4)$ arises   
when one sums over all
possible values of m.

 From these  terms (fig.2 (i)-(v)),
we obtain the pole terms $\lambda_1 \lambda_3 x_4/(x_1 - x_2)$.
Thus adding these terms, one finds the relation
\be\label{41}
2 C_{[\rm I]} + (4 + 2 + 2(k-4))C_{[\rm I,\rm I]} + 2 C_{[\Lambda]} = 0
\ee
\begin{picture}(150,100)
\put(84,42){\line(2,3){16}}
%\put(100,70){\line(2,-5){12}}
\put(85,70){\dots{1}}
\put(78,70){2}
\put(80,30){3}
%\put(110,30){4}
\put(90,15){(i)}
\put(150,42){\line(2,3){18}}
\put(146,70){\line(2,0){20}}
\put(144,70){2}
\put(166,70){1}
\put(150,30){3}
\put(150,15){(ii)}
\put(207,42){\line(2,3){18}}
\put(225,70){\line(2,-5){12}}
\put(207,70){\dots{1}}
\put(204,70){2}
\put(210,30){3}
\put(228,30){m}
\put(210,15){(iii)}
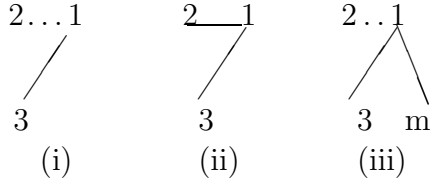
\put(35,-5){{\bf Fig.1}, three contributions of pole terms from the  
graph $C_{[\Lambda]}$.}
\put(35,-20){(i)  $-2 (\lambda_1 - \lambda_2)\tau_{13}$, (ii)
  $-4 (\lambda_1 - \lambda_2)\tau_{13}$,
(iii) $ -2 (\lambda_1 - \lambda_m)\tau_{13}$.}
\end{picture}
\vskip 10mm

 From this equation we find
\be
C_{[\rm I,\rm I]} = \frac{k-2}{k(k-1)^2}
\ee
\vskip 3mm
%***************************************************
\begin{picture}(150,100)
\put(34,45){\line(2,0){18}}
%\put(100,70){\line(2,-5){12}}
\put(35,70){\dots{1}}
\put(26,70){2}
\put(26,40){3}
\put(53,40){4}
\put(40,15){(i)}
%\end{picture}
%\begin{picture}
\put(100,45){\line(1,0){23}}
\put(100,70){\line(1,0){23}}
\put(90,70){2}
\put(125,70){1}
\put(90,40){3}
\put(125,40){4}
\put(105,15){(ii)}
%\end{picture}
%\begin{picture}
\put(185,70){\line(1,-6){4}}
\put(165,70){\dots{1}}
\put(160,45){\line(2,0){27}}
\put(155,70){2}
\put(152,40){3}
\put(189,40){4}
\put(165,15){(iii)}
\put(240,70){\dots{1}}
\put(225,40){\line(2,5){12}}
\put(260,70){\line(2,-5){12}}
\put(220,40){3}
\put(230,70){2}
\put(270,40){4}
\put(235,15){(iv)}
\put(310,70){\dots{1}}
\put(300,70){2}
\put(330,70){\line(2,-3){14}}
\put(345,45){m}
\put(310,40){\line(2,0){20}}
\put(300,40){4}
\put(332,40){3}
\put(320,15){(v)}
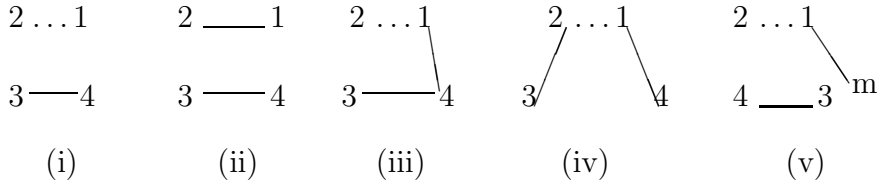
\put(20,-5){{\bf Fig.2}, five contributions of pole terms from the  
graph $C_{[\Lambda]}$
and $C_{[\rm I,\rm I]}$.}
\put(20,-27){(i)  $-2 (\lambda_1 - \lambda_2)\tau_{34} C_{[\rm I,\rm  
I]}$,
(ii) $-4 (\lambda_1 - \lambda_4)\tau_{34}C_{[\Lambda]},$
(iii)  $-2(\lambda_1-\lambda_4)\tau_{34}C_{[\Lambda]}$,}
\put(20,-49){(iv)  $-2 (\lambda_2 - \lambda_3)\tau_{14}C_{[\rm I,\rm  
I]}$,
(v)  $-2 (\lambda_1 - \lambda_m)\tau_{34}.$}
\end{picture}

\vskip 30mm

{\it at third order:}

At order three , there are five different graphs contributing to   
$\psi_3$.
\be
\psi_3 = \psi_{[\rm Y]} + \psi_{[\rm N]} + \psi_{[\Delta]} +  
\psi_{[\Lambda,\rm I]}
+ \psi_{[\rm I,\rm
I,\rm I]}
\ee
The method for obtaining the coefficients $C_{[\rm Y]},C_{[\rm  
N]},C{[\Delta]},C_{[\Lambda,\rm I]},
C_{[\rm I,\rm I,\rm I]}$ is the same as for the second order  
calculation.
For instance, to obtain $C_{[Y]}$, we look at the pole  $\lambda_1  
\lambda_3 \lambda_4
x_1^2/(x_1 - x_2)$. This singularity is obtained from the three graphs  
of fig.3.
Adding these three terms, one finds

\be
- 2 C_{[\Lambda]} - 4 C_{[\rm Y]} - 2 (k-4) C_{[\rm Y]} = 0
\ee
which leads to
\be
C_{[\rm Y]} = - \frac{1}{k-2} C_{[\Lambda]} = - \frac{1}{k(k-1)(k-2)}
\ee

\begin{picture}(150,100)
\put(34,45){\line(2,3){18}}
\put(50,70){\line(2,-5){11}}
\put(35,70){\dots{1}}
\put(26,70){2}
\put(24,40){3}
\put(61,40){4}
\put(40,15){(i)}
\put(94,45){\line(2,3){18}}
\put(110,70){\line(2,-5){12}}
\put(95,70){\line(2,0){18}}
\put(115,70){1}
\put(90,70){2}
\put(84,40){3}
\put(118,38){4}
\put(100,15){(ii)}
\put(154,45){\line(2,3){18}}
\put(170,70){\line(1,-6){4}}
\put(155,70){\dots{1}}
\put(150,70){2}
\put(146,40){3}
\put(172,40){4}
\put(172,70){\line(3,-5){15}}
\put(185,40){m}
\put(160,15){(iii)}
\put(20,0){{\bf Fig.3}, graphs for the determination of $C_{[\rm Y]}$.}
\end{picture}

\vskip 15mm
For $C_{[N]}$, we concentrate over the combination
$\lambda_1 \lambda_3^2 x_1 x_4$,  contained in
the term $(\lambda_1 - \lambda_2)\tau_{31}\tau_{34}$ for
instance. The relevant graphs are

\begin{picture}(150,100)
\put(34,45){\line(2,3){18}}
\put(34,45){\line(2,0){15}}
\put(35,70){\dots{1}}
\put(26,70){2}
\put(24,45){3}
\put(46,45){4}
\put(40,15){(i)}
\put(84,45){\line(2,3){18}}
\put(84,45){\line(2,0){16}}
\put(85,70){\line(2,0){18}}
\put(98,70){1}
\put(76,70){2}
\put(74,45){3}
\put(101,45){4}
\put(95,15){(ii)}
\put(144,45){\line(2,3){18}}
\put(160,70){\line(2,-5){10}}
\put(144,45){\line(2,0){18}}
\put(144,70){\dots{1}}
\put(140,70){2}
\put(140,40){3}
\put(168,40){m}
\put(158,40){4}
\put(140,15){(iii)}
\put(190,70){\line(2,-5){10}}
\put(197,45){\line(2,3){15}}
\put(213,70){\line(1,-3){8}}
\put(190,70){\dots{1}}
\put(188,70){2}
\put(195,37){3}
\put(220,37){4}
\put(195,15){(iv)}
\put(20,0){{\bf Fig.4}, graphs for the determination of $C_{[\rm N]}$.}
\end{picture}
\vskip 10mm

These four graphs give $-2(\lambda_1 - \lambda_2)\tau_{31}\tau_{34}$,
$-4(\lambda_1-\lambda_2)\tau_{13}\tau_{34}$, $-2(\lambda_1 -  
\lambda_m)\tau_{13}
\tau_{34}$($m>4$),and $2(\lambda_2-\lambda_3)\tau_{31}\tau_{14}$,  
respectively.
We extract the pole term $\lambda_1 \lambda_3^2 x_1 x_4/(x_1 - x_2)$   
from
these contributions, and the cancellation yields
\be
-2 C_{[\Lambda]} - 4 C_{[\rm N]} - 2 (k-4) C_{[\rm N]}- 2 C_{[\rm N]}=0
\ee
or
\be
C_{[\rm N]} = - \frac{1}{k-1} C_{[\Lambda]} = - \frac{1}{k(k-1)^2}
\ee
For $C_{[\Delta]}$, we consider the cancellation of the pole
terms with the residue $\lambda_1 \lambda_3^2 x_1 x_2$.
The cancellation gives
\be
2 C_{[\Lambda]}+ 4 C_{[\Delta]} + 2 (k-3) C_{[\rm N]} = 0\ee
which determines $C_{[\Delta]}$
\be
C_{[\Delta]} = -\frac{1}{k(k-1)^2}
\ee

For $C_{[\Lambda,\rm I]}$, the second order term gives $- 2(\lambda_1 -  
\lambda_2)
\tau_{34}\tau_{35}$ as  residue. Therefore, we look at $\lambda_1  
\lambda_3^2 x_4 x_5$
for residue. There are six different graphs which contribute to this  
residue.

Adding these terms, the cancellation gives
\be
-2 C_{[\Lambda]} - 4 C- 2 (k-5) C_{[\Lambda,\rm I]} - 4 C_{[\Lambda,\rm  
I]} + 4 C_{[\rm N]}
- 4 C_{[\rm N]} - 2 C_{[\rm Y]} = 0
\ee
which determines $C_{[\Lambda,\rm I]}$
\be
   C_{[\Lambda,\rm I]} = - \frac{k-3}{k(k-1)^2(k-2)}
\ee

The last coefficient $C_{[\rm I,\rm I,\rm I]}$ needs 7 graphs when
we consider the cancellation of the residue $\lambda_1 \lambda_3  
\lambda_4 x_5 x_6$.
The cancellation reads
\be
- 4 C_{[\rm I,\rm I]} - 4 (k-2) C_{[\rm I,\rm I,\rm I]} -  
C_{[\Lambda,\rm I]} = 0
\ee
which determines $C_{[\rm I,\rm I,\rm I]}$
\ba
C_{[\rm I,\rm I,\rm I]} &=& \frac{1}{k-2}(- C_{[\rm I,\rm I]} - 2  
C_{[\Lambda,\rm I]})\nonumber\\
&=& - \frac{k^2 - 6 k + 10}{k(k-1)^2 (k-2)^2}
\ea
Note that this coefficient $C_{[\rm I,\rm I,\rm I]}$ has a nontrivial  
numerator, which is not a
product of simple factors. This term is characterized as a  
non-intersecting, three lines graph.
When the graphs are made of non-intersecting lines, the expression for  
$C$ becomes more
complex, and there is no obvious simple  structure. In other words, for
  graphs with non-intersecting lines, the decomposition rule found in  
ref.\cite{BHc} for n=4 does
not hold. For  graphs with connecting lines, we find that the  
decomposition rule does  hold.
\\{\it at fourth order:}\\ we have computed the coefficients of every  
graph up to order four.
However wed have not been able to find an a priori rule giving the  
weight of an arbitrary graph.
  In Table A, the coefficients $C$ are listed.
\vskip 30mm
\begin{picture}(150,100)
\put(30,140){\bf Table A: WKB-expansion coefficients for $\beta=4$}
\put(30,120){\line(2,0){300}}
\put(74,100){$C_{[\rm I]} = -\frac{1}{k}$}
\put(34,100){l=1}
\put(34,70){l=2}
\put(74,70){$C_{[\Lambda]} = \frac{1}{k(k-1)}, \hskip 3mm C_{[\rm I,\rm  
I]} =
\frac{k-2}{k(k-1)^2}$}
\put(34,40){l=3}
\put(74,40){$C_{[\Delta]} = C_{[\rm N]} = - \frac{1}{k (k-1)^2}, \hskip  
3mm C_{[\rm Y]} = - \frac{1}{k
(k-1)(k-2)}$}
\put(74,10){$C_{[\Lambda,\rm I]} = -\frac{k-3}{k(k-1)^2(k-2)}, \hskip  
3mm C_{[\rm I,\rm I,\rm I]}
= - \frac{k^2 - 6 k + 10}{k (k-1)^2 (k-2)^2}$}
\end{picture}
\vskip 4mm
\begin{picture}(150,100)
\put(74,100){$C_{[\Box]} = C_{[\unrhd]}= C_{[\angle\angle]}=\frac{1}{k  
(k-1)^2 (k-2)}$}
\put(34,100){l=4}
\put(74,70){$C_{[\rm M]} = \frac{k-3}{k(k-1)^2(k-2)^2}, \hskip 3mm  
C_{[\rm X]} =
\frac{1}{k(k-1)(k-2)(k-3)}$}
\put(74,40){$C_{[\bigtriangleup,\rm I]} = C_{[N,\rm I]} =   
\frac{(k-3)^2}{k (k-1)^2 (k-2)^3},
  \hskip 3mm C_{[\rm Y,\rm I]} =  \frac{k-4}{k (k-1)^2(k-2)(k-3)}$}
\put(74,10){$C_{[\wedge,\wedge]} = \frac{k-4}{k(k-1)^2(k-2)^2}$}
\put(74,-20){$C_{[\wedge,\rm I,\rm I]} = \frac{k^3 - 10 k^2 + 34 k -  
38}{k (k-1)^2 (k-2)^3
(k-3)}$}
\put(74,-50){$C_{[\rm I,\rm I,\rm I,\rm I]}= \frac{k^4 - 14 k^3 + 76  
k^2 - 188 k + 174}{k (k-1)^2
(k-2)^3 (k-3)^2}$}
%\put(34,-80){\line(2,0){300}}
\end{picture}
\vskip 30mm

%**************************************************************

\section{The $\tau$-expansion
for $\beta $= 2m, (m=2,3,4,$\cdots$)}

Hereabove we have dealt with $\beta=4$. For  $\beta= 2m$, m=2,3,4,...,  
again we
found that the WKB
expansion for $f$ terminates after a finite number of terms.
However a given $\tau_{ij}$ may now appear non linearly,  up to
degree
$(m-1)$.
Therefore, the graphical
representations may have now multiple lines connecting the two points  
(i) and (j) (with a
mulitiplicity at most equal to $(m-1)$).Therefore
there are now new types of  graphs which did not occur  for $\beta=4  
(m=2)$.

 From the equation
\be
\hat P =  \sum_{i=1}^k \frac{\partial^2}{\partial x_i^2} - \beta  
(\frac{\beta}{2}
  -1) \sum_{i<j}
\frac{1}{(x_i - x_j)^2} - \sum_{i=1}^k \lambda_i^2 .
\ee
\be \hat P \psi = 0
\ee
We expand $\psi$ as
\be
\psi = \frac{e^{\sum \lambda_i  
x_i}}{[\Delta(x)\Delta(\lambda)]^{\beta/2 -1}}[ 1 + C_{[\rm I]}
( \tau_{12} + \cdots) + C_{[\parallel]}(\tau_{12}^2 + \tau_{13}^2 +  
\cdots) + C_{[\wedge]} (
\tau_{12}\tau_{13} + \cdots) + \cdots]
\ee

$\hat P \psi = 0$ connects the terms in $\psi$ of degree (n-1) to   
those of degree n.

 From $\hat \psi_0$, we have a pole term of
\be
\hat P \psi_0 =\frac{C_0}{[\Delta(\lambda)\Delta(x)]^{\beta/2-1}}[ - 2  
(\frac {\beta}{2} - 1)
\frac{\lambda_1 - \lambda_2}{x_1
- x_2}]
\ee

 From $\hat P \psi_{[\rm I]}$, we obtain
\be
\hat P \psi_{[\rm I]} = \frac{C_{[\rm  
I]}}{[\Delta(\lambda)\Delta(x)]^{\beta/2-1}}
[ - 4 (\frac{\beta}{2} - 1) - 2
(\frac{\beta}{2}- 1) (k-2)] \frac{\lambda_1}{x_1 - x_2}
\ee

Thus we obtain the first order result,

\be
C_{[\rm I]} = - \frac{1}{k} .
\ee

It is remarkable that this result is the same as for $\beta =4$ but now  
it holds for  arbitrary
$\beta$. Every residue had the same factor $(\frac{\beta}{2} - 1)$ and  
thus $\beta$ drops from
the equation.
This property holds to all orders, and this characteristics will become  
important when we discuss
  the duality relation for  zonal polynomials.
Therefore, the  cancellation of the pole terms gives the same results  
as for
$\beta=4$, except that there are now  new terms coming with the  
multiple lines.

We list here the results of the cancellation conditions which determine  
   the coefficients
characterizing the
$\tau$-expansion. They are valid for arbitrary $\beta$, and in fact  
they do not depend on
$\beta$.\\
{\it Second order}:
\be\label{60}
  C_{[\rm I]} + (k-1)C_{[\rm I,\rm I]} + C_{[\Lambda]} = 0
\ee
\be\label{61}
C_{[\rm I]} + (k-1) C_{[\Lambda]} + 2 C_{[\rm II]} = 0
\ee
(\ref{60}) is same as (\ref{41}), and (\ref{61}) is different from  
(\ref{40}) by the
double line term $C_{[\rm II]}$.\\
{\it Third order}:
\be
C_{[\rm I,\rm I]} + (k-2)C_{[\rm I,\rm I,\rm I]} + 2 C_{[\Lambda,\rm  
I]} = 0
\ee

\be
C_{[\rm II]} + 3 C_{[\rm III]} + (k-1)C_{[\hskip 2mm \underline{\angle}  
\hskip 2mm]} = 0
\ee

\be
C_{[\Lambda]} + (k-1)C_{[\Lambda,\rm I]} + C_{[\rm Y]} = 0
\ee

\be
C_{[\Lambda]} + (k-2)C_{[\rm Y]} + 4 C_{[\hskip  
1mm\underline{\angle}\hskip 1mm]} =0
\ee

\be\label{resid3a}
C_{[\Lambda]} + (k-2)C_{[\rm N]} + C_{[\bigtriangleup]} + C_{[\rm  
II,\rm I]} +
C_{[\hskip 1mm\underline{\angle}\hskip 1mm]} =0
\ee

\be\label{resid3b}
C_{[\Lambda]} - C_{[\rm II]} + 2 C_{[\bigtriangleup]} - C_{[\hskip  
1mm\underline{\angle}\hskip 1mm]}
  + (k-3)C_{[\rm N]}
- (k-3)C_{[\rm II,\rm I]}=0
\ee\\
{\it Fourth order}:

\be\label{max1}
C_{[\rm I,\rm I,\rm I]} + (k-3)C_{[\rm I,\rm I,\rm I,\rm I]} + 3  
C_{[\Lambda,\rm I,\rm I]} = 0
\ee

\be\label{max2}
C_{[\Lambda,\rm I]} + (k-2)C_{[\Lambda,\rm I,\rm I]}+ C_{[\rm Y,\rm I]}  
+ C_{[\Lambda,\Lambda]}=0
\ee

\be\label{max3}
C_{[\Lambda,\rm I]} + (k-3) C_{[\Lambda,\Lambda]} + 2 C_{[\rm M]} + 2  
C_{[\Lambda,\rm II]} = 0
\ee

\ba\label{max4}
&&2 C_{[\rm I,\rm I,\rm I]} + 8 C_{[\rm N,\rm I]} - C_{[\Lambda,\rm I]}  
+ 4 C_{[\rm II,\rm I,\rm
I]}
\nonumber\\
&&- C_{[\Lambda,\Lambda]} - 3 C_{[\rm Y,\rm I]} + (k-6)C_{[\Lambda,\rm  
I,\rm I]} = 0
\ea

These equations are obtained from the cancellation conditions of the  
residues of the
  pole $\frac{1}{x_1 - x_2}$ proportional to
$\lambda_1\lambda_3\lambda_4\lambda_5 x_6 x_7 x_8$, $\lambda_1  
\lambda_3 \lambda_4\lambda_5 x_6^2
x_7$,
$\lambda_1^2 \lambda_3^2 x_4 x_5 x_6$,  
$\lambda_1^2\lambda_2\lambda_3x_4 x_5 x_6$, respectively.

Further, from the residue of the form
$\lambda_1^2\lambda_3^2 x_4 x_5^2$, one finds

\ba\label{max5}
&&C_{[\rm II,\rm I]} + C_{[\rm N]} + 3 C_{[\sqsupseteq]}+ 2C_{[\rm  
II,\rm II]} +
C_{[\Box]}\nonumber\\
&&+(k-2)C_{[\Lambda,\rm II]} + (k-2)C_{[\rm M]} =0
\ea

 From the  vanishing of the residue containing $\lambda_1^2 \lambda_3  
\lambda_4 x_5^3$,
one obtains

\be
C_{[\rm Y]} + 2 C_{[\unrhd]} - 2 C_{[\underline{\amalg}]} +
2 C_{[\hskip 1mm\underline{\angle},\rm I]} + 2 C_{[\models]} +  
(k-3)C_{[\angle \angle]} = 0
\ee
and we have

\be
C_{[\hskip 1mm \underline{\angle},\rm I]} = C_{[\hskip 1mm  
\underline{\amalg}\hskip 1mm]}
\ee

However, starting with order three and higher, the coefficients of the  
$\tau$ expansion are
not determined uniquely, since  the $\tau$ variables are not  
independent. Starting with degree
three there are identities, i.e. polynomials vanishing identically.  In  
the appendix C,
we discuss the origin of these identities and derive the cubic identity  
and the quartic identities.

The cubic one is

\be\label{identity22}
I_3 =[\rm II,\rm I] - [\rm N] + (k-3) [\bigtriangleup] = 0.
\ee

and the quartic ones are

\be\label{identity4ff}
2 [\rm II,\rm I,\rm I] - [\rm N,\rm I] + (k-5)[\bigtriangleup,\rm I] = 0
\ee

and

\be\label{identity4aa}
  [\rm III,\rm I] + 2[\rm II,\rm II] - [\unrhd] + (k-3)[\hskip
1mm\underline{\bigtriangleup}\hskip 1mm]
  - 4 [\Box]
- [\hskip 1mm \underline{\amalg}\hskip 1mm]=0
\ee

and

\be\label{identity4bb}
[\Lambda, \rm II]  - [\rm M]+ 2 [\bigtriangleup, \rm I] + (k-4)[\Box] -  
\frac{1}{2}
(k-4) [\rm II,\rm II] =0
\ee

  and

\be\label{identity4cc}
  [\underline{\angle},\rm I] - 2 [ \angle\angle] - 2 [  
\bigtriangleup,\rm I]
+ (k-4) [\hskip 1mm
\unrhd\hskip 1mm]
  - 2(k-4)[\Box] + (k-4) [\rm II,\rm II]
= 0
\ee

Those identities give some freedom in writing the above expansion.  
Indeed one may add one of those
identities multiplied by an
arbitrary coefficient to the $\tau$ expansion of the HIZ-integral.  For  
the third order, the
coefficients $C_{\rm II,\rm I}$,$C_{\rm N}$ and $C_{\bigtriangleup}$
may be shifted as

\ba
  C_{\rm II,\rm I} &\rightarrow& C_{\rm II,\rm I} + \alpha\nonumber\\
  C_{\rm N} &\rightarrow& C_{\rm N} - \alpha\nonumber\\
C_{\bigtriangleup} &\rightarrow& C_{\bigtriangleup} + (k-3)\alpha
\ea

where $\alpha$ is arbitrary.

At order four, from (\ref{identity4ff}),  
(\ref{identity4aa}),(\ref{identity4bb}) and (\ref{identity4cc}),
the following shifts leave the expansion unchanged

\ba
   C_{\rm II,\rm I,\rm I} &\rightarrow& C_{\rm II,\rm I,\rm I} + 2  
\gamma_4\nonumber\\
   C_{\rm N,\rm I} &\rightarrow& C_{\rm N,\rm I} - \gamma_4\nonumber\\
   C_{\rm III,\rm I} &\rightarrow& C_{\rm III,\rm I} +  
\gamma_1\nonumber\\
  C_{\rm II,\rm II} &\rightarrow& C_{\rm II,\rm II} + 2\gamma_1 -  
\frac{1}{2}(k-4)\gamma_2 +
(k-4)\gamma_3
\nonumber\\
C_{\unrhd} &\rightarrow& C_{\unrhd} -\gamma_1 + (k-4)\gamma_3\nonumber\\
C_{\underline{\bigtriangleup}} &\rightarrow&  
C_{\underline{\bigtriangleup}} + (k-3)\gamma_1\nonumber\\
C_{\Box} &\rightarrow& C_{\Box} - 4 \gamma_1 + (k-4)\gamma_2 - 2  
(k-4)\gamma_3\nonumber\\
C_{\bigtriangleup,\rm I} &\rightarrow& C_{\bigtriangleup,\rm I} +  
2(k-4)\gamma_2 - 2 (k-4)\gamma_3
+ (k-5)\gamma_4
\nonumber\\
C_{\underline{\amalg}} &\rightarrow& C_{\underline{\amalg}} -  
\gamma_1\nonumber\\
C_{\rm M} &\rightarrow& C_{\rm M} - \gamma_2\nonumber\\
C_{\underline{\angle},\rm I} &\rightarrow& C_{\underline{\angle},\rm I}  
+ \gamma_3\nonumber\\
C_{\angle\angle} &\rightarrow& C_{\angle\angle} - \gamma_3
\ea

where $\gamma_1$,$\gamma_2$ and $\gamma_3$ are arbitrary numbers,  
arising as multiplicative
factors of the quartic identities.

In Table B, we give the coefficients $C_{[graphs]}$. For the  
coefficients,
which have ambiguities due to the above identities, we use a fixing  
rule  which
will be discussed in the next section.

The values of Table B are also derived systematically by the  
$\tau$-trasformation of the Jack polynomial
series as discussed later in Appendix B and Appendix D. In Table B, we  
use the parameter $\alpha$ instead
of $\beta$. The parameter $\alpha$ is a parameter of Jack polynomial,  
and in Table B, $\alpha$ is given by
$\alpha= \frac{2}{2-\beta}$. For $\beta=4$, $\alpha$ becomes $-1$, and  
the values in Table B coincide
with the values in Table A.

\vskip 20mm

\begin{picture}(150,150)
\put(50,120){\bf Table B: WKB-expansion coefficients for $\beta= 2m$}
\put(30,90){\line(2,0){300}}
\put(74,60){$C_{[\rm I]} = -\frac{1}{k}$}
\put(34,60){l=1}
\put(34,30){l=2}
\put(74,30){$C_{[\wedge]} = \frac{1}{k (k+ \alpha)}$}
\put(74,0){$ C_{[\rm I,\rm I]} =
\frac{1}{k (k+\alpha)} ( 1+ \frac{\alpha}{k-1})$}
\put(74,-30){$C_{[\rm II]}= \frac{1+\alpha}{2 k (k+\alpha)}$}
\put(34,-60){l=3}
\put(74,-60){$C_{[\rm Y]}=- \frac{1}{k (k+\alpha)(k+2 \alpha)}$}
\put(74,-90){$C_{[\wedge,\rm I]} = - \frac{1}{k (k+\alpha)(k+2  
\alpha)}(1 + \frac{2 \alpha}{
k-1})$}
\put(74,-120){$C_{[\rm III]}= - \frac{(1+\alpha)(1+2 \alpha)}{6 k  
(k+\alpha)(k+2\alpha)}$}
\put(74,-150){$C_{[\hskip 1mm{\underline{\angle}}\hskip 1mm]}= -  
\frac{1+\alpha}{2 k (k+\alpha)
(k+2\alpha)}$}
\put(74,-180){$C_{[\rm I,\rm I,\rm I]}= - \frac{1}{k (k+\alpha)(k+  
2\alpha)}(1 + \frac{3 \alpha}{
k-1}+\frac{2 \alpha^2}{ (k-1)(k-2)})$}
\put(74,-210){$C_{[\rm II,\rm I]}= - \frac{1+\alpha}{2 k (k+\alpha)(k +  
2\alpha)}
( 1 + \frac{2 \alpha}{ k-1})$}
\put(74,-240){$C_{[\rm N]}= - \frac{1}{k (k+\alpha) (k+2\alpha)}( 1+   
\frac{\alpha}{
k-1})$}
\put(74,-270){$C_{[\bigtriangleup]} = - \frac{1}{k (k+\alpha)(k+  
2\alpha)}
( 1- \frac{\alpha^2}{k-1})$}
\put(40,-295){(The last three coefficients $C$ have arbitrariness due  
to the cubic identity)}

\end{picture}

\vskip 88mm

\newpage
\begin{picture}(150,127)
\put(20,90){l=4}
\put(40,90){$C_{[\rm X]}= \frac{1}{k (k+\alpha)(k+2 \alpha)(k+  
3\alpha)}$}
\put(40,60){$C_{[\rm Y,\rm I]}  = \frac{1}{k (k+\alpha) (k+2  
\alpha)(k+3 \alpha)}( 1 + \frac{3\alpha}{
k-1})$}
\put(40,30){$C_{[\Lambda,\Lambda]}= \frac{1}{k (k+\alpha)(k+2  
\alpha)(k+3 \alpha)}( 1 + \frac{\alpha}{
k + \alpha-1})(1+ \frac{3\alpha}{k-1})$}
\put(40,0){$C_{[\Lambda,\rm I,\rm I]} = \frac{1}{k (k +  
\alpha)(k+2\alpha)(k+3\alpha)}( 1 - \frac{2\alpha}{
k+\alpha-1} + \frac{7\alpha}{k-1} +  
\frac{6\alpha^2}{(k-1)(k-2)}-\frac{2\alpha^2}{(k-2)(k+\alpha-1)})$}
\put(40,-30){$C_{[\rm I,\rm I,\rm I,\rm I]} = \frac{1}{k (k +  
\alpha)(k+2\alpha)(k+3\alpha)} (1 +
\frac{6\alpha}{k+\alpha-1} + \frac{9\alpha^2}{(k-1)(k+\alpha-1)} +  
\frac{8\alpha^2}{(k-2)(k+\alpha-1)}$}
\put(70,-50){$+ \frac{25\alpha^3}{(k-1)(k-3)(k+\alpha-1)} -  
\frac{8\alpha^3}{(k+\alpha-1)(k-2)(k-3)}
+\frac{6\alpha^4}{(k-1)(k-2)(k-3)(k+\alpha-1)})$}
\put(40,-80){$C_{[\sqsupseteq]}= \frac{1+\alpha}{2 k (k +  
\alpha)(k+2\alpha)(k+3\alpha)}(1 +
\frac{2\alpha}{k-1})$}
\put(40,-110){$C_{[\ll]}=\frac{(1+\alpha)^2}{4 k (k +  
\alpha)(k+2\alpha)(k+3\alpha)}$}
\put(40,-140){$C_{[{\models}]}=\frac{1+\alpha}{2 k (k +  
\alpha)(k+2\alpha)(k+3\alpha)}$}
\put(40,-170){$C_{[\hskip 1mm{\underline{\underline{\angle}}}\hskip  
1mm]} =
\frac{(1+\alpha)(1+ 2\alpha)}{ 6 k (k + \alpha)(k+2\alpha)(k+3\alpha)
}$}
\put(40,-200){$C_{[\hskip 1mm \rm IIII \hskip 1mm]} =
\frac{(1+\alpha)(1+ 2\alpha)(1 + 3\alpha)}{24 k (k +  
\alpha)(k+2\alpha)(k+3\alpha)}$}
\put(40,-230){(These 10 coefficients are determined without  
ambiguities)}

\end{picture}

\newpage

\begin{picture}(150,127)
\put(74,90){$C_{[\hskip 1mm{\underline{\amalg}}\hskip 1mm]}=  
\frac{1+\alpha}{2 k(k+\alpha)(k+2\alpha)
(k+3\alpha)}( 1+ \frac{\alpha}{k-1})$}
\put(74,60){$C_{[\hskip 1mm{\underline{\bigtriangleup}}\hskip 1mm]}
= \frac{1+\alpha}{2 k (k+\alpha)(k+2\alpha)(k+3\alpha)}( 1-  
\frac{2\alpha^2}{
k-1})$}
\put(74,30){$C_{[{\angle \angle}]}= \frac{1}{k  
(k+\alpha)(k+2\alpha)(k+3\alpha)}(1 +
\frac{2\alpha}{k-1})$}
\put(74,0){$C_{[{\unrhd}]}= \frac{1}{k (k+\alpha)(k+2\alpha)(k+3\alpha)  
}( 1-
\frac{\alpha(\alpha-1)}{k-1})$}
\put(74,-30){$C_{[\hskip 1mm{\underline{\angle},\rm I}]} =  
\frac{1+\alpha}{2 k (k+\alpha)(k+2\alpha)
(k+3\alpha)}( 1 + \frac{3\alpha}{
k-1})$}
\put(74,-60){$C_{[\Lambda,\rm II]}= \frac{1+\alpha}{2 k  
(k+\alpha)(k+2\alpha)(k+3\alpha)}
( 1 + \frac{\alpha}{k+\alpha-1})(1+\frac{3\alpha}{k-1})$}
\put(74,-90){$C_{[\rm II,\rm I, \rm I]}= \frac{1+\alpha}{2 k  
(k+\alpha)(k+2\alpha)(k+3\alpha)}
( 1 -  
\frac{2\alpha}{k+\alpha-1}+\frac{7\alpha}{k-1}+\frac{6\alpha^2}{(k 
-1)(k-2)}
-\frac{2\alpha^2}{(k-2)(k+\alpha-1)})$}
\put(74,-120){$C_{[\rm II,\rm II]}= \frac{(1+\alpha)^2}{4 k  
(k+\alpha)(k+2\alpha)(k+3\alpha)}(
1 + \frac{4\alpha}{k-1} + \frac{2\alpha^2(2+\alpha)}{(1+\alpha)  
(k-1)(k+\alpha-1)})$}
\put(74,-150){$C_{[\rm III,\rm I]} = \frac{(1+\alpha)(1+2\alpha)}{6 k  
(k+\alpha)(k+2\alpha)(k+3\alpha)}
( 1 + \frac{3\alpha}{k-1})$}
\put(74,-180){$C_{[\Box]} = \frac{1}{k  
(k+\alpha)(k+2\alpha)(k+3\alpha)}( 1+ \frac{2\alpha}{k-1})$}
\put(74,-210){$C_{[\rm M]}=
\frac{1}{k(k+\alpha)(k+2\alpha)(k+3\alpha)}( 1+ \frac{\alpha}{k+\alpha-1})
(1+\frac{2\alpha}{k-1})$}
\put(74,-240){$C_{[\rm N,\rm I]}= \frac{1}{k  
(k+\alpha)(k+2\alpha)(k+3\alpha)}( 1+
\frac{8\alpha}{k-1}-\frac{4\alpha}{k-2}-\frac{\alpha^2}{(k+\alpha-1)(k 
-2)}
+\frac{\alpha(1+\alpha)(4 k +3\alpha -4)}{(k-1)(k-2)(k+\alpha-1)})$}
\put(74,-270){$C_{[\bigtriangleup,\rm I]} = \frac{1}{k  
(k+\alpha)(k+2\alpha)(k+3\alpha)}
( 1- \frac{\alpha(\alpha-3)}{k+\alpha-1} -  
\frac{\alpha^2(4\alpha-3)}{(k-2)(k+\alpha-1)}
- \frac{\alpha^2(3\alpha^2- 2\alpha + 3)}{(k-1)(k-2)(k+\alpha-1)})$}
\put(65,-300){(These 13 coefficients involve arbitrariness due to 4  
quartic identities.}
\put(65,-320){The determinations of these coefficients are given in  
Appendix D.)}
\end{picture}

\vskip 120mm

%\newpage

\begin{picture}(150,100)
%\put(30,100){\line(2,0){150}}
\put(74,80){\line(2,3){8}}
\put(76,80){\line(2,3){8}}
\put(78,80){\line(2,3){8}}
\put(87,93){\line(2,-3){8}}
\put(80,50){[\hskip 1mm$\underline{\underline{\angle}}$\hskip 1mm]}
\put(150,80){\line(2,3){8}}
\put(157,94){\line(2,-3){8}}
\put(160,94){\line(2,-3){8}}
\put(169,80){\line(2,3){8}}
\put(155,50){[\hskip 1mm$\underline{\amalg}$\hskip 1mm]}
\put(210,80){\line(2,3){8}}
\put(213,80){\line(2,3){8}}
\put(213,80){\line(2,0){17}}
\put(220,94){\line(2,-3){8}}
\put(215,50){[\hskip 1mm$\underline{\bigtriangleup}$\hskip 1mm]}
\put(280,80){\line(2,3){8}}
\put(283,80){\line(2,3){8}}
\put(290,94){\line(2,-3){8}}
\put(290,94){\line(2,0){17}}
\put(286,50){[$\models$]}
\put(350,80){\line(2,3){8}}
\put(353,80){\line(2,3){8}}
\put(360,94){\line(2,-3){8}}
\put(376,94){\line(2,-3){8}}
\put(355,50){[\hskip 1mm${\underline{\angle}},\rm I$]}
\put(74,-6){\line(2,3){8}}
\put(80,6){\line(2,-3){8}}
\put(80,6){\line(2,0){12}}
\put(90,6){\line(2,-3){8}}
\put(80,-20){[${\angle \angle}$]}
\put(150,-6){\line(2,3){8}}
\put(158,6){\line(2,-3){8}}
\put(150,-6){\line(2,0){14}}
\put(158,6){\line(2,0){14}}
\put(155,-20){[${\unrhd}$]}
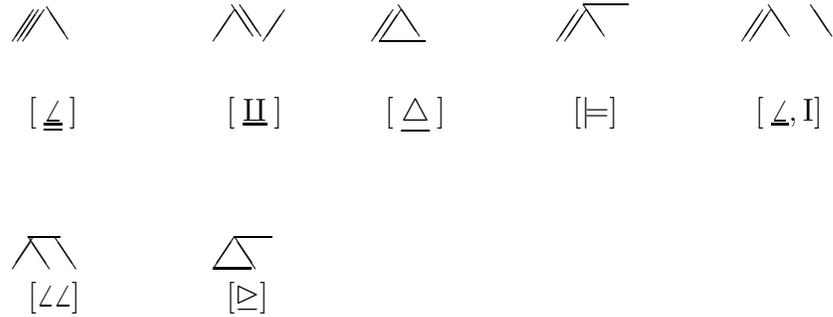
\put(40,-45){{\bf Fig.5.} symbols for the several graphs.}
\end{picture}
%*********************************************

\vskip 60mm

\section{The  $\tau$-expansion from  zonal polynomials
for $\beta=1$}

  The  Table B shows the  first coefficients of the WKB-expansion   for  
general
$\beta$. It was derived pertubatively assuming that $\beta$ was an even  
integer.
However the expression may be analytically continued to  all integers.
In this section we consider  $\beta=1$, which is of practical importance
since it is the case of a measure invariant under the orthogonal group  
(the
matrices
$X$ and
$\Lambda$ being real and symmetric).

The   HIZ integral (\ref{I}) may be
expanded in a series involving products of zonal polynomials $Z_{p}(X)$  
and $Z_{p}(\La)$, as

\ba\label{zonal1}
I_{\beta=1} &=& \int_{O(k)}[
\exp{{\rm{tr}} (X g \Lambda g^{-1})} dg\nonumber\\ &=&  
\sum_{m=0}^{\infty}
\frac{1}{m!}\int_{O(k)}[\tr (X g \Lambda g^{-1})]^{m} dg\nonumber\\
&=& \sum_{m=0}^{\infty} \frac{1}{m!}\frac{2^m m!}{(2m)!}\sum_{p(m)}
\chi_p(1)\int_{O(k)}Z_p(Xg\Lambda g^{-1})dg\nonumber\\
&=& \sum_{m=0}^\infty \frac{2^m}{(2m)!}\sum_{p(m)}
\chi_p(1) \frac{Z_p(X)Z_p(\Lambda)}{Z_p(I_k)}
\ea
where $p(m)$ is the partition of order m ;  $Z_p(X)$, the zonal  
polynomial,  is a symmetric
homogeneous polynomial of degree p in the $k$ eigenvalues of $X$  
\cite{James}.
The third equality in (\ref{zonal1}) follows from the identity
\be
({\rm tr} M)^q = c_q \sum_{p} \chi_p(1) Z_p(M)
\ee
where $c_q= \sum \chi_p(1)$. The sum over $p$ runs over the partitions  
of the number $q$, i.e.
over all Young tableaux with $q$ boxes. For the orthogonal group
$c_q = 2^q q!/(2q)!$. The function $Z_p(M)$ is a symmetric function of  
the eigenvalues of $M$ of
degree q.

In the unitary case, a similar expression in terms of a character  
expansion is well known ; it is
used explicitly in  \cite{Itzykson-Zuber} ;  there  $\chi_p(1)$ is the
dimension of the representation of the permutation group of $p$ objects  
corresponding to a given
Young tableau. (In an appendix, we give the general expression of this  
integral for arbitary
$\beta$, including the unitary case.)

The zonal polynomial $Z_p(X)$ beeing an homogeneous symmetric function  
of $x_1,...,x_k$, may be
expressed in terms of the sums
$s_n$ (n=1,2,3,...),
\be
s_n = {\rm tr} X^n = x_1^n + x_2^n + \cdots + x_k^n.
\ee
The zonal polynomial $Z_{p}(X)$, the constants $\chi_{p}(1)$, and
  $Z_{p}(\rm I)$, which is obtained by setting all $x_i=1$,
  are listed up to order five in the Appendix A, when one takes
$\alpha=\frac{2}{\beta}=2$.

 From (\ref{expf}) and (\ref{zonal1}), we have the relation,
\be\label{beta1}
  e^{  \sum_{i=1}^k x_i \lambda_i} f_{\beta=1} =  I_{\beta=1}(x_i,
\lambda_i)
\ee
For $\beta$=1,the Vandermonde
factor disappears, since it is raised to the power is $\beta - 1$.
For arbitrary $\beta$ there is a sum in (\ref{I}) over the $k!$  
permutations of the
$\lambda_i$. However, if one expands in powers both the exponential  
term, together with
$f_{\beta=1}$, it turns out that each term of given  order is a  
symmetric function of
the $\lambda_i$ and of the
$x_i$. Therefore the  sum over permutations in (\ref{expf}) is  not  
necessary for $\beta=1$.

In the previous section, using the coefficients $C$ of the Table B,  
setting $\beta=1$,we have
obtained the
$\tau$-expansion for
$f_{\beta=1}$. The same result may be derived from the zonal polynomial  
expansion for
$\beta=1$.

\ba\label{A1}
f_{\beta=1} &=& 1 - \frac{1}{k}(\tau_{12} + \cdots)
  + \frac{3}{2k ( k+2)}
(\tau_{12}^2 +
\cdots)
  + \frac{1}{k(k+2)} (\tau_{12}
\tau_{23} + \cdots)\nonumber\\
  &+& \frac{k+1}{k(k+2)(k-1)}(\tau_{12}\tau_{34} +
\cdots) + O(x^3)
\ea
%&& - \frac{5}{2k (k+2)(k+4)} (\tau_{12}^3 +
%\cdots)
%\nonumber\\
%&& - \frac{3}{2k (k+2)(k+4)} (\tau_{12}^2 \tau_{23} + \cdots )  
\nonumber\\
%&& -\frac{k+1}{k(k+2)(k+4)(k-1)}
%(\tau_{12}\tau_{23}\tau_{34} +
%\cdots)\nonumber\\
%&& - \frac{1}{k(k+2)(k+4)}(\tau_{12}\tau_{13}\tau_{14} +
%\cdots)\nonumber\\
%&& - \frac{k-5}{k(k+2)(k+4)(k-1)}
%(\tau_{12}\tau_{13}\tau_{23} + \cdots)\nonumber\\
%&& - \frac{3(k+3)}{2k(k+2)(k+4)(k-1)}
%(\tau_{12}^2 \tau_{34} + \cdots)\nonumber\\
%&& - \frac{k+3}{k(k+2)(k+4)(k-1)}
%(\tau_{12}\tau_{23} \tau _{45} + \cdots)\nonumber\\
%&& - \frac{k^2+3k-2}{k(k+2)(k+4)(k-1)(k-2)}
%(\tau_{12}\tau_{34}\tau_{56} + \cdots)
%+ O(\tau^4)
%\ea

Indeed, using the explicit values of the
character $\chi_p(1)$,the zonal polynomial expansion  (\ref{zonal1})  
reads
\ba\label{zonal2}
I_{\beta=1} &=& 1 + \frac{s_1(x)s_1(\lambda)}{k} + \frac{1}{6}[
\frac{((s_1(x))^2+2 s_2(x))((s_1(\lambda))^2+2 s_2(\lambda))}{k(k+2)}  
\nonumber\\
&+& 2 \frac{((s_1(x))^2- s_2(x))((s_1(\lambda))^2-  
s_2(\lambda))}{k(k-1)}] \nonumber\\
&+& \frac{1}{90}[ \frac{Z_{[3]}(x)Z_{[3]}(\lambda)}{k(k+2)(k+4)}
+ 9 \frac{Z_{[21]}(x)Z_{[21]}(\lambda)}{k(k+2)(k-1)} + 5
\frac{Z_{[1^3]}(x)Z_{[1^3]}(\lambda)}{k(k-1)(k-2)}]\nonumber\\
&+&
\cdots
\ea
For a comparison of this expansion with our earlier expressions  
$f_{\beta=1}$ , one must still
expand the exponential factor $e^{\sum x_i \lambda_i}$.
But this exponential factor has no explicit k dependence ; thus the  k  
-dependent coefficients in
$f_{\beta=1}$ will not be modified when one expands the exponential.  
For instance, at  first
order, the coefficient
$\frac{1}{k}$ is present both in (\ref{A1}) and (\ref{zonal2}) ;    
expanding the
exponential
$e^{\sum x_i
\lambda_i}$, we do find that the two expansions coincide.
At order two, one writes the coefficient of $\tau_{12}\tau_{34}$ as
\be\label{second}
\frac{k+1}{k(k+2)(k-1)} = \frac{1}{3k}(\frac{1}{k+2} + \frac{2}{k-1}),
\ee
then the terms of degree two in $e^{\sum x_i \lambda_i} f_{\beta=1}$  
have
coefficients which are either $\frac{1}{k(k+2)}$ or $\frac{1}{k(k-1)}$  
; they are
exacly  the inverse of the dimensional constant $Z_{[p]}(\rm I)$ in  
(\ref{zonal2}).

At order three, consider for instance $\tau_{12}\tau_{34}\tau_{56}$  
(three non-intersecting
lines). Taking in Table B the
coefficient of this term, one may decompose it as
\ba
   \frac{k^2 + 3 k - 2}{k(k+2)(k+4)(k-1)(k-2)} &=&\frac{1}{15 k  
(k+2)(k+4)} + \frac{9}{15
k(k+2)(k-1)}
\nonumber\\
&+& \frac{5}{15 k ( k-1)(k-2)}
\ea
which is the sum of $$\frac{1}{15 Z_{[3]}(I_k)} +  
\frac{9}{15Z_{[21]}(I_k)} +
\frac{5}{15 Z_{[1^3]}(I_k)}.$$ The numerators of  these terms are in  
the ratios 1:9:5, which are
   also the ratios of the characters  
$\chi_{[3]}(1):\chi_{[21]}(1):\chi_{[1^3]}(1)$.
(We will soon find the reason for this coincidence.)

Let us return to the derivation of the $\tau$-expansion from zonal  
polynomials.
We first shift   the diagonal matrices $X$ and $\Lambda$ to make them  
traceless,
\be
\Lambda = \frac{1}{k} {\rm tr} \Lambda + \tilde \Lambda
\ee
\be
X = \frac{1}{k}{\rm tr} X + \tilde X
\ee
In terms of eigenvalues, it is
\be
\tilde \lambda_a = \frac{1}{k}\sum_b (\lambda_a - \lambda_b)
\ee
Noting that tr $\tilde \Lambda$ = tr $\tilde X$ = 0, the HIZ integral is
\ba
I_{\beta=1} &=& e^{\frac{1}{k}({\rm tr} \Lambda)( {\rm tr} X)}\int {\rm  
d}g e^{{\rm tr}\tilde
\Lambda g \tilde X g^T}
\nonumber\\
&=& e^{\frac{1}{k}({\rm tr} \Lambda)( {\rm tr} X)}\sum_{m=0}^\infty  
\frac{2^m}{(2m)!}\sum
\chi_p(1)
\frac{Z_p(\tilde \Lambda) Z_p(\tilde X)}{Z_p(\rm I)}
\ea

 From the expressions of  zonal polynomials in terms of  the symmetric  
functions $s_n$ given in
Appendix A, noting that $s_1= \sum \tilde x_a=0$, and using simple  
identities such as
$\frac{1}{k}(\sum \lambda_i)(\sum x_i) = \sum \lambda_i x_i -  
\frac{1}{k}\sum_{i<j} (\lambda_i -
\lambda_j)(x_i - x_j)$, the integral becomes
\ba\label{tilde}
I &=& e^{\sum \lambda_i x_i - \frac{1}{k}\sum_{i<j}\tau_{ij}}[ 1 +  
\frac{1}{(k+2)(k-1)}s_2(\tilde
x)
s_2(\tilde \lambda)\nonumber\\
&+& \frac{4k}{3 (k+2)(k+4)(k-1)(k-2)}s_3(\tilde x)s_3(\tilde  
\lambda)\nonumber\\
&+& \frac{2k(k^2 + k + 2)}{(k+1)(k+2)(k+4)(k+6)(k-1)(k-2)(k-3)}  
s_4(\tilde x)s_4(\tilde
\lambda)\nonumber\\
&+& \frac{k^4 + 5 k^3 - 6 k^2 - 36 k + 72}{2 k  
(k+1)(k+2)(k+4)(k+6)(k-1)(k-2)(k-3)}(s_2(\tilde
x)s_2(\tilde \lambda))^2
\nonumber\\
&-& \frac{2 (2 k^2 + 3 k  
-6)}{(k+1)(k+2)(k+4)(k+6)(k-1)(k-2)(k-3)}\nonumber\\
&\times& ((s_2(\tilde x))^2 s_4(\tilde
\lambda) +
s_4(\tilde x)(s_2(\tilde \lambda))^2)
+
   (x^5)].
\ea

The paired product  $s_2(\tilde x) s_2(\tilde \lambda)$ is expressed in  
terms of
the $\tau_{ij}$ as
\ba\label{s2}
s_2(\tilde x)s_2(\tilde \lambda) &=&  
[\frac{1}{k}\sum_{i<j}(x_i-x_j)^2][\frac{1}{k}\sum_{i<j}
(\lambda_i - \lambda_j)^2]\nonumber\\
&=&  \frac{(k-1)^2}{k^2}\sum_{i<j} \tau_{ij}^2 -\frac{2(k-1)}{k^2}
   \sum \tau_{ij}\tau_{ik} + \frac{2}{k^2} \sum
\tau_{ij}\tau_{kl}
\ea
The second sum is restricted to $j<k$ and $i,j,k$ all different.
The last sum is restricted to $i<j,k<l$ and $i,j,k,l$ all different.
The identity (\ref{s2}) holds
for arbitrary $x_i$ and $\lambda_i$.
In order to fix the coefficients of this identity, one may  choose  
simple values of $X$ and
$\Lambda$.
Let us, for instance, take $X=\Lambda=(1,1,...,1,0,...0)$, where the  
eigenvalue one is
q-fold degenerate, and zero $(k-q)$.
Then we find
\be
s_2(\tilde x)s_2(\tilde \lambda) = \frac{q^2(k-q)^2}{k^2}
\ee
Note that  $\displaystyle{\frac{q^2(k-q)^2}{k^2}}$ is invariant under the replacement
$q\rightarrow k-q$.
The three terms of the products of $\tau$ in the r.h.s. of (\ref{s2})  
become $q(k-q)$,
$q(k-q)(k-2)/2$ and
$q(q-1)(k-q)(k-q-1)/2$
respectively. These numbers are also invariant by $q\rightarrow k-q$.
Then with three unknown coefficients, a,b and c, we write
\be\label{qk}
\frac{q^2(k-q)^2}{k^2} = a [q(k-q)] + b [\frac{q (k-q)(k-2)}{2}] + c
[\frac{q(q-1)(k-q)(k-q-1)}{2}] .
\ee
This relation fixes the three unknown coefficients :
$\displaystyle{a=\frac{(k-1)^2}{k^2},b=-\frac{2(k-1)}{k^2}}$ and $\displaystyle{
c=\frac{2}{k^2}}$.

In the Appendix B, we derive this identity with  the help of  
differential operators.
In the same appendix, we discuss the general expression of product of  
symmetric functions
$s_n(\tilde x) s_n(\tilde
\lambda)$  in terms of
$\tau$.

Remarkably, as may be checked  in the explicit expressions hereabove  
for the second order,
although each of the three quadratic expressions in $\tau$ in  
(\ref{s2}) is not invariant
under the permutations of $\lambda_i$, the  combination of the three  
types of  products of
$\tau_{ij}$  is a totally symmetric function as it should. To study  
further this
invariance under permutation, we consider
\ba
X &=& (1,...,1,0,......,0)\nonumber\\
\Lambda &=& (0,......,0,1,...,1)
\ea
where the eigenvalue one is q-fold degenerate, and we choose $q<k$.  
This is a particular
permutation of the previous choice of $X=\Lambda = (1,...1,0,...,0)$.  
We assume
$q< k/2 $. The various $\tau$  are not the same as in (\ref{qk}) ; the  
three quadratic terms are
now respectively
$[q^2],[q^2(q-1)]$ and $[q^2 (q-1)^2/2]$. For this particular  
permutation of the $\lambda_i$,
the  identity
\be\label{qq}
\frac{q^2(k-q)^2}{k^2} = \frac{(k-1)^2}{k^2}[q^2] -  
\frac{2(k-1)}{k^2}[q^2 (q-1)] + \frac{2}{k^2}
[\frac{q^2(q-1)^2}{2}]
\ee
is indeed consistent with (\ref{s2}).

The equation (\ref{qq}) is interesting since, in this choice of  
permutation, the value of
the $\tau$'s  are all function of q and not of k.If we set k=q, this  
equation shows a
non-trivial cancellation for three terms, since the left hand side  
vanishes ; in this case the
right hand side of ((\ref{qq}) vanishes as
\be
(1 - 2 + 1)q^2 (q-1)^2 = 0
\ee
  To summarize we have found a method to write $s_2(\tilde x)s_2(\tilde  
\lambda)$ in terms
of $\tau$'s. Given that it is a function of the $\tau$'s the method  
consists of (i) evaluation of
$s_2(\tilde x)s_2(\tilde
\lambda)$ for the choice $X=(1,...,1,0,...,0)$,  
$\Lambda=(0,...,0,1,...,1)$, where 1 is q-fold
degenerate.(ii) Decompose this value as
$\frac{q^2}{k^2}[(k-1)^2 - 2 (q-1)(k-1) + (q-1)^2]$. (iii)Evaluate the  
$\tau$ terms for this
same choice  as a function of q. (iv) The comparison with the expansion  
of (iii) fixes the
unknown  coefficients.

Similar identities expressing the symmetric functions $s_n(\tilde  
X)s_n(\tilde \La) $ in terms of
  $\tau$'s hold at  higher order.
At order three
since $s_1(\tilde X)=0$, $Z_p(\tilde X)$ is given  only by $s_3(\tilde  
X)$ (see Appendix A).
In order to express  $s_3(\tilde x)s_3(\tilde \lambda)$, into the  
$\tau_{ij}=(x_i-
x_j)(\lambda_i - \lambda_j)$ polynomials,
in the Appendix B, we present a method for deriving these identities
based on a  differential operator $D_{l,m,n}^{i,j,k}$. Let us for  
instance quote the result for
the third order
\ba\label{s3}
s_3(\tilde X)s_3(\tilde \Lambda) &=& - \frac{1}{k^4}(k-1)^2 (k-2)^2 (  
\tau_{12}^3 +
\cdots)\nonumber\\
&+& 3 \frac{(k-1)(k-2)^2}{k^4}(\tau_{12}^2 \tau_{23} +  
\cdots)\nonumber\\
&+& 3 \frac{(k-2)^3}{k^4} (\tau_{12}\tau_{23}\tau_{13} + \cdots  
)\nonumber\\
&-& 12 \frac{(k-1)(k-2)}{k^4} (\tau_{12}\tau_{13}\tau_{14} +  
\cdots)\nonumber\\
&-& 6 \frac{(k-2)^2}{k^4}(\tau_{12}\tau_{23}\tau_{34} +  
\cdots)\nonumber\\
&-& 3 \frac{(k-2)^2}{k^4} (\tau_{12}^2 \tau_{34} + \cdots)\nonumber\\
&+& 12 \frac{(k-2)}{k^4}(\tau_{12}\tau_{23}\tau_{45} +  
\cdots)\nonumber\\
&-& \frac{24}{k^4}(\tau_{12}\tau_{34}\tau_{56} + \cdots).\nonumber\\
\ea
In this expression we have used the freedom given by the cubic identity  
among the $\tau$'s to
remove the ambiguities.

However the result may be obtained  also by the direct method which was  
used above for the
quadratic case. One first choose
$X=\Lambda=(1,..,1,0,...,0)$ with eignevlue one q-fold degenerate ;  
then $s_3(\tilde
X)s_3(\tilde
\Lambda) =q^2(q-k)^2( 2 q - k)^2/k^4$, which is invariant by the  
substitution of $q\rightarrow
k-q$. For instance, the last term of non-intersecting graph  
of$(\tau_{12}
\tau_{34}\tau_{56} + perm.)$ becomes  
$q(q-1)(q-2)(k-q)(k-q-1)(k-q-2)/6$.  Each of the  8 terms
is invariant by  the substitution of $q\rightarrow k-q$. Therefore the  
coefficients have to be
functions of k only.
Then, we apply again the permutation
$\Lambda = (0,...,0,1,...,1)$.
The
  various $\tau$ terms are  polynomial in  q. The remarkable
factorization identity,
\ba\label{6term}
&&\frac{q^2(q-k)^2(2q-k)^2}{k^4} = -\frac{24}{k^4}[ \frac{1}{6}q^2  
(q-1)^2 (q-2)^2]\nonumber\\
&+& 12 \frac{k-2}{k^4}[q^2(q-1)^2(q-2)]
 - 9\frac{(k-2)^2}{k^4}[q^2(q-1)^2]\nonumber\\
&-& 12 \frac{(k-1)(k-2)}{k^4}[\frac{1}{3}q^2 (q-1)(q-2)]
 + 3\frac{(k-1)(k-2)^2}{k^4}
[ 2 q^2(q-1)]\nonumber\\
&-& \frac{(k-1)^2(k-2)^2}{k^4}[q^2]
\ea
extension of the analogous quartic identity, fixes the unknown  
coefficients.

As discussed in the previous section,  the cubic identity
\be\label{iid}
I_3=[\tau_{12}^2\tau_{34}] - [\tau_{12}\tau_{23}\tau_{34}] + (k-3)[
\tau_{12}\tau_{13}\tau_{24}]=0
\ee
where $[\tau_{12}^2\tau_{34}]= \tau_{12}^2\tau_{34} + \cdots$ leads to  
an ambiguity, since one
may add $I_3$ with an  arbitrary coefficient
$\alpha$.
At order three, one has from (\ref{tilde}),
\ba\label{I3}
&&\tilde I^{(3)} = \frac{1}{6 k (k+2)(k+4)(k-1)(k-2)}\nonumber\\
&&\times[
(k^2 + 3 k -2) q^6 + (-12 k -24)q^5 + (6 k^2 + 12 k + 56)q^4 -48 q^3 +  
8 k^2 q^2]\nonumber\\
\ea
By taking the choice $X=(1,...,1,0,....,0)$ and  
$\Lambda=(0,...,0,1,...,1)$, we evaluate
$[\tau_{12}\tau_{34}\tau_{56}]= \frac{1}{6}q^2(q-1)^2(q-2)^2$. By  
comparing the highest order of
q,
$q^6$, we find
\be
C_{[\rm I,\rm I,\rm I]} = - \frac{k^2 + 3 k -2}{k(k+2)(k+4)(k-1)(k-2)}
\ee
Subtracting this $C_{[\rm I,\rm I,\rm I]}[\tau_{12}\tau_{34}\tau_{56}]$  
from $\tilde I^{(3)}$,
we have
\ba\label{Iq53}
  \tilde I_{q^5}^{(3)} &=& \frac{1}{6 k (k+2)(k+4)(k-1)(k-2)}\nonumber\\
&\times&[(6 k +18)q^5 + (-7 k - 41) q^4 + (12 k +12)q^3 + (4 k-4)q^2]
\ea
The quantity $[\tau_{12}\tau_{13}\tau_{45}]$ becomes $q^2(q-1)^2(q-2)$  
and it is order $q^5$.
Thus we determine its coefficient $C_{[\Lambda,\rm I]}$ from the $q^5$  
term in (\ref{Iq53}),
\be
C_{[\Lambda,\rm I]} = - \frac{(k+3)}{k(k+2)(k+4)(k-1)}
\ee
We subtract $C_{[\Lambda,\rm I]}[\tau_{12}\tau_{13}\tau_{45}]$ from  
(\ref{Iq53}), and
we obtain
\be\label{Iq43}
\tilde I_{q^4}^{(3)} = - \frac{(17 k + 31)q^4 + (-18k -78)q^3 + (16 k +  
32)q^2}{6 k (k+2)(k+4)(k-1)}
\ee
The terms, which give order $q^4$, are $[\tau_{12}^2 \tau_{34}]=  
q^2(q-1)^2$,
$[\tau_{12}\tau_{13}\tau_{34}]=q^2(q-1)^2$ and
$[\tau_{12}\tau_{13}\tau_{14}]=\frac{1}{3}q^2(q-1)(q-2)$.
($[\tau_{12}\tau_{23}\tau_{13}]$ becomes vanishing.)
Thus we have
\be
\frac{1}{3}C_{[\rm Y]} + C_{[\rm II,\rm I]} + C_{[\rm N]} = - \frac{17  
k + 31}{6k(k+2)(k+4)(k-1)}
\ee
Since $C_{[\rm Y]}$ is
\be
C_{[\rm Y]}= - \frac{1}{k(k+2)(k+4)}
\ee
we have
\be
C_{[\rm II,\rm I]} + C_{[\rm N]} = - \frac{5 k + 11}{2 k  
(k+2)(k+4)(k-1)}
\ee
We write
\be\label{910}
\frac{5 k + 11}{2 k (k+2)(k+4)(k-1)} =
\frac{9}{10 k (k+2)(k+4)} + \frac{8}{5 k (k+2)(k-1)}
\ee
In principle, the coefficients $C_{[\rm II,\rm I]}$, $C_{[\rm N]}$ and  
$C_{[\bigtriangleup]}$
have ambiguities, since there is an identity of (\ref{iid}).
One method of fixing this ambiguity may be large k assumption:
We assume
\be\label{c21}
C_{[\rm II,\rm I]} = -\frac{3}{2}[\frac{a}{k(k+2)(k+4)} +  
\frac{1-a}{k(k+2)(k-1)}]
\ee
and
\be\label{cN}
C_{[\rm N]} = -[\frac{c}{k(k+2)(k+4)} + \frac{1-c}{k(k+2)(k-1)}]
\ee
We assume that $C_{[\rm II,\rm I]}\sim \frac{3}{2}\frac{1}{k^3}$ in the  
large k limit The factor
$\frac{3}{2}$ is the degereracy factor $=(\beta/4-1)/(\beta/2-1)$ for  
$\beta=1$ (
in the next section, we will discuss this factor by the substitution of  
$\alpha = 2/(2 - \beta)$)
  . And
$C_{[\rm N]}\sim \frac{1}{k^3}$ also. Then, the comparison with  
(\ref{910}) gives
\be\label{ac}
\frac{3}{2} a + c = \frac{9}{10}
\ee
Still one parameter $a$ or $c$ remains undetermined.
The coefficient $C_{[\bigtriangleup]}$ is determined by the
residual equation of (\ref{c21}).
However, the ambiguity constant $a$ remains.
When we consider the case $\beta=4$, we have a definite value of  
$C_{[\bigtriangleup]}$,
since
there is no double line in $\beta=4$, and no cubic equation exists.
For the general value of $C_{[\bigtriangleup]}$, extending the case of  
$\beta=4$, we write
\be\label{triangleassump}
C_{[\bigtriangleup]} = - \frac{1}{k (k+\alpha)(k+2\alpha)}[1 -  
\frac{\alpha^2}{k-1}]
\ee
Then, using the residual equation of (\ref{resid3a}), and  
(\ref{resid3b}), we determine the
constant
$a$ and $c$ in (\ref{c21}) and (\ref{cN}) as
\be
a= \frac{1}{5}, \hskip 5mm c=\frac{3}{5}
\ee
which also satisfy the large k limit condition (\ref{ac}). In other  
words, we have assumed
the value of $C_{[\bigtriangleup]}$ as (\ref{triangleassump}), since we  
are free to choose
of the coefficients of the cubic identity.

Note that when we subtract $[C_{[\rm II,\rm
I]} + C_{[\rm N]}]
q^2(q-1)^2$ from  (\ref{Iq43}), we obtain
\be
\Delta I^{(3)}= \frac{2 q^4 + 12 q^3 + q^2}{6 k (k+2)(k+4)}
\ee
where there is no $\frac{1}{k-1}$ factor.
This quantity coincides with  the sum of $C_{[\rm III]}[\tau_{12}^3]$,  
$C_{[\hskip 1mm
\underline{\angle}\hskip 1mm]}[\tau_{12}^2\tau_{13}]$
and $C_{[\rm Y]}[\tau_{12}\tau_{13}\tau_{14}]$, since they are
\be
C_{[\rm III]} = -\frac{5}{2}\frac{1}{k(k+2)(k+4)}
\ee
\be
C_{[\hskip 1mm\underline{\angle}\hskip 1mm]} =  
-\frac{3}{2}\frac{1}{k(k+2)(k+4)}
\ee
\be
C_{[\rm Y]} = -\frac{1}{k(k+2)(k+4)}
\ee
where $[\tau_{12}^3]=q^2$,$[\tau_{12}^2\tau_{34}]=2 q^2(q-1)$ and  
$[\tau_{12}\tau_{13}\tau_{14}
]=\frac{1}{3}
q^2(q-1)(q-2)$.

%%%%%%%%%%%%%%%%%%%%%%%%%%%%%%%%%%%%%%%%%%%%%%%%%%%%%%%%%%%%%%%%%%%%%%%% 
%%%%%%%%%

It may be interesting to investigate the fourth order term.
In (\ref{tilde}), we use the choice that $X=\Lambda=(1,...,1,0,...,0)$
where 1 enries q-times, 0 entries (k-q) times. Then, noting that  
$\displaystyle{s_2(\tilde x)s_2(\tilde \lambda) =
\frac{q^2(k-q)^2}{k^2}}$,\\ $\displaystyle{s_3(\tilde x)s_3(\tilde \lambda) =  
\frac{q^2(q-k)^2(2 q-k)^2}{k^4}}$,
$\displaystyle{s_4(\tilde x)s_4(\tilde \lambda) = \frac{q^2(k-q)^2(k^2 - 3 k q + 3  
q^2)^2}{k^6}}$,\\
$\displaystyle{[s_2(\tilde x)s_2(\tilde \lambda)]^2 = \frac{q^4(k-q)^4}{k^4}}$,\\
  $\displaystyle{(s_2(\tilde x))^2 s_{4}(\tilde \lambda)
+ s_4(\tilde x)(s_2(\tilde \lambda))^2=  \frac{2q^3 (k-q)^3 (k^2 - 3 k  
q + 3 q^2)}{k^5}}$, \\we have
the fourth term together with  the expansion of  
$\displaystyle{e^{-\frac{1}{k}\sum_{i<j}\tau_{ij}} =
e^{\frac{q^2}{k}}}$,

\ba\label{order4}
\tilde I^{(4)} &=&\frac{1}{24 k  
(k+1)(k+2)(k+4)(k+6)(k-1)(k-2)(k-3)}\nonumber\\
&\times&[(k^4 + 7 k^3 + k^2 - 35 k - 6)q^8
+(-24 k^3 - 144 k^2 - 72 k + 144)q^7\nonumber\\
  &+& (12 k^4 + 72 k^3 + 308 k^2 + 1064 k + 864)q^6\nonumber\\
&+&(-240 k^3 -1008 k^2 -2304 k - 1728)q^5 \nonumber\\
&+& (44 k^4 + 188 k^3 + 1944 k^2 + 2064 k + 96)q^4\nonumber\\
&+&(-576 k^3 - 672 k^2 - 192 k )q^3 + (48 k^4 + 48 k^3 + 96 k^2)q^2]
\ea

There are 23 terms in the fourth order for the $\tau$ expansion as  
shown in Table B.
For the choice of $X=(1,...,1,0,...,0)$, $\Lambda=(0,...,0,1,...,1)$,
each $\tau$ terms are evaluated  as  they are  polynomials
of $q$, and each terms have different orders of q. The highest order of  
q is obtained from
the term $[\tau_{12}\tau_{34}\tau_{56}\tau_{78}]$, (here $[\tau  
\cdots]$ means the sum of the
permutation of the indecies). It becomes
\be
[\tau_{12}\tau_{34}\tau_{56}\tau_{78}] =  
\frac{1}{4!}q^2(q-1)^2(q-2)^2(q-3)^2
\ee
The order $q^8$ term is only this term. Then, from (\ref{order4}), we  
obtain the coefficient of
this term $[\tau_{12}\tau_{34}\tau_{56}\tau_{78}]$, as
\ba\label{ciiii}
&&C_{[{\rm I,\rm I,\rm I,\rm I}]} =
\frac{(k+3)(k^2 + 6 k + 1)}{k (k+1)(k+2)(k+4)(k+6)(k-1)(k-3)}\nonumber\\
&=& \frac{1}{k(k+2)(k+4)(k+6)}[ 1 + \frac{12}{k-1} +  
\frac{40}{(k-1)(k-3)}
+ \frac{4}{(k+1)(k-3)}]\nonumber\\
\ea
This result is expressed by the use of the dimensional constant  
$Z_p(\rm I)$ and by the constant
$\chi_p (1)$, which appear in the zonal polynomial expansion,
\be
C_{[{\rm I,\rm I,\rm I,\rm I}]} = \frac{1}{105} \sum_p  
\frac{\chi_p(1)}{Z_p(\rm I)}
\ee

The next highest order term of $q^7$ is  
$[\tau_{12}\tau_{13}\tau_{45}\tau_{67}]$, which has a
factor $C_{[\Lambda, \rm I,\rm I]}$. The value of  
$[\tau_{12}\tau_{13}\tau_{45}\tau_{67}]$ is
$\frac{1}{2}q^2 (q-1)^2 (q-2)^2(q-3)$. Then from the coefficient of  
$q^7$ in
(\ref{order4}), we obtain the coefficient $C_{[\Lambda, \rm I,\rm I]}$,
\ba\label{clii}
&&\frac{1}{2}C_{[\Lambda, \rm I,\rm I]}- \frac{1}{2}C_{[\rm I,\rm I,\rm  
I,\rm I]}=
\frac{1}{24 k (k+1)(k+2)(k+4)(k+6)(k-1)(k-2)(k-3)}\nonumber\\
&& \times (-24 k^3 - 144 k^2 - 72 k + 144)
\ea
which reads
\ba
&&C_{[\Lambda, \rm I,\rm I]} = \frac{k^3 + 8 k^2 + 13 k  
-2}{k(k+1)(k+2)(k+4)(k+6)(k-1)(k-2)}
\nonumber\\
&=& \frac{1}{k(k+2)(k+4)(k+6)}[1 + \frac{10}{k-1} +  
\frac{20}{(k-1)(k-2)} + \frac{4}{(k+1)(k-2)}]
\nonumber\\
\ea

We note here that the following relation of (\ref{max1}), which was  
derived for the
arbitrary value of $\beta$ in the section 5, is satisfied by above
two expressions,
\be
  C_{[\rm I,\rm I,\rm I]} + 3 C_{[\Lambda,\rm I,\rm I]} + (k-3)C_{[\rm  
I,\rm I,\rm I,\rm I]} = 0
\ee
since we have from Table B, $(\beta=1)$,
$C_{[\rm I,\rm I,\rm I]} = - \frac{1}{k(k+2)(k+4)}( 1 + \frac{6}{k-1} +  
\frac{8}{(k-1)(k-2)})$.
The pole term $\frac{1}{k+1}$ in $C_{[\rm I,\rm I,\rm I,\rm I]} $ is  
cancelled by the pole term of
$C_{[\Lambda,\rm I,\rm I]}$. This $\frac{1}{k+1}$ factor comes only  
from the dimension of the zonal
polynomial $Z_{[2^2]}$, and it appears first in the fourth order terms.

For the terms, which give order $q^6$, new 4 terms
$[\tau_{12}\tau_{13}\tau_{34}\tau_{56}]= q^2(q-1)^2(q-2)^2$,
$[\tau_{12}\tau_{13}\tau_{14}\tau_{56}]=\frac{1}{3}q^2(q-1)^2(q-2)(q 
-3)$,
$[\tau_{12}\tau_{13}\tau_{45}\tau_{46}]= \frac{1}{4}q^2(q-1)^2(q-2)^2(2  
q -5)$,
$[(\tau_{12})^2 \tau_{34}\tau_{56}]= \frac{1}{2}q^2(q-1)^2(q-2)^2$  
appear.

  From (\ref{order4}), after the subtraction of  the contribution of
$[\tau_{12}\tau_{34}\tau_{56}\tau_{78}]$ and  
$[\tau_{12}\tau_{13}\tau_{45}\tau_{67}]$, the
$q^6$ part of (\ref{order4}) becomes
\be\label{cq6}
C_{q^6}=\frac{31 k^3 + 196 k^2 + 119 k - 310}{12 k  
(k+1)(k+2)(k+4)(k+6)(k-1)(k-2)}
\ee
which have to be the sum of these four terms.
to determine these four coefficients, we fist fix two coefficients  
$C_{[\rm Y,\rm I]}$ and
$C_{[\Lambda,\Lambda]}$ by the residual equation of (\ref{max2}).

These are satisfied by the values of
\be
C_{[\rm Y,\rm I]}= \frac{1}{k(k+2)(k+4)(k+6)}( 1 + \frac{6}{k-1})
\ee
\be
C_{[\Lambda,\Lambda]}= \frac{1}{ k(k+2)(k+4)(k+6)}( 1 + \frac{8}{k-1} +
\frac{8}{(k-1)(k+1)})
\ee

Then, we subract these two contributions from (\ref{cq6}), and the  
remaining two coefficients are
\be\label{NIII}
C_{[\rm N,\rm I]} + \frac{1}{2}C_{[\rm II,\rm I,\rm I]} =
\frac{7k^3 + 48 k^2 + 51 k - 30}{4 k (k+1)(k+2)(k+4)(k+6)(k-1)(k-2)}
\ee

 From the residual equation of (\ref{max4}), we are able to confirm  
above result.

There are ambiguities in the fourth order by the 4 quartic identities
as shown in appendix C ((\ref{quarticI}) $\sim$ (\ref{quarticIV})).
Due to the quartic identity (\ref{quarticI}), it is free to choose the  
value of $C_{[\rm II,\rm I,\rm I]}$.
  Here we assume that $C_{[\rm II,\rm I,\rm I]}$ is same as
$C_{[\Lambda,\rm I,\rm I]}$ except a factor of the multiple line,  
$\frac{3}{2}$.
\be
C_{[\rm II, \rm I,\rm I]} = (\frac{3}{2})\frac{k^3 + 8 k^2 + 13 k  
-2}{k(k+1)(k+2)(k+4)(k+6)(k-1)(k-2)}
\ee
Then we have
\be
C_{[N,\rm I]}=\frac{k^3 + 6 k^2 + 3 k -  
6}{k(k+2)(k+4)(k+6)(k+1)(k-1)(k-2)}
\ee
These two values satisfy (\ref{NIII}).

The limit of $q\rightarrow 1$ in (\ref{order4}) determines uniquely the  
coefficient of the term
$(\tau_{12})^4$, which is four line degeneracy, and it has a value  
$q^2$. Other terms are
proportional to $q-1$. The coefficient, therefore, becomes
\be
C_{[\rm IIII]} = \frac{35}{8 k(k+2)(k+4)(k+6)}.
\ee

The limit $q \rightarrow 2$, in (\ref{order4}), after the subtractions  
of
$[\tau_{12}]^4$,
becomes
\be
\tilde I^{(4)}=\frac{125 k^2 + 320 k +291}{2 k  
(k+1)(k+2)(k+4)(k+6)(k-1)}
\ee
which is the sum of $[\tau_{12}\tau_{23}\tau_{14}\tau_{34}]$,
  $[\tau_{12}(\tau_{13})^2\tau_{34}]$,
$[(\tau_{12})^2\tau_{23}\tau_{34}]$,
$[(\tau_{12}\tau_{13})^2]$, $[(\tau_{12})^3\tau_{13}]$,
$[(\tau_{12})^3 \tau_{34}]$,$[(\tau_{12})^2(\tau_{34})^2]$, which have  
values of
$\frac{1}{4}q^2(q-1)^2$,$q^2(q-1)^2$,$2  
q^2(q-1)^2$,$q^2(q-1)$,$2q^2(q-1)$,$q^2(q-1)^2$, and
$\frac{1}{2}q^2(q-1)^2$,respectively. By the summation of these terms  
with the coefficients
we have for q=2,
\ba
&&\tilde I^{(4)}=\frac{1}{k(k+2)(k+4)(k+6)}[(1 +\frac{4}{k-1}) + 6(1 +  
\frac{2}{k-1})\nonumber\\
&+& 12(1 + \frac{4}{k-1})+ 9 + 20 + 10 (1 + \frac{6}{k-1}) +  
\frac{9}{2}(1 + \frac{8}{k-1} +
  \frac{32}{3(k-1)(k+1)})]\nonumber\\
\ea
where 7 terms are added respectively.
We derive the following expression, which involves $\frac{1}{k+1}$  
pole. Other terms
are easily derived.
\be
C_{[\rm II,\rm II]} = \frac{9}{4}\frac{1}{k(k+2)(k+4)(k+6)}( 1 +  
\frac{8}{k-1} + \frac{32}{3(k-1)(k+1)})
\ee
and
\be\label{3-1}
C_{[\rm III,\rm I]} = \frac{5}{2k(k+2)(k+4)(k+6)}(1 + \frac{6}{k-1})
\ee
We have following equations,
\be
C_{[\rm III]} + 2 C_{[\hskip 1mm \underline{\underline{\angle}}\hskip  
1mm]}
- C_{[\hskip 1mm \underline{\bigtriangleup}\hskip 1mm]}
+ C_{[\hskip 1mm \underline{\amalg}\hskip 1mm]} + (k-2)C_{[\rm III,\rm  
I]} = 0
\ee
\be
C_{[\hskip 1mm \underline{\angle}\hskip 1mm]}- C_{[\rm III]}
+ (k-3)C_{[\hskip 1mm \underline{\amalg}\hskip 1mm]} - (k-3)C_{[\rm  
III,I]} + 2 C_{[\hskip 1mm
\underline{\bigtriangleup}\hskip 1mm]} = 0
\ee
One could eliminate the coefficientof $C_{[\hskip 1mm  
\underline{\bigtriangleup}\hskip 1mm]}$,
  the triangle diagram with one double
line. Then, we know that $C_{[\rm III,\rm I]}$ has no pole of  
$\frac{1}{k+1}$. This is consistent
with the expression (\ref{3-1}).

We have up to now confirmed 14 terms at fourth order. The number of   
fourth order graphs
is 23, and
among them, three graphs exist, which contain the triangles. These  
terms are vanishing for the
present choice of $X$ and $\Lambda$. Thus, 6 terms still have not yet  
been determined.
These 6 terms are  
$[\tau_{12}\tau_{13}\tau_{14}\tau_{15}]$,$[\tau_{12}\tau_{13}\tau_{14}
\tau_{45}]$,
$[\tau_{12}\tau_{23}\tau_{34}\tau_{45}]$,$[(\tau_{12})^2\tau_{13}\tau_{1 
4}]$,$[(\tau_{12})^2 \tau_{13}
\tau_{45}]$ and $[\tau_{12}\tau_{13}(\tau_{45})^2]$. They have values  
of $\frac{1}{12}q^2(q-1)
(q-2)(q-3)$, $q^2 (q-1)^2(q-2)$,$q^2(q-1)^2(q-2)$,$q^2(q-1)(q-2)$,  
$2q^2(q-1)^2(q-2)$, and
$2q^2(q-1)^2(q-2)$. The sum of these 6 terms should be
the value of (\ref{order4}), once one subtracts the subtractions the  
previous 14 terms.

 From the equations of the residues, we obtain
\be\label{lambdaM}
C_{[\Lambda,\rm II]} + C_{[\rm M]} = \frac{(k+3)(5 k + 21)}{2 k  
(k+1)(k+2)(k+4)(k+6)}
\ee
which satisfy the residue equation of (\ref{max3}). Note that $C_{[\rm  
M]}$ is not uniquely
determined in views of  the quartic identity discussed in appendix C.

For  fixing this ambiguity,
we take the value of $C_{[\Lambda, \rm II]}$ to be the same as  
$C_{[\Lambda,\Lambda]}$
within a factor $\frac{3}{2}$,
\be
C_{[\Lambda,\rm II]} = \frac{3}{2} \frac{k^3 + 6 k^2 - k -  
38}{k(k+1)(k+2)(k+4)(k+6)}
\ee
Note there is no $\frac{1}{k-2}$ factor in the expression as same as  
(\ref{lambdaM}).
 From (\ref{lambdaM}), $C_{[\rm M]}$ is determined as
\be
C_{[\rm M]} = \frac{(k+3)^2}{k(k+2)(k+4)(k+6)(k+1)(k-1)}
\ee

In the next section and in Appendix D, we discuss other terms for  
general $\beta$ (or $\alpha$).

As a summary of this section,  from the zonal polynomial expansion of  
the HIZ-integral,
  we have obtained the series given by the symmetric polynomials of  
(\ref{tilde}).
  By writing  these
symmetric polynomials $s_n(\tilde x)$ in terms of $\tau$, we have  
obtained
the $\tau$-expansion of the
HIZ-integral in the form  $e^{\sum x_i \lambda_i} f$, where $f$ is  
given as (\ref{A1}).
We have found that the coefficients are expressed by  sums of the  
inverse of the
dimensional constants.
We have shown that $e^{\sum x_i \lambda_i} f$ is invariant under the  
permutations of $\lambda_i$.
Therefore, the summation of the permutation contained in (\ref{DD2}) is  
not necessary for $\beta=1$.

\section{ Character expansion for general $\beta$ and a duality}

In this section, we generalize the results of the previous section  to  
arbitrary $\beta$, and
we discuss the derivations of the $\tau$-expansion using a dual  
representation.

We write the zonal (Jack) polynomial expansion for the integral  
$I_{\beta}$,
\be
I_{\beta} = e^{\sum x_i \lambda_i -\frac{1}{k}\sum_{i<j} \tau_{ij}}[
\sum_{m=0}^\infty \frac{1}{m!}\frac{1}{\prod_{q=0}^{m-1}(1 +  q  
\alpha)}\sum_p \chi_p(1) \frac{
Z_p(\tilde X)Z_p(\tilde \Lambda)}{Z_p(\rm I)}]
\ee
where we have shifted $x_i \rightarrow \tilde  x_i = x_i -  
\frac{1}{k}\sum_{i=1}^k x_i$
(subtraction of the mean).

The generalized zonal polynomial $Z_p(X)$ has a parameter $\alpha$,  
which is equal to the present
  $\frac{2}{\beta}$.
This polynomial is called as Jack polynomial.
As for  $\beta=1$, we expand in a power series for small $X$ and  
$\Lambda$.
  The quantity, which is an generalization of the integral (\ref{I}) to  
arbitrary $\beta$,
is a symmetric function by interchange of the $x_i$ and $\lambda_i$.
We shall write this quantity $I_{\beta}$ in terms of $\tau_{ij}= (x_i -  
x_j)(\lambda_i - \lambda_j)$.

\be\label{withoutperm}
I_{\beta} = e^{\sum_{i=1}^k x_i \lambda_i}[ 1 +  D_{[\rm I]}  
[\tau_{12}] + D_{[\rm I,\rm I]}
[\tau_{12}\tau_{34}] + \cdots]
\ee

This expansion is different from the form discussed in  section 2,  
which was
\be\label{withperm}
I_{\beta} = \sum_{perm.} \frac{e^{\sum x_i \lambda_i}}{\prod_{i<j}^k  
(\tau_{ij})^{\beta-1}}
[ 1 + C_{[\rm I]} [\tau_{12}] + C_{[\rm I,\rm I]}[\tau_{12}\tau_{34}] +  
\cdots]
\ee
The differences are that (\ref{withperm}) has a Vandermonde  
determinant, and also it has a sum
about the permutations of the $\lambda_i$. The expression
   (\ref{withoutperm}) has no Vandermonde term and no
  sum over permutations.
Therefore, the coefficients $C_{[graph]}$ and $D_{[graph]}$ are not the  
same  in these
  two expansion.  (The $\beta=1$
case is exceptional since there the two expansions coincide).

To obtain the coefficients $D$ in the expansion (\ref{withoutperm}), we
take $X$ and $\Lambda$ as
$X=(1,...,1,0,...,0)$ and $\Lambda = (0,...,0,1,...,1)$, where 1 is  
q-fold degenerate, and
$q < k/2$.
Since $\sum_{i<j}\tau_{ij}$ becomes
$- q^2$, with this choice, we have
up to  second order,
\ba\label{g1}
I &=& e^{\sum x_i \lambda_i} [ 1 + \frac{q^2}{k} + \frac{q^4}{2 k^2} +  
\cdots]\nonumber\\
   &\times& [ 1 + \frac{1}{2} \frac{1}{(1 + \alpha)}( \chi_{[2]}(1)  
\frac{Z_{[2]}(\tilde X)Z_{[2]}(
\tilde \Lambda)}{
Z_{[2]}(\rm I)} + \chi_{[1^2]}(1) \frac{Z_{[1^2]}(\tilde  
X)Z_{[1^2]}(\tilde \Lambda)}{Z_{[1^2]}(\rm I)}) + \cdots]
\nonumber\\
&=& e^{\sum x_i \lambda_i}[ 1 + \frac{q^2}{k} + \frac{q^4}{2 k^2} +  
\frac{\alpha}{2 ( k+ \alpha)(k-1)}
s_2(\tilde x) s_2(\tilde \lambda) + O(x^3)]
\ea

The symmetric functions $s_n(x)= \sum x_i^n$ are easily evaluated  
within this choice of $X$,
  and we
obtain  $\displaystyle{s_2(\tilde x) = q ( 1 - \frac{q}{k})^2 + (k-q)  
(-\frac{q}{k})^2 = \frac{q(k-q)}{k}}$,
and thus
$\displaystyle{s_{2}(\tilde x) s_{2}(\tilde \lambda) = \frac{q^2(k-q)^2}{k^2}}$.

Thus we have
\be\label{comp}
  I = e^{\sum x_i \lambda_i}[ 1 + \frac{q^2}{k} +  
[\frac{q^4}{2k^2}+\frac{\alpha q^2(q-k)^2}{2 (k+\alpha)
(k-1) k^2}] + O(x^3)]
\ee

We denote
$\sum_{perm.}( \tau_{12} + \cdots)$ by $[\tau_{12}]$. Under this choice  
of $X$ and $\Lambda$,
we find that
$[\tau_{ij}\tau_{kl}\cdots]$ are polynomials in q.

We evaluate them at order two as
$[\tau_{12}\tau_{34}]= \frac{1}{2}q^2(q-1)^2$, $[\tau_{12}\tau_{13}]=  
q^2(q-1)$, $
[(\tau_{12})^2]= q^2$. From (\ref{withoutperm}) and (\ref{g1}), we write
\ba
&&1 + D_{[\rm I]}[\tau_{12}] + D_{[\rm I,\rm I]}[\tau_{12}\tau_{34}] +
  D_{[\Lambda]}[\tau_{12}\tau_{13}]
+ D_{[\rm II]}[(\tau_{12})^2] + \cdots \nonumber\\
&& = 1 + D_{[\rm I]}(-q^2) + D_{[\rm I,\rm I]}(\frac{1}{2}q^2(q-1)^2) +
D_{[\Lambda]} q^2(q-1)
+ D_{[\rm II]}q^2 + \cdots \nonumber\\
\ea

By comparing the powers $q^4$,$q^3$ and $q^2$ in this expression
with the expansion of (\ref{comp}),
  we find,
\be\label{202}
D_{[\rm I]} = - \frac{1}{k}
\ee
\be\label{203}
D_{[\rm I,\rm I]} = \frac{1}{1 + \alpha} ( \frac{1}{k(k+ \alpha)} +  
\frac{\alpha}{k(k-1)} )
\ee
\be\label{204}
D_{[\Lambda]} =\frac{1}{k (k + \alpha)}
\ee
\be\label{205}
D_{[\rm II]} = \frac{1+ \alpha}{2 k ( k + \alpha)}
\ee
When $\alpha=2$, they reduce to the previous $\beta=1$ result.
These coefficients are written in terms of the inverse dimensional  
constants $\frac{1}{Z_{[p]}(\rm I)}$.
The coefficients $1$ and $\alpha$ in (\ref{203}) are the characters  
$\chi_{p}(1)$ as shown in the
appendix A.

As we remarked, these coefficients $D_{[\rm I,\rm I]}$, $D_{[\Lambda]}$  
and
$D_{[\rm II]}$ are the coefficients of (\ref{withoutperm}), and not of  
(\ref{withperm}).
For instance, the value of $D_{[\Lambda]}$ at $\beta=4$ is different  
from the value
of $C_{[\Lambda]}$ of (\ref{withperm}) as
\be\label{206}
D_{[\Lambda]}= \frac{1}{k (k+ \alpha)}|_{\beta=4} = \frac{1}{k(k+  
\frac{1}{2})}\neq
\frac{1}{k(k-1)}
\ee
The residual recursion equations, which gives the recursive relations  
between the coefficients,
  are derived
for the coefficients of (\ref{withperm}). However, we find that the  
following
relations are also valid for the expressions of (\ref{202}) -  
(\ref{205}), which
are the coefficients
in (\ref{withoutperm}),
\be
D_{[\rm I]} + (k-1) D_{[\rm I,\rm I]} + D_{[\Lambda]} = 0
\ee
\be
D_{[\rm I]} + (k-1)D_{[\Lambda]} + 2 D_{[\rm II]} = 0
\ee
The reason for the remarkable fact that the residual equations are  
satisfied,  is that
the residual equation is independent of the value of $\beta$, and thus  
of  the value of $\alpha$.
It means that the residual equation holds independently of the  
existence of the
Vandermonde factor in (\ref{withperm}).

Therefore, for an arbitrary parameter $\alpha$, we have found  
expressions for the coefficients
which satisfy the residual equation. The coefficients $C_{[graph]}$ in  
(\ref{withperm})
satisfies the same recursive equation. We have assumed that $\alpha =  
\frac{2}{\beta}$.
However, if we take this value of $\alpha$ as
\be\label{duality}
\alpha =  \frac{2}{2 - \beta}
\ee
the coefficients $D_{[graph]}$ in (\ref{withoutperm}) for the
case of $\beta$ become the coefficient $C_{[graph]}$ in  
(\ref{withperm}) for the same $\beta$.

For instance, the previous   $D_{[\Lambda]}$ becomes, after  
substitution of
$\alpha =  \frac{2}{2 - \beta}$,

\be\label{177}
D_{[\Lambda]} = \frac{1}{k(k+\alpha)} = \frac{\frac{\beta}{2}-1}{k  
((\frac{\beta}{2} -1)k - 1)}
\ee
which is indeed the value of $C_{[\Lambda]}$ in (\ref{withperm}) as  
shown in
Table B. Similarly, we find that $D_{[\rm I, \rm I]}$ in (\ref{203})  
and $D_{[\rm II]}$ in
(\ref{205}) are  the expressions given in Table B after the  
substitution of
$\alpha = \frac{2}{2-\beta}$,

\be\label{178}
D_{[\rm I,\rm I]} = \frac{1}{1 + \alpha}[\frac{1}{k(k+\alpha)} +  
\frac{\alpha}{k(k-1)}]
  = \frac{(\frac{\beta}{2}-1)}{k((\frac{\beta}{2}-1)k - 1)}[
1 - \frac{1}{(\frac{\beta}{2}-1)(k-1)}]
\ee
\be\label{179}
D_{[\rm II]} = \frac{1 + \alpha}{2 k (k+ \alpha)} =  
\frac{(\frac{\beta}{4}- 1)}{k ((\frac{\beta}{2}-1)k
- 1)}
\ee

The relation (\ref{duality}) is   a duality relation, and by this  
relation,
we obtain explicit expressions of the WKB expansion of the integral.
  This transformation $\alpha= \frac{2}{\beta} \rightarrow
\frac{2}{2 - \beta}$ becomes an identity for the case of $\beta=1$. For  
the value of $\beta=2$,
this transformation becomes singular, and
it needs special consideration (We will discuss this case
in section nine).

Ar order three, we have to consider the next order in (\ref{g1}). It  
becomes
\ba
&&\frac{q^6}{6 k^3} + \frac{q^2}{k}\frac{\alpha}{2  
(k+\alpha)(k-1)}s_2(\tilde x)
s_2(\tilde \lambda) \nonumber\\
&&+ \frac{1}{6(1 + \alpha)(1 + 2  
\alpha)}(\chi_{[3]}(1)\frac{Z_{[3]}(\tilde x)
Z_{[3]}(\tilde \lambda)}{Z_{[3]}(\rm I)} +  
\chi_{[21]}(1)\frac{Z_{[21]}(\tilde x)Z_{[21]}(\tilde \lambda)}
{Z_{[21]}(\rm I)} \nonumber\\
&&+ \chi_{[1^3]}(1)\frac{Z_{[1^3]}(\tilde x)Z_{[1^3]}(\tilde  
\lambda)}{Z_{[1^3]}(\rm I)})
\nonumber\\
&&= \frac{q^6}{6 k^3} + \frac{q^2}{k} \frac{\alpha}{2 (k + \alpha)(1 +  
2 \alpha)}s_2(\tilde x)s_2(\tilde
\lambda)\nonumber\\
&&+ \frac{\alpha^2 k}
{3  (k+ \alpha)(k+ 2 \alpha) (k-1)(k-2)} s_3(\tilde x)s_3(\tilde  
\lambda)
\ea

Using the values of $\displaystyle{s_2(\tilde x) = \frac{q(k-q)}{k}}$, $\displaystyle{
s_{3}(\tilde  
x)= \frac{q^2(q-k)^2(2 q-k)^2}{k^4}}
$, it becomes
\ba\label{I33}
\tilde I^{(3)} &=&\frac{1}{6 k (k+\alpha)(k+2 \alpha)(k-1)(k-2)}
q^2(2 \alpha^2 k^2 - 12 \alpha^2 k q + 14 \alpha^2 q^2 \nonumber\\
&&- 6 \alpha k q^2 + 6 \alpha^2 k q^2 + 3 \alpha k^2 q^2 \nonumber\\
&&+ 12 \alpha q^3 - 12 \alpha^2 q^3 - 6 \alpha k q^3\nonumber\\
&&+ 2 q^4 - 6 \alpha q^4 + 2 \alpha^2 q^4 - 3 k q^4 + 3 \alpha k q^4 +  
k^2 q^4)
\ea
When we put $\alpha = 2$ in the above quantity, it becomes the same as  
(\ref{I3}).
This quantity $\tilde I^{(3)}$ should be equal to the following sum of  
8 terms,
\ba\label{I333}
\tilde I^{(3)} &=& D_{[\rm I,\rm I,\rm I]}\cdot  
[\tau_{12}\tau_{34}\tau_{56}] +
D_{[\Lambda,\rm I]}\cdot [\tau_{12}\tau_{13}\tau_{45}]\nonumber\\
&+& D_{[\rm Y]}\cdot[\tau_{12}\tau_{13}\tau_{14}] + D_{[\rm II,\rm  
I]}\cdot[\tau_{12}^2 \tau_{34}]
+ D_{[\rm N]}\cdot[\tau_{12}\tau_{13}\tau_{34}]\nonumber\\
&+& D_{[\rm III]}\cdot[\tau_{12}^3] + D_{[\hskip 1mm \underline{\angle}  
\hskip 1mm]}\cdot[
\tau_{12}^2 \tau_{13}] +  
D_{[\bigtriangleup]}\cdot[\tau_{12}\tau_{23}\tau_{13}]\nonumber\\
&=&-[ D_{[\rm I,\rm I,\rm I]}\cdot \frac{1}{6}q^2 (q-1)^2(q-2)^2 +
D_{[\Lambda,\rm I]}\cdot q^2(q-1)^2(q-2)\nonumber\\
&+& D_{[\rm Y]}\cdot\frac{1}{3}q^2(q-1)(q-2) + D_{[\rm II,\rm I]}\cdot  
q^2(q-1)^2
+ D_{[\rm N]}\cdot q^2 (q-1)^2\nonumber\\
&+& D_{[\rm III]}\cdot q^2 + D_{[\hskip 1mm \underline{\angle} \hskip  
1mm]}\cdot 2 q^2 (q-1) +
D_{[\bigtriangleup]}\cdot 0]
\ea

Comparing $q^6$ term in (\ref{I333}) with (\ref{I33}), we have
\be
D_{[\rm I,\rm I,\rm I]} = -\frac{k^2 + 3 (\alpha-1) k + 2 (\alpha^2 - 3  
\alpha + 1)}{
k (k+\alpha)(k + 2 \alpha)(k-1)(k-2)}
\ee
By the dimensional constants $Z_p(\rm I)$ for general $\alpha$, it is  
expressed as
\be
D_{[\rm I,\rm I,\rm I]} = -\frac{1}{(1 + \alpha)(1 + 2 \alpha)} [  
\frac{\chi_{[3]}(1)}{Z_{[3]}(\rm I)}
+ \frac{\chi_{[21]}(1)}{Z_{[21]}(\rm I)} +
\frac{\chi_{[1^3]}(1)}{Z_{[1^3]}(\rm I)}]
\ee
where the characters $\chi_{[p]}(1)$ and the dimensional constant  
$Z_p(\rm I)$ are given in Table D.
($\displaystyle{\chi_{[3]}(1) = 1, \chi_{[21]}(1)= \frac{6 \alpha(1 + \alpha)}{2 +  
\alpha}, \chi_{[1^3]}(1)=
\frac{\alpha^2(1 + 2 \alpha)}{2 + \alpha}}$).

By the dual transformation of (\ref{duality}), this expression becomes
\ba\label{185}
D_{[\rm I,\rm I,\rm I]} &=& -  
\frac{(\frac{\beta}{2}-1)^2}{k((\frac{\beta}{2}  -  
1)k-1)((\frac{\beta}{2}
-1)k-2)}[1 - \frac{3}{(\frac{\beta}{2}-1)(k-1)}\nonumber\\
  &+& \frac{2}{(\frac{\beta}{2}-1)^2(k-1)(k-2)}]
\ea
which coincide, here also with the previous values listed in Table B.

By the comparison of $q^5$ terms in (\ref{I333}) and (\ref{I33}), we  
obtain the expression of
$D_{[\Lambda,\rm I]}$ as
\be
D_{[\Lambda,\rm I]} = -\frac{k+ 2 \alpha -1}{k(k+\alpha)(k + 2  
\alpha)(k-1)}
\ee
  By the replacement of $\alpha = 2/(2 - \beta)$, it becomes
\be\label{187}
D_{[\Lambda,\rm I]} = -  
\frac{(\frac{\beta}{2}-1)^2}{k((\frac{\beta}{2}k-1)((\frac{\beta}{2}
-1)k -2)}[1 - \frac{2}{(\frac{\beta}{2}-1)(k-1)}]
\ee
which coincides with the result in Table B.

By putting $q=1$ in (\ref{I33}) and (\ref{I333}), we obtain the  
expression for $D_{[\rm III]}$ as
\be
D_{[\rm III]}= -\frac{(1 + \alpha)(1 + 2 \alpha)}{6 k ( k + \alpha)(k +  
2 \alpha)}
\ee
and by the dual transformation of (\ref{duality}), it becomes
\be\label{189}
D_{[\rm III]} = -  
\frac{(\frac{\beta}{4}-1)(\frac{\beta}{6}-1)}{k((\frac{\beta}{2}-1)k 
-1)((\frac{\beta}{2}-1)
k -2)}
\ee
which agrees with the value in Table B. Note that for $\beta=4$, this  
quantity is vanishing, and
it means that for $\beta=4$, there are no multiple line graphs.

Extracting the contributions of $D_{[\rm I,\rm I,\rm  
I]}$,$D_{[\Lambda,\rm I]}$ and $D_{[\rm III]}$ from
(\ref{I33}) and (\ref{I333}), we have
\ba
&&D_{[\rm Y]}\cdot\frac{1}{3}q^2(q-1)(q-2) + D_{[\rm II,\rm I]}\cdot  
q^2(q-1)^2
+ D_{[\rm N]}\cdot q^2 (q-1)^2\nonumber\\
&&  + D_{[\hskip 1mm \underline{\angle} \hskip 1mm]}\cdot 2 q^2 (q-1)  
\nonumber\\
&&= -\frac{q^2(q-1)}{6 k (k+\alpha)(k+ 2 \alpha)(k-1)}[ 3 \alpha k q +  
11 k q + 6 \alpha^2 q\nonumber\\
&& + 9 \alpha q - 11 q + 3 \alpha k - 7 k - 6 \alpha^2 -15 \alpha + 7]
\ea
Dividing both sides by $q^2(q-1)$ factor,  we obtain by putting $q=1$,
\be
- \frac{1}{3} D_{[\rm Y]} + 2 D_{[\hskip 1mm \underline{\angle}\hskip  
1mm]}
= -\frac{6 \alpha + 4}{k (k + \alpha)(k + 2 \alpha)}
\ee
 From the simple structure of $[\rm Y]$ (no multiple lines), we assume  
that
\be\label{192}
D_{[\rm Y]} = -\frac{1}{k ( k + \alpha)(k + 2 \alpha)} = -  
\frac{(\frac{\beta}{2}- 1)^2}{k ((\frac{\beta}{2}
k - 1)((\frac{\beta}{2}- 1)k - 2)}
\ee
which is also consistent with the known value for $\beta=4$, $k=4$
($D_{[\rm Y]} = - \frac{1}{24}$
  \cite{BHb}).
Then, we get
\be\label{193}
D_{[\hskip 1mm \underline{\angle}\hskip 1mm]}
= -\frac{1 + \alpha}{2 k ( k + \alpha)(k + 2 \alpha)}
\ee

The sum of the remaining two terms is
\be\label{227}
D_{[\rm II,\rm I]} + D_{[\rm N]} = -\frac{1}{k (k + \alpha)(k + 2  
\alpha)(k-1)} [ \frac{1}{2}(k-1)(3 + \alpha) + \alpha ( 2 + \alpha)]
\ee
Since $D_{[\rm II,I]}$ is a coefficient of the term which has a double  
line in the
graphic representation, it should be proportional to the factor $(1 +  
\alpha)/2$,
which corresponds to the
double line multiple factor. Therefore, the sum of (\ref{227}) is  
divided into two parts,
\ba\label{195}
D_{[\rm II,\rm I]} &=& -\frac{1 + \alpha}{2 k ( k + \alpha)(k +  
2\alpha)}[ 1 + \frac{2 \alpha}{k-1}]
\nonumber\\
&=& -\frac{1}{1 + 2\alpha}[ \frac{1 + \alpha}{2 Z_{[3]}(\rm I)}
+ \frac{\alpha(1 + \alpha)}{Z_{[21]}(\rm I)}]
\ea
\ba\label{196}
D_{[\rm N]} &=& -\frac{1}{k ( k + \alpha)(k + 2 \alpha)} [ 1 +  
\frac{\alpha}{k-1}]\nonumber\\
&=& - \frac{1}{(1 + 2 \alpha)}[ \frac{1 + \alpha}{Z_{[3]}(\rm I)} +
\frac{\alpha}{Z_{[21]}(\rm I)}]
\ea
These two expressions coincide with the values of Table B by the duality
  transformation (\ref{duality}).
Since there is a cubic identity equation for three terms $[\tau_{12}^2  
\tau_{34}]$, $
[\tau_{12}\tau_{23}\tau_{34}]$ and $[\tau_{12}\tau_{23}\tau_{13}]$, the  
coefficients of these
terms $D_{[\rm II,\rm]}$,$D_{[\rm N]}$ and $D_{[\bigtriangleup]}$ have  
ambiguities.
We have to fix these ambiguities. Here we used a fixing assumption for  
the form
of $D_{[\rm II,\rm I]}$, namely
that it has  $\frac{1+ \alpha}{2}$ as  overall factor, and a pole at  
k=1.

The last coefficient $D_{[\bigtriangleup]}$ is not determined by the  
present choice of $X$ and $\Lambda$.
One method to determine $D_{[\bigtriangleup]}$ is to use the residual  
equation,
\be
D_{\Lambda} + (k-2)D_{[\rm N]} + D_{[\bigtriangleup]} + D_{[\rm II,\rm  
I]} + D_{[\underline{
\angle}]} = 0
\ee
Using the previous expressions, we obtain
\be
D_{[\bigtriangleup]} = - \frac{1}{k(k+\alpha)(k+ 2 \alpha)}[ 1-  
\frac{\alpha^2}{k-1}]
\ee
This reads to
\be\label{199}
D_{[\bigtriangleup]} = -\frac{1}{1 + 2 \alpha}[ \frac{(1 +  
\alpha)^2}{Z_{[3]}(\rm I)}-
\frac{\alpha^2}{Z_{[21]}(\rm I)}]
\ee
This expression coincides with the value in Table B after the duality  
transformation  (\ref{duality}).

In Appendix D, we have derived the expressions of the coefficients $D$  
up to order fourth,
from the
extended zonal polynomial expansions. The results are shown in Table B.

\vskip 8mm
%%%%%%%%%%%%%%%%%%%%%%%%%%%%%%%%%%%%%%%%%%%%%%%%%%%%%%%%%%%%%%%%%%%%%%
\section{ Series expansion for $\beta= 4$ }
\vskip 5mm

We have found in the previous section that the integral  (\ref{I}) may  
be expressed in two
different dual ways,
\be\label{D1}
I_{\beta} = e^{\sum x_i \lambda_i-\frac{1}{k}\sum_{i<j}  
\tau_{ij}}\sum_{m=0}^\infty \frac{1}{m!}\frac{1}{\prod_{q=0}^{m-1}(1 +  
q \alpha)}
\sum_p \chi_p(1) \frac{Z_p(\tilde X)Z_p(\tilde \Lambda)}{Z_p(\rm I)}
\ee
and
\ba\label{D2}
I_{\beta} &=& \sum_{perm.} [\frac{e^{\sum x_i \lambda_i}}{\prod_{i<j}
[(x_i-x_j)(\lambda_i - \lambda_j)]^{\beta-1}}
\nonumber\\
&\times&e^{-\frac{1}{k}\sum_{i<J} \tau_{ij}}\sum_{m=0}^\infty  
\frac{1}{m!}\frac{1}{\prod_{q=0}^{m-1}(1 + q \alpha^{\prime})}
\sum_p \chi_p(1) \frac{Z_p(\tilde X)Z_p(\tilde \Lambda)}{Z_p(\rm I)}]
\ea
where $\displaystyle{\tilde x_i= x_i - \frac{1}{k}\sum_{j=1}^k x_j}$, and $\displaystyle{
\alpha=  
\frac{2}{\beta}}$,
$\displaystyle{\alpha^{\prime} = - 1/(\frac{\beta}{2}-1)}$. The polynomials $Z_p(x)$  
are the extended zonal
polynomials with  parameters $\alpha$ in (\ref{D1}) and  
$\alpha^{\prime}$ in (\ref{D2}),
and $\chi_p(1)$ are the characters.

In the previous section, the zonal polynomials were further expressed  
as polynomials in
$\tau_{ij}$ as (\ref{withoutperm}) and (\ref{withperm}).

The duality means that (\ref{D1}) and (\ref{D2}) give the same  
expression for the integral
$I_{\beta}$ under the relation of
\be
    \alpha = \frac{2}{\beta}, \hskip 10 mm \alpha^{\prime} =  \frac{2}{2  
- \beta}
\ee
or equivalently,
\be
    \beta = \frac{2}{\alpha} = 2 - \frac{2}{\alpha^{\prime}}.
\ee

In the case $\beta=4,6,...,2m$ for the even integers, $\alpha^{\prime}$  
becomes
$-1,-\frac{1}{2},...$,\\$-\frac{1}{m-1}$. The zonal polynomial for  
$\alpha^{\prime}=-1$
was discussed before in a different context \cite{Macdonald}.
In such cases, the value of $\alpha^{\prime}$ is negative, and the  
factor in the denominator
$\displaystyle{\prod_{q=0}^{m-1}(1 + q \alpha^{\prime})}$ in (\ref{D2}) vanishes,  
although
the whole expression remains finite. Therefore, we need a special  
treatment for such
$\beta = 2m$ case, and we briefly discuss the case  $\beta=4$ here. The  
dimensional
constants $Z_p(\rm I)$ become degenerate in such cases.

In the case $\beta=4$, the parameter $\alpha^{\prime}$ is $-1$. The  
dimensional
constants show the degeneracy ;  for instance,  $Z_{[3]}(\rm I) =
k (k-1)(k-2)$, and $Z_{[1^3]}(\rm I)= k(k-1)(k-2)$. Such degeneracies  
can be
seen in the Table A.
In addition to this degeneracy,
the multiple line factors are proportional to $(1 + \alpha)$, and there  
are no
multiple line graphs in $\beta=4$ in the representation of (\ref{D2}),  
and
the expansion becomes rather simple for $\beta=4$ as shown in  
\cite{BHb,BHc}.

Here we study the zonal polynomial expansion in the form  (\ref{D2})  
for the
$\beta=4$ case,  without writing it as a function of $\tau$.

The expression (\ref{D2}) may be converted to more useful series in the  
case $\alpha=-1$.
In the previous section, we have written it in terms of the symmetric  
functions $s_n$ for
general $\alpha$. Although there is a divergent factor $\frac{1}{1 +  
\alpha}$ in
(\ref{D2}), this divergence is cancelled by the sum of the
partitions of $p$. We investigate here  the case $\alpha=-1 (\beta=4)$  
in higher
orders, and obtain a useful expression for the integral $I$.

We define $\phi(x,\lambda)$ as

\be\label{phi}
I_{\beta} = \sum_{perm.} [\frac{e^{\sum x_i \lambda_i}}{\prod_{i<j}
[(x_i-x_j)(\lambda_i - \lambda_j)]^{\beta-1}}
e^{-\frac{1}{k}\sum_{i<J} \tau_{ij}}
\phi(x,\lambda)]
\ee
\be
\phi(x,\lambda)=\sum_{m=0}^\infty  
\frac{1}{m!}\frac{1}{\prod_{q=0}^{m-1}(1 + q \alpha)}
\sum_p \chi_p(1) \frac{Z_p(\tilde X)Z_p(\tilde \Lambda)}{Z_p(\rm I)}
\ee
Note that we have $f= e^{^\frac{1}{k}\sum \tau_{ij}}\phi(x,\lambda)$.
Using the table of  extended zonal polynomials and  characters given in  
the appendix,
we write the zonal polynomials in terms of the symmetric functions
$s_n$, and  find the expression for $\phi$, by noting that  
$s_{1}(\tilde x)= 0$.

\ba
\phi(x,\lambda) &=& 1 + \frac{\alpha}{2 (k+ \alpha)(k-1)} s_2(\tilde  
x)s_2(\tilde \lambda)\nonumber\\
  &+&
  \frac{ \alpha^2 k}{3 ( k +\alpha)(k+ 2 \alpha)(k-1)(k-2)}s_3(\tilde  
x)s_3(\tilde \lambda)\nonumber\\
&+& \frac{\alpha^2 }{8 k (k-1)(k-2)(k-3)(k+ \alpha-1)(k+\alpha)
(k + 2\alpha)(k+ 3 \alpha)}\nonumber\\
&\times& [ 2 \alpha k^2 (k^2 + \alpha k - k + \alpha) s_4(\tilde  
x)s_4(\tilde y)\nonumber\\
&&   + 2 \alpha k ( 2 k^2 + 3 \alpha k - 3 k - 3 \alpha)(s_4(\tilde  
x)s_2^2(\tilde y) + s_2^2(\tilde x)
s_4(\tilde y))\nonumber\\
&&   + ( k^4 + 5 \alpha k^3 - 5 k^3 + 6 \alpha^2 k^2 - 18 \alpha k^2 +  
6 k^2 - 18 \alpha^2 k\nonumber\\
&&  + 18 \alpha k + 18 \alpha^2) s_2^2(\tilde x)s_2^2(\tilde y)  
]\nonumber\\
&+& O(x^5)
\ea

There is no divergence in the limit $\alpha \rightarrow -1$ in this  
expression.
Thus we obtain for $\beta=4$, from the table in the appendix, up to  
order six ,
\ba\label{phi6}
   \phi &=& 1 - \frac{1}{2 (k-1)^2}s_2(\tilde x)s_2(\tilde \lambda) +  
\frac{k}{3 (k-1)^2(k-2)^2}s_3(\tilde x)s_3(\tilde
\lambda)\nonumber\\
&+&\frac{1}{8 k(k-1)^2(k-2)^3(k-3)^2}[- 2 k^2(k^2 - 2 k -1) s_4(\tilde  
x)s_4(\tilde \lambda)\nonumber\\
&& + 2 k (2 k^2 - 6 k + 3)(s_4(\tilde x)s_2^2(\tilde \lambda) +  
s_2^2(\tilde x)s_4(\tilde \lambda)
\nonumber\\
&& + (k^4 - 10 k^3 + 30 k^2 - 36 k + 18) s_2^2(\tilde x)s_2^2(\tilde  
\lambda)]\nonumber\\
&+& \frac{k^2(k^2 - 2 k - 5)}{5 (k-1)^2  
(k-2)^3(k-3)^2(k-4)^2}s_5(\tilde x)s_5(\tilde \lambda)\nonumber\\
&-& \frac{k(k^2 - 4 k + 2)}{(k-1)^2(k-2)^3(k-3)^2(k-4)^2}[s_2(\tilde  
x)s_3(\tilde x)s_5(\tilde \lambda)
+ s_5(\tilde x)s_2(\tilde \lambda)s_3(\tilde \lambda)]\nonumber\\
&+& \frac{k^4 - 14 k^3 + 48 k^2 - 48 k + 24}{6  
(k-1)^2(k-2)^3(k-3)^2(k-4)^2}
s_2(\tilde x)s_3(\tilde x)s_2(\tilde \lambda)s_3(\tilde  
\lambda)\nonumber\\
&+&\frac{1}{{\left( -5 + k \right) }^2\,{\left( -4 + k \right) }^2\,
     {\left( -3 + k \right) }^4\,{\left( -2 + k \right) }^3\,{\left( -1  
+ k \right) }^2}
\nonumber\\
&\times&
\{ s_6(\tilde x)s_6(\tilde \lambda)(-\frac{1}{6}k )[ 8 - 44\,k -  
135\,k^2 + 76\,k^3 + 6\,k^4 - 8\,
k^5 + k^6 ]\nonumber\\
&+&(s_2(\tilde x)s_4(\tilde x)s_6(\tilde \lambda)+s_2(\tilde  
\lambda)s_4(\tilde \lambda)
s_6(\tilde x))
[-20 + 70\,k - 61\,k^2 - 29\,k^3 \nonumber\\
&& + 38\,k^4 - 11\,k^5 + k^6]\nonumber\\
&+&(s_2(\tilde x)s_4(\tilde x)s_2(\tilde \lambda)s_4(\tilde \lambda))
\frac{1}{8k}[-2400 + 3600\,k - 1740\,k^2 - 240\,k^3 + 1099\,k^4  
\nonumber\\
&& - 708\,k^5
+ 196\,k^6 - 24\,k^7 + k^8]
\nonumber\\
&+&(s_3(\tilde x)^2s_6(\tilde \lambda)+ s_3(\tilde \lambda)^2s_6(\tilde  
x))
\frac{1}{6}[80 - 200\,k + 251\,k^2 - 284\,k^3 \nonumber\\
&& + 154\,k^4 - 36\,k^5 + 3\,k^6]\nonumber\\
&+&(s_3(\tilde x)^2s_2(\tilde \lambda)s_4(\tilde \lambda)+s_3(\tilde  
\lambda)^2
s_2(\tilde x)s_4(\tilde x))
(-\frac{1}{2k})[-400 + 200\,k + 525\,k^2 \nonumber\\
&&- 690\,k^3 + 322\,k^4 - 66\,k^5
+ 5\,k^6]
\nonumber\\
&+&(s_3(\tilde x)^2s_3(\tilde \lambda)^2)\frac{1}{18k}[-2400 - 1200\,k  
+ 7890\,k^2
\nonumber\\
&&- 8310\,k^3 + 4461\,k^4 - 1416\,k^5
+ 264\,k^6 - 26\,k^7 + k^8]
\nonumber\\
&+&(s_2(\tilde x)^3s_6(\tilde \lambda)+s_2(\tilde \lambda)^3s_6(\tilde  
x))
\frac{1}{6}[240 - 769\,k + 882\,k^2 - 424\,k^3 + 90\,k^4 -  
7\,k^5]\nonumber\\
&+&(s_2(\tilde x)^3s_2(\tilde \lambda)s_4(\tilde \lambda)+s_2(\tilde  
x)^3s_2(\tilde \lambda)s_4(\tilde \lambda))
\frac{1}{8k}[4800 - 11180\,k + 11480\,k^2 \nonumber\\
&& - 6775\,k^3 + 2384\,k^4 - 481\,k^5 + 50\,k^6 - 2\,k^7]\nonumber\\
&+&(s_2(\tilde x)^3s_3(\tilde \lambda)+s_2(\tilde \lambda)^3s_3(\tilde  
x))
\frac{1}{3k}[-1200 + 2270\,k \nonumber\\
&&- 1725\,k^2 + 640\,k^3 - 115\,k^4 + 8\,k^5]
\nonumber\\
&+&s_2(\tilde x)^3s_2(\tilde \lambda)^3\cdot\frac{1}{48k}[-25200 +  
60960 k - 64030 k^2 + 38192 k^3
\nonumber\\
&&   -13976 k^4 + 3170 k^5 - 431 k^6 + 32 k^7 - k^8]\} \nonumber\\
&+& O(x^7)
\ea

We first check the simple $k=3$ case, given in the introduction.
  Multiplyiong by  $e^{-\frac{1}{3}(\tau_{12}+\tau_{23}+\tau_{13})}$ to  
$\phi$, we
  recover the known result,
\ba\label{k3}
e^{-\frac{1}{3}(\tau_{12}+\tau_{23}+\tau_{13})}\phi &=& 1 -
\frac{1}{3}(\tau_{12}+\tau_{23}+\tau_{13})\nonumber\\
&+&\frac{1}{6}(\tau_{12}\tau_{23}+\tau_{23}\tau_{13}+\tau_{13}\tau_{12}) 
\nonumber\\
&-& \frac{1}{12}\tau_{12}\tau_{23}\tau_{13}
\ea

where the last term is obtained from the identity
\ba
&&- \frac{1}{3! 3^3}(\tau_{12}+\tau_{23}+\tau_{13})^3 +  
\frac{1}{24}(\tau_{12}+ \tau_{23} +
\tau_{13}) s_2(\tilde x)s_2(\tilde \lambda)\nonumber\\
&&+ \frac{1}{4}s_3(\tilde x)s_3(\tilde \lambda) =  
-\frac{1}{12}\tau_{12}\tau_{23}\tau_{13}.
\ea
The fourth order term in (\ref{phi6}) has a divergent coefficient  
$\displaystyle{\frac{1}{(k-3)^2}}$. However
the fourth order term is finite in the limit $k\rightarrow 3$, and with  
the exponential term,
it is vanishing at the end. In the fifth order, the same situation  
occurs. Therefore,
we have recovered exactly (\ref{k3}) to all orders.

The expression (\ref{phi6}) is lengthy, however its large k limit  
becomes simple.
In the large k limit, for each order $x^l$ , the coefficients of the  
products of the
symmetric functions are  of order  $\displaystyle{\frac{1}{k^l}}$. We take these  
leading terms,

\ba\label{large1k}
\phi &\sim& 1 -\frac{1}{2 k^2}s_2(\tilde x)s_2(\tilde \lambda)+  
\frac{1}{3 k^3}
s_3(\tilde x)s_3(\tilde \lambda)\nonumber\\
&-& \frac{1}{4 k^4}s_4(\tilde x)s_4(\tilde \lambda)+ \frac{1}{8  
k^4}s_2(\tilde x)^2s_2(\tilde \lambda)^2
\nonumber\\
&+& \frac{1}{5 k^5}s_5(\tilde x)s_5(\tilde \lambda)+  
\frac{1}{6k^5}s_2(\tilde x)s_3(\tilde x)s_2(\tilde \lambda)
s_3(\tilde \lambda)\nonumber\\
&-& \frac{1}{6k^6}s_6(\tilde x)s_6(\tilde \lambda)+ \frac{1}{8  
k^6}s_2(\tilde x)s_4(\tilde x)s_2(\tilde
\lambda)s_4(\tilde \lambda)+ \frac{1}{18 k^6}s_3(\tilde x)^2s_3(\tilde  
\lambda)^2\nonumber\\
&-& \frac{1}{48 k^6}s_2(\tilde x)^3 s_2(\tilde \lambda)^3\nonumber\\
&+& O(x^7)
\ea

There is a rule for the coefficients in (\ref{large1k}).
Only the combination of the same symmetric functions for $\tilde x$ and  
$\tilde \lambda$
give the leading terms in the large k limit. For instance, $s_2(\tilde  
x)^3s_2(\tilde \lambda)$
and $s_2(\tilde x)s_4(\tilde x)s_2(\tilde \lambda)s_4(\tilde \lambda)$  
are of  order
$\displaystyle{\frac{1}{k^6}}$. The coefficient of $s_2(\tilde x)^3s_3(\tilde  
\lambda)^2$, which has different
symmetric functions, is of  order $\displaystyle{\frac{1}{k^9}}$.

The coefficients of (\ref{largek}) is obtained as follows.
For the term of $s_n(\tilde x)^p s_m(\tilde x)^t$, $(n\ne m)$, the  
coefficient  becomes
\be
   C= (-1)^{n^p+m^t}\frac{1}{p! n^p t! m^t}\frac{1}{k^{n^p+m^t}}
\ee
For the general case, $s_{n_1}^{p_1} s_{n_2}^{p_2} \cdots  
s_{n_j}^{p_j}$, the
coefficient is proportional to \\
$\displaystyle{\frac{1}{p_1!p_2!\cdots p_j!  
n_1^{p_1}n_2^{p_2}\cdots n_{j}^{p_j}}}$.
The order of $\tilde x$ is given by $n_1^{p_1}+\cdots +n_j^{p_j}$, and  
if this order is even,
the minus sign has to be included.
Thus, we have the large k expression for $\phi$.

In the large k limit for fixed $x_i,\lambda_j$, $f$ is given by  
$e^{-\frac{1}{k}\sum \tau_{ij}}$,
and $\phi(x,\lambda)=1$, for all values of $\alpha$, as shown in  
Appendix D. However,
in some problems, in which $s_n(\tilde x)$ is not order of one, the  
above large k formula might be
important.

%%%%%%%%%%%%%%%%%%%%%%%%%%%%%%%%%%%%%%%%%%%%%%%%%%%%%%%%%%%%%%%%%%%%%%
%%%%%%%%%%%%%%%%%%%%%%%%%%%%%%%%%%%%%%%%%%%%%%%%%%%%%%%%%%%%%%%%%%%%%%%

\section{ The HIZ-integral for $\beta=2$ and the character expansion}
\vskip 5mm

We use here  the same zonal polynomial method as for $\beta=1$.
We define $\tilde x_a = x_a - \frac{1}{k}\sum_{b=1}^k x_b$.
The integral becomes ($\alpha=\frac{2}{\beta}= 1$) from the expression  
(\ref{withoutperm}),
\be
I = e^{\sum x_i \lambda_i - \frac{1}{k}\sum \tau_{ij}}[ \sum_m  
\sum_{p(m)}
\frac{1}{m!}\frac{1}{\prod_{q=0}^{m-1} ( 1 + q)} \frac{Z_p(\tilde  
x)Z_p(\tilde \lambda)}{
Z_p(\rm I)}]
\ee
This series expansion has to reduce to the simple closed form
\be\label{IZform}
I_{\beta=2} = \sum_{perm.of \lambda_i} \frac{e^{\sum_{i=1}^k x_i  
\lambda_i}}{\prod_{i<j}\tau_{ij}}
\ee
in order to reproduce the original HIZ formula  
\cite{Harish-Chandra,Itzykson-Zuber}..

In other words one must prove  the identity,
\be\label{proposition}
\sum_{perm.} \frac{e^{\frac{1}{k}\sum \tau_{ij}}}{\prod \tau_{ij}} =
\sum_m \sum_p \frac{1}{m!}\frac{1}{\prod_{q=0}^{m-1} ( 1 + q)}   
\frac{Z_p(\tilde x)Z_p(\tilde \lambda)}{Z_p(\rm I)}
\ee

The proof of this identity is easily done by writing the zonal  
polynomials, which are
Shur functions,
as the ratio of the determinants,
\be
    Z_p(x) = \frac{{\rm det}[ x_i^{l_j} ]}{{\rm det}[ x_i^{j-1}]}
\ee
The product of the determinants is also a determinant, it is simply  
${\rm det}[e^{x_i \lambda_j}]$,
from the Binet-Cauchy theorem \cite{Zinn}.

This identity leads to  interesting equations.
The right hand side of (\ref {proposition}) is written by the terms of  
$\tau$,
as shown in the case of $\beta=1$.

For the case k=2, the proposition (\ref{proposition}) is easily proved  
by expanding the exponent.
For k=3, we have the following identity : for instance,
\be
\sum_{perm.}\frac{(\tau_{12} + \tau_{23} +  
\tau_{13})^p}{\tau_{12}\tau_{23}\tau_{13}}=0
\ee
for p=1,2 and 4. Let us remind the reader here, that the permutations
in this sum are interchanges of the
$\lambda_i$'s for fixed $x_j$.  For p=3, we have
\be\label{p3}
\sum_{perm.}\frac{(\tau_{12} + \tau_{23} +  
\tau_{13})^3}{\tau_{12}\tau_{23}\tau_{13}}=81
\ee
For p=5, it becomes

\ba\label{p5}
\sum_{perm.} \frac{(\tau_{12} + \tau_{13} +  
\tau_{23})^5}{\tau_{12}\tau_{23}\tau_{13}} &=&
\frac{3645}{4}s_2(\tilde x)s_2(\tilde \lambda)\nonumber\\
&=&405 [(\tau_{12}^2 + \tau_{23}^2 + \tau_{13}^2) - (\tau_{12}\tau_{23}  
+ \tau_{12}\tau_{13} + \tau_{23}
\tau_{13})]\nonumber\\
\ea
where the second equality comes from the expression of $s_2(\tilde x)$  
of (\ref{s2}).
For p=6, we have
\be
\sum_{perm.} \frac{(\tau_{12} + \tau_{13} +  
\tau_{23})^6}{\tau_{12}\tau_{23}\tau_{13}} =
(81)^2 s_3(\tilde x)s_3(\tilde \lambda)
\ee
For p=7,
\be
\sum_{perm.} \frac{(\tau_{12} + \tau_{13} +  
\tau_{23})^7}{\tau_{12}\tau_{23}\tau_{13}} =
1701(\frac{3}{2})^4 [s_2(\tilde x)s_2(\tilde \lambda)]^2
\ee
where for k=3, we have $s_4(\tilde x) = \frac{1}{2} [s_2(\tilde x)]^2$.
For p=8,
\be
\sum_{perm.} \frac{(\tau_{12} + \tau_{13} +  
\tau_{23})^8}{\tau_{12}\tau_{23}\tau_{13}} =
648 \times (\frac{27}{2})^2 [s_2(\tilde x) s_3(\tilde x) s_2(\tilde  
\lambda)s_3(\tilde \lambda)]^2.
\ee
  From these results, one checks the identity (\ref{proposition}).

To understand these identities, we divide the left hand side of  
(\ref{p5})
into five different types of terms. They are
all represented by $s_2(\tilde x)s_2(\tilde \lambda)$.
\ba
  \sum_{perm.} \frac{\tau_{12}^5 + \tau_{23}^5 +  
\tau_{13}^5}{\tau_{12}\tau_{23}\tau_{13}}
&=& \frac{450}{4}s_2(\tilde x)s_2(\tilde \lambda)\nonumber\\
\sum_{perm.} \frac{\tau_{12}^4(\tau_{13} +\tau_{23}) +  
\cdots}{\tau_{12}\tau_{23}\tau_{13}}
&=& \frac{225}{4}s_2(\tilde x)s_2(\tilde \lambda)\nonumber\\
\sum_{perm.}(\tau_{12}^2 + \tau_{23}^2 + \tau_{13}^2) &=& 18 s_2(\tilde  
x)s_2(\tilde \lambda)\nonumber\\
\sum_{perm.} \frac{(\tau_{12}^3 \tau_{23}^2 +  
\cdots)}{\tau_{12}\tau_{23}\tau_{13}}
&=&\frac{9}{4}s_2(\tilde x)s_2(\tilde \lambda)\nonumber\\
\sum_{perm.}(\tau_{12}\tau_{23} + \cdots) &=& \frac{9}{2}s_2(\tilde  
x)s_2(\tilde \lambda)
\ea
After summing over permutations, the $\tau$-series are expressible by  
symmetric
functions of $\tilde x$ and $\tilde \lambda$, and are given by  
homogeneous polynomials
in $\tau$.

We thus have  shown that the zonal (character) polynomial series can  
give the expression for $f$,
although $f$ is just one.

The dual representation of (\ref{DD2}) is singular for $\beta=2$. The  
transformation
$\alpha = \frac{2}{2-\beta}$ becomes divergent at $\beta=2$.
However, if we write $f$ as a $\tau$ expansion for fixed $k$, and let
  $\alpha\rightarrow \infty$, we have an
infinite series, as can be seen in Table B.  There are non-vanishing  
terms
in the limit $\alpha=
\infty$. For instance, at  first order, $C_{\rm I}=-\frac{1}{k}$.
This looks quite strange, since we know that $f$ is one. In the  
following, we consider
this puzzling problem.

Collecting the non-vanishing terms in the limit $\alpha \rightarrow
\infty$ from  Table B, we find the  following expression,
\ba\label{nontrivial}
I_{\beta=2} &=& \sum_{perm. of \lambda_i}\frac{1}{\prod  
\tau_{ij}}e^{\sum x_i \lambda_i}
[ 1 - \frac{1}{k}(\tau_{12} + \cdots) +  
\frac{1}{k(k-1)}(\tau_{12}\tau_{34} + \cdots)
\nonumber\\
&+& \frac{1}{2k}(\tau_{12}^2 + \cdots) + O(x^3)].
\ea

According to this expression, $f_{\beta=2}$ is not one. This  
discrepancy may be
understood by looking at the simple example k=2. In this k=2 case, we  
have
\ba
I_{\beta=2} &=& \sum_{perm. of \lambda_i}\frac{e^{x_1 \lambda_1 + x_2  
\lambda_2}}{\tau_{12}}( 1 - \frac{1}{2}\tau_{12}
+ \frac{1}{4}\tau_{12}^2 + \cdots)\nonumber\\
&=& \frac{e^{x_1 \lambda_1 + x_2 \lambda_2}}{\tau_{12}}( 1 -  
\frac{1}{2}\tau_{12}
+ \frac{1}{4}\tau_{12}^2 + \cdots)- \frac{e^{x_1 \lambda_2 + x_2  
\lambda_1}}{\tau_{12}}( 1 + \frac{1}{2}\tau_{12}
+ \frac{1}{4}\tau_{12}^2 + \cdots) \nonumber\\
\ea
This expansion is divided into two parts;
\ba\label{zero}
I_{\beta=2} &=& \frac{e^{x_1 \lambda_1 + x_2 \lambda_2}-e^{x_1  
\lambda_2 + x_2 \lambda_1}}{\tau_{12}}
\nonumber\\
&-& \frac{1}{2}[e^{x_1 \lambda_1 + x_2 \lambda_2}(1 -  
\frac{1}{2}\tau_{12} + \cdots)+
e^{x_1 \lambda_2 + x_2 \lambda_1}(1 + \frac{1}{2}\tau_{12} + \cdots)]
\ea
Remarkably, if one expands this expression in powers of the $x$'s and  
the $\lambda$'s,
one finds cancellations between the first and second term between  
brackets, and agreement with
  (\ref{IZform}).

Thus we find that the limit $\beta=2$ is somehow singular, and
the $\tau$-expansion, with the coefficients of Table B for $\beta=2$,  
gives
one  for the value of the function $f_{\beta=2}$ after  summing over  
the permutations.

We have thus recovered the ordinary HIZ formula of (\ref{IZform}).
The $\tau$-expansion for $\beta=2$ of (\ref{nontrivial}) may be useful  
for a check
  of the coefficients
for  general $\beta$, since it should reduce to one for  $f_{\beta=2}$.

%%%%%%%%%%%%%%%%%%%%%%%%%%%%%%%%%%%%%%%%%%%%%%%%%%%%%%%%%%%%%%%%

\section {Expansion of $f_\beta$ for large $\beta$, and for large k and  
fixed $\beta$}

We note that the final result for $f_\beta$ in the large $\beta$ limit  
is simple.
The coefficients $C$ involve the products of the inverse of the  
multiplicity $l!$.
For instance, the term of $\tau_{12}^2 \tau_{24}^3  
\tau_{23}\tau_{13}\tau_{34}$
has a coefficient as
\be
   \frac{(-1)^8}{2!3!k^8}\tau_{12}^2 \tau_{24}^3  
\tau_{23}\tau_{13}\tau_{34}
\ee
Namely $f_\beta$ is
\be\label{multi}
   f_\beta = \sum \frac{1}{l_1! l_2! \cdots} \frac{1}{k^n}(-1)^n \prod  
\tau_{ij}
\ee
where n is the total number of the bonds $\tau_{ij}$, and $l_j$ is the  
multiplicity.
  The expression (\ref{multi}) is simply equl to
\be\label{expo}
f_\beta = e^{-\frac{1}{k}\sum_{i < j} \tau_{ij}}
\ee
This formula is valid for large $\beta$ and for arbitrary k.
It may be interesting to refine above expression.
For  fixed  $\beta$ and in the large k limit, we have an expansion in  
powers of
  $\tau$. We may thus expand (\ref{DD2}) for large k.
$f$ is given by
\be
f_{\beta} =
e^{-\frac{1}{k}\sum_{i<j} \tau_{ij}}\sum_{m=0}^\infty
\frac{1}{m!}\frac{1}{\prod_{q=0}^{m-1}(1 + q \alpha)}
\sum_p \chi_p(1) \frac{Z_p(\tilde X)Z_p(\tilde \Lambda)}{Z_p(\rm I)}]
\ee
and as shown in Appendix D, or shown in the Table B, it is transformed  
into a $\tau$-series.
We
find that $f_\beta$ is given by
\be\label{expression}
   f_\beta = \prod_{i<j}^k
[ 1 - \frac{\tau_{ij}}{(\frac{\beta}{2}-1)k}]^{\frac{\beta}{2}-1}
\ee
  Of course it reduces  again to (\ref{expo}) in the large $\beta$ limit.
The above expression (\ref{expression}) automatically satisfies the  
condition
that the series of $f_\beta$ stops at the order
$\frac{\beta}{2}-1$ in $\tau$ when $\beta$ is an even integer.
Thus we have obtained the improved expression of (\ref{expression})
for $f_\beta$, which is valid  in the large
$\beta$ limit or in the large k limit. This asymptotic form  
(\ref{expression})
may be useful for
finite fixed $\beta$ (for instance, $\beta=1$) and  large k.

In the Appendix D, the large k limit of the coefficients $C$ at $l$-th  
order is investigated.
They are given by
\be
C = (-1)^l \frac{g}{\prod_{m=0}^{l-1}(k+ m \alpha)}(1 +  
O(\frac{\alpha}{k}))
\ee
where $g$ is a degeneracy factor for the multiple lines. It is  
remarkable that the only
dimensional
factor  which appears in the
large k limit,  is the first row of the  Young tableau, which has the  
form of $(k+ m \alpha)$.
For instance $(k-1)$,$(k+\alpha-1)$,..,which are the second row  
factors, do not appear in the large
k limit.

\section {Summary and discussions}

In this article, we have given the expression of the integral   
(\ref{I}) as series
in the variables $\tau_{ij}=
(x_i - x_j)(\lambda_i - \lambda_j)$ for the function $f$ defined in
(\ref{eq2}). As discussed in section seven, from the extended zonal  
polynomial expansion
of the integral $I$, and the use of the dual representation of  
(\ref{duality}), we
have obtained  expressions for this same function $f$. The coefficients  
$D_{[graph]}$
of this $\tau$ expansion are expressed through the dimensional constant  
$Z_{[p]}(\rm I)$ of the
extended zonal polynomials (Jack polynomials).
The results for these $C_{[graph]}$  coincide with the direct  
perturbational
calculations developed in the section five, where the residual  
equations among various
coefficients $C_{[graph]}$ are used. It is remarkable that these  
recursive residual equations
are indpendent of the parameter $\beta$. We have found the explicit  
expression
for the  WKB expansion  by this duality.

In this paper, we have proved that the extended  
Harish-Chandra-Itzykson-Zuber integral
for the general $\beta$ case is expressed by the variables of  
$\tau_{ij}$. The proof is given
in  two stages; the first is the expression of the  
Harish-Chandra-Itzykson-Zuber integral
by the zonal polynomial expansion with the parameter $\alpha$, which is  
$\frac{2}{2 - \beta}$
by the duality. The second is the transformations of the products of the
symmetric functions $s_n(\tilde x)$ in the zonal polynomial
expansion into the $\tau$ variables. This transformation is discussed
in Appendix B, and in Appendix D in detail.
We found that there are identities among the $\tau$ terms.
For instance, we have found explicitly these cubic and quartic  
identities in the appendix C.
These identities
give  a sort of gauge freedom to choose the values of the coefficients  
of the $\tau$ expansion.
   We have considered a fixing of these ambiguities from the large k  
behavior by imposing
definite asymptotic forms. For the $\beta=4$ case, only single line  
graphs appear,
and for this reason, there are
no ambiguities for the coefficients $C$, which are uniquely determined.

The integral (\ref{I}) is important for the investigation of the random  
matrix theory,
specially in the presence of an external matrix source, as shown for  
the $\beta=2$ in \cite{BHa,BHd,BHe,BHf}.
We will discuss in a separate paper the applications of the present  
results \cite{BHg}.

\vskip 8mm
%%%%%%%%%%%%%%%%%%%%%%%%%%%%%%%%%%%%
{\bf Acknowledgements}
\vskip 2mm
We thank Dr. A. Okounkov who pointed out us the Baker-Akhiezer formula  
by Berest.
S.H. is supported by  a Grant-in-Aid for Scientific Research (B) by  
JSPS.
%%%%%%%%%%%%%%%%%%%%%%%%%%%%%%%%%%%%%%%%%%%%
\vskip 5mm
{\bf Appendix A: Extended zonal polynomials (Jack polynomials)}
\vskip 5mm

%************************************************************
  We  have uses the expansions of the extended zonal polynomials ( named  
Jack polynomials)
in section seven, and we have derived the expression for $f$ from the
extended zonal polynomial expansion of the HCIZ-integral $I_{\beta}$ by
duality. Therefore, in this appendix, we give the needed important  
quantities,
the characters $\chi_p(1)$, and the dimensional constants $Z_p(\rm I)$
\cite{Jack,Jack2,Macdonald}.

The lower Jack symmetric polynomials $Z_{[p]}(X)$ and their dimensions
$Z_{[p]}(\rm I)$ are
\ba\label{Jackconstant}
   &&Z_{[1]}(X) = s_1, \hskip 10mm Z_{[1]}(\rm I) = k, \hskip 2mm  
\chi_{[1]}(1)=1\nonumber\\
   && Z_{[2]}(X) = s_1^2 + \alpha s_2, \hskip 10mm Z_{[2]}(\rm I) = k
(k + \alpha),\hskip 2mm \chi_{[2]}(1)=1\nonumber\\
   && Z_{[1^2]}(X) = s_1^2 - s_2, \hskip 10mm Z_{[1^2]}(\rm I) =  
k(k-1),\hskip 2mm
\chi_{[1^2]}(1)=\chi_{[1^2]}(1) = \alpha\nonumber\\
   && Z_{[3]}(X) = s_1^3 + 3 \alpha s_1 s_2 + 2 \alpha^2 s_3,\nonumber\\
       && \hskip 30mm
       Z_{[3]}(\rm I) = k(k + \alpha)
(k + 2 \alpha),
\hskip 2mm \chi_{[3]}(1)=1\nonumber\\
   && Z_{[21]}(X) = s_1^3 + (\alpha - 1)s_1 s_2 - \alpha s_3,
\nonumber\\
&& \hskip 30mm
Z_{[21]}(\rm I) = k (k + \alpha)(k-1),\hskip 2mm \chi_{[21]}(1) =  
\frac{6\alpha ( 1 + \alpha)}{
2 + \alpha}\nonumber\\
   && Z_{[1^3]}(X) = s_1^3 - 3 s_1 s_2 + 2 s_3,\hskip 5mm Z_{[1^3]}(\rm  
I) = k(k-1)(k-2)\nonumber\\
&& \hskip 30mm \chi_{[1^3]}(1) = \frac{\alpha^2(1 + 2 \alpha)}{2 +  
\alpha}
\ea
where $\alpha = \beta/2$. The classical symmetric functions are denoted  
by $s_n$,  $s_n= \sum x_i^n$.

Then one has the relation,
\be\label{sumrule2}
  (\rm tr X)^q = \frac{1}{\prod_{m=1}^{q-1}( 1 + m \alpha)}[\sum_p  
\chi_p(1) Z_p(x)]
\ee
where $p$ is a partition of the integer $q$, when $\alpha=2/\beta=2$.

The dimensional constants $Z_p(\rm I)$ are obtained by putting $X=I$ in  
the zonal polynomials
$Z_p(X)$. They are factorized as a polynomial in k.

 From the constants given in (\ref{Jackconstant}), and  the sum rule
(\ref{sumrule2}), we find the HIZ-integral for
general  $\beta$ :
\be
I_{\beta} = \sum_{m=0}^\infty \frac{1}{m!}\frac{1}{\prod_{q=0}^{m-1}(1  
+ q \alpha)}
\sum_p \chi_p(1) \frac{Z_p(X)Z_p(\Lambda)}{Z_p(\rm I)}
\ee
If $\Lambda$=I, it becomes
\be
I_{\beta} = e^{{\rm tr} X}
\ee
which is the correct expression by definition. The values in the  
following tables
of  coefficients of the zonal polynomials come from \cite{Jack2} (with   
a minor correction).
The characters $\chi_p(1)$, are then evaluated on the basis of these  
values.
  It agrees with \cite{James} when $\alpha=2$ up to  sixth order.

\vskip 30mm
{\bf The extended zonal polynomial (Jack polynomial)$ Z_p(x)$ with a  
parameter $\alpha$ and
its coefficient of the symmetric functions}\\
$\chi_p(1)$ is a character and $Z_p(\rm I)$ is a dimensional constant.

\vskip 2mm
{$ l=1$}
\vskip 2mm
\begin{tabular}{|   l||   l||   c|| c |}
\hline
      & $s_1$ & $\chi_p(1)$& $Z_{p}(\rm I)$ \\
\hline
      $Z_{[1]}$    &    1   &  1& $k$ \\
\hline
\end{tabular}

\vskip 4mm
{$l=2$}
\vskip 2mm
\begin{tabular}{|   l||   l| l ||   c|| c |}
\hline
      & $s_1^2$ & $ s_2$ & $\chi_p(1)$& $Z_p(\rm I)$ \\
\hline
      $Z_{[2]}$    &    1   &  $\alpha$ & 1& $k(k+\alpha)$ \\
      $Z_{[1^2]}$ & 1 & -1 & $\alpha$ & $k(k-1)$\\
\hline
\end{tabular}

\vskip 4mm
{$l=3$}
\vskip 2mm
\begin{tabular}{|   l||   l| l| l ||   c|| c |}
\hline
      & $s_1^3$ & $s_1 s_2$ & $s_3$ & $\chi_p(1)$ & $Z_p(\rm I)$ \\
\hline
      $Z_{[3]}$    &    1 & 3 $\alpha$   &  2 $\alpha^2$  &   1  &  
$k(k+\alpha)(k+ 2 \alpha)$ \\
      $Z_{[2,1]}$ & 1 & $\alpha-1$  &  $-\alpha$ & $\frac{6 \alpha (1 +  
\alpha)}{2 + \alpha}$&
$k(k+ \alpha)(k-1)$ \\
      $Z_{[1^3]}$ & 1 & - 3 & 2 & $\frac{\alpha^2 ( 1 + 2 \alpha)}{2 +  
\alpha}$ & $k(k-1)(k-2)$ \\
\hline
\end{tabular}
\vskip 8mm

\vskip 4mm
{$l=4$}
\vskip 2mm
\begin{tabular}{|   l||  l| l | l| l| l ||   c|}
\hline
      & $s_1^4$ & $s_1^2 s_2$ & $s_2^2$ & $s_1s_3$ & $s_4$ & $\chi_p(1)$  
\\
\hline
      $Z_{[4]}$    &    1 & 6 $\alpha$   &  $3 \alpha^2$& $8\alpha^2$ &  
$6 \alpha^3$  &   1
        \\
      $Z_{[3,1]}$  & 1    &  $3 \alpha-1$   & $-\alpha$ & $2 \alpha^2 -  
2 \alpha$ & $-2 \alpha^2$
         & $\frac{6 \alpha (1 + 2 \alpha)}{1 + \alpha}$  \\
      $Z_{[2^2]}$ & 1     & $2 \alpha - 2$ & $\alpha^2 + \alpha + 1$      
& $-4 \alpha$ & $\alpha-\alpha^2$
         & $\frac{6 \alpha^2(1 + 3 \alpha)}{(1 + \alpha)(2 + \alpha)}$
   \\
      $Z_{[2 1^2]}$ & 1 & $\alpha-3$ & $-\alpha$ & $2 - 2 \alpha$ & $2  
\alpha$
         & $\frac{6 \alpha^2 ( 1+2 \alpha)(1 + 3 \alpha)}{(1+ \alpha)(3  
+ \alpha)}$
      \\
      $Z_{[1^4]}$   & 1  & $-6$ & 3 & 8 & $- 6$
& $\frac{\alpha^3(1 + 2 \alpha)(1 + 3 \alpha)}{(2 + \alpha)(3 +  
\alpha)}$
     \\
\hline
\end{tabular}
\vskip 3mm
\begin{tabular}{| l || l |}
\hline
   & $Z_p(\rm I)$ \\
\hline
$Z_{[4]}$ & $k (k+ \alpha)(k+ 2 \alpha)( k + 3 \alpha)$ \\
$Z_{[3,1]}$& $ k (k+ \alpha)(k+ 2 \alpha)(k-1)$ \\
$Z_{[2^2]}$ & $ k (k + \alpha)(k + \alpha -1)(k-1)$ \\
$Z_{[2 1^2]} $ & $k(k+ \alpha)(k-1)(k-2)$ \\
$Z_{[1^4]} $ & $k(k-1)(k-2)(k-3)$\\
\hline
\end{tabular}
\vskip 8mm

{$l=5$}
\vskip 2mm
\begin{tabular}{|   l||  l| l | l| l| l | l| l|}
\hline
      & $s_1^5$ & $s_1^3 s_2$ & $s_1s_2^2$ & $s_1^2 s_3$ & $s_2 s_3$ &  
$s_1s_4$&$s_5$ \\
\hline
      $Z_{[5]}$    &    1 & 10 $\alpha$   &  $15 \alpha^2$& $20\alpha^2$  
& $20 \alpha^3$  &  30 $\alpha^3$ &
               24 $\alpha^4$  \\
      $Z_{[4,1]}$  & 1    &  $6 \alpha-1$   & $3\alpha(\alpha-1)$ &  
$\alpha (8 \alpha - 3)$
            & $-5 \alpha^2$   &  $6\alpha^2(\alpha-1)$ & $-6\alpha^3$ \\
      $Z_{[32]}$  &   1    & $2(2\alpha-1)$  &$3\alpha^2-\alpha + 1$  
&$2\alpha(\alpha-3)$
            & $2\alpha(\alpha^2 + 1)$      &  $-\alpha(7\alpha-1)$   &   
$-2\alpha^2(\alpha-1)$    \\
      $Z_{[31^2]}$ & 1     & $3(\alpha-1)$ & $-5\alpha$     &  
$2(\alpha-1)^2$ & $-2\alpha(\alpha-1)$
          & $-4\alpha(\alpha-1)$& $4\alpha^2$ \\
      $Z_{[2^2, 1]}$ & 1 & $2(\alpha-2)$ & $\alpha^2 -\alpha+3$ &  
$-2(3\alpha-1)$ & $-2(\alpha^2+1)$
              &$-\alpha(\alpha-7)$ &  $2 \alpha(\alpha-1)$\\
      $Z_{[21^3]}$   & 1  & $\alpha-6$ & $-3(\alpha-1)$ & $-(3\alpha-8)$  
& $5\alpha$ & $6(\alpha-1)$&
           $-6 \alpha$\\
      $Z_{[1^5]}$ &1 & $-10$& 15& 20 & $-20$ &$-30$ & 24\\
\hline
\end{tabular}
\vskip 2mm
\vskip 2mm
\begin{tabular}{|   l||  l| l | }
\hline
      & $\chi_p(1)$ & $Z_p(\rm I)$  \\
\hline
      $Z_{[5]}$    & $1$
                     &  $k(k+\alpha)(k+2\alpha)(k+3\alpha)(k+4\alpha)$   
\\
      $Z_{[4,1]}$  & $\frac{20\alpha(1+3\alpha)}{2 + 3\alpha}$
                  &  $k(k+\alpha)(k+2\alpha)(k+3\alpha)(k-1)$   \\
      $Z_{[32]}$  & $\frac{30\alpha^2(1+4\alpha)}{(1+\alpha)(2+\alpha)}$
                    & $k(k+\alpha)(k+ 2\alpha)(k-1)(k+\alpha-1)$  \\
      $Z_{[31^2]}$ &  
$\frac{30\alpha^2(1+2\alpha)(1+3\alpha)(1+4\alpha)}{(1+\alpha)(3+2\alpha 
)(2+3\alpha)}$
                   & $k(k+\alpha)(k+2\alpha)(k-1)(k-2)$ \\
      $Z_{[2^2, 1]}$ &  
$\frac{30\alpha^2(1+3\alpha)(1+3\alpha)(1+4\alpha)}{(1+\alpha)(2+\alpha) 
(3+\alpha)}$
                     & $k(k+\alpha)(k-1)(k+\alpha-1)(k-2)$ \\
      $Z_{[21^3]}$   &  
$\frac{20\alpha^3(1+2\alpha)(1+3\alpha)(1+4\alpha)}{(2+\alpha)(4+\alpha) 
(3+2\alpha)}$
                     & $k(k+\alpha)(k-1)(k-2)(k-3)$ \\
      $Z_{[1^5]}$ &   
$\frac{\alpha^4(1+2\alpha)(1+3\alpha)(1+4\alpha)}{(2+\alpha)(3+\alpha)(4 
+\alpha)}$
                     & $k(k-1)(k-2)(k-3)(k-4)$ \\
\hline
\end{tabular}
\vskip 2mm

\vskip 3mm
l=6
\vskip 2mm
\begin{tabular}{|   l||  l| l | l| l| l |  }
\hline
      & $s_1^6$ & $s_1^4 s_2$ & $s_1^2s_2^2$ & $s_2^3$ & $s_1^3 s_3$ \\
\hline
      $Z_{[6]}$    &    1 & 15 $\alpha$   &  $45\alpha^2$      &  
$15\alpha^3$ & $40\alpha^2$   \\
      $Z_{[5,1]}$  & 1    &  $10\alpha-1$ & $3\alpha(5\alpha-2)$ &  
$-3\alpha^2$
            & $4\alpha(5\alpha-1)$    \\
      $Z_{[42]}$  &   1    & $7\alpha-2$  &$9\alpha^2-5\alpha + 1$  
&$\alpha(3\alpha^2+\alpha+1)$
            & $8\alpha(\alpha-1)$           \\
      $Z_{[3^2]}$ & 1     & $3(2\alpha-1)$ & $3(3\alpha^2-\alpha+1)$      
& $-(5\alpha^2+3\alpha+1)$ & $4\alpha(\alpha-3)$
         \\
      $Z_{[41^2]}$ & 1 & $6\alpha-3$ & $3\alpha(\alpha-4)$ &  
$-3\alpha^2$ & $2(4\alpha^2-3\alpha+1)$
            \\
      $Z_{[321]}$   & 1  & $4(\alpha-1)$ & $3(\alpha-1)^2$ &  
$-\alpha(\alpha-1)$ & $2\alpha^2-9\alpha+2$
           \\
      $Z_{[31^3]}$ &1 & $3\alpha-6$&$-3(4\alpha-1)$& $3\alpha $&  
$2(\alpha^2-3\alpha+4)$ \\
      $Z_{[2^3]}$  &1  & $3(\alpha-2)$    &$3(\alpha^2-\alpha+3)$    
&$\alpha(\alpha^2+ 3\alpha+5)$    &$ -4(3\alpha-1)$          \\
      $Z_{[2^21^2]}$ &1 &$2\alpha-7$    &$\alpha^2-5\alpha + 9$    
&$-(\alpha^2+\alpha+3)$    & $-8(\alpha-1)$             \\
      $Z_{[21^4]}$   &1 & $\alpha-10$   & $-3(2\alpha-5)$  & $3\alpha$    
& $-4(\alpha-5)$            \\
      $Z_{[1^6]}  $  &1 & -15   &45   & -15   &  40        \\
\hline
\end{tabular}
\vskip 2mm
(continued)
\vskip 2mm
\begin{tabular}{|   l||  l| l | l| l|  }
\hline
      & $s_1s_2s_3$& $s_3^2$ & $s_1^2 s_4$ & $s_2s_4$   \\
\hline
      $Z_{[6]}$    & $120\alpha^3$    & 40 $\alpha^4$   &  $90\alpha^3$&  
$90\alpha^4$  \\
      $Z_{[5,1]}$  & $20\alpha^2(\alpha-1)$    &  $-8\alpha^3$   &  
$6\alpha^2(5\alpha-2)$ & $-18\alpha^3$
              \\
      $Z_{[42]}$  & $4\alpha(2\alpha-1)(\alpha-1)$      &  
$-2\alpha^2(\alpha-1)$  &$\alpha(6\alpha^2-17\alpha+1)$  
&$\alpha^2(6\alpha^2-\alpha+5)$
                    \\
      $Z_{[3^2]}$ & $12\alpha(\alpha^2+1)$     &  
$2\alpha^2(2\alpha^2+3\alpha+3)$ & $-3\alpha(7\alpha-1)$     &  
$-3\alpha(4\alpha^2+\alpha+1)$
          \\
      $Z_{[41^2]}$ & $-6\alpha(3\alpha-1)$ & $4\alpha^2$ &  
$6\alpha(\alpha-1)^2$ & $-6\alpha^2(\alpha-1)$
       \\
       $Z_{[321]}$   &$(\alpha-1)(\alpha-2)(2\alpha-1)$   &  
$-\alpha(2\alpha^2+\alpha+2)$ & $-9\alpha(\alpha-1)$ &  
$-2\alpha(\alpha-1)^2$
           \\
      $Z_{[31^3]}$ &$-6\alpha(\alpha-3)$ & $4\alpha^2$&  
$-6(\alpha-1)^2$& $6\alpha(\alpha-1)$ \\
      $Z_{[2^3]}$  &$-12(\alpha^2+1)$  & $2(3\alpha^2+3\alpha+2)$    &  
$-3\alpha(\alpha-7)$  & $-3\alpha(\alpha^2+\alpha+4)$            \\
      $Z_{[2^21^2]}$ &$-4(\alpha-1)(\alpha-2)$ &$2\alpha(\alpha-1)$    &  
$-(\alpha^2-17\alpha+6)$  & $5\alpha^2-\alpha+6$            \\
      $Z_{[21^4]}$   &$20(\alpha-1)$ & $-8\alpha$   & $6(2\alpha-5)$  &  
$-18\alpha$                \\
      $Z_{[1^6]}  $  & -120& 40   & -90  &  90                \\
\hline
\end{tabular}
(continued)
\vskip 2mm
\begin{tabular}{|   l|| l| l|| l |}
\hline
                 &$s_1s_5$  &   $s_6$    &$\chi_p(1)$         \\
\hline
      $Z_{[6]}$   &144$\alpha^4$ &  $ 120 \alpha^5$        &             
1      \\
      $Z_{[5,1]}$  &$24\alpha^3(\alpha-1)$&  $-24\alpha^4$          &  $  
  \frac{15\alpha(1 + 4\alpha)}{1 + 2 \alpha} $\\
      $Z_{[42]}$  &$-4\alpha^2(5\alpha-1)$ &   $-6\alpha^3(\alpha-1)$
       & $\frac{90 \alpha^2(1 + 2 \alpha)(1 + 5\alpha)}{  (1 +  
\alpha)^2(2 + 3 \alpha)} $                \\
      $Z_{[3^2]}$ &$-12\alpha^2(\alpha-1)$ &   
$-2\alpha^2(\alpha-1)(2\alpha-1)$
      &  $ \frac{30 \alpha^3(1 + 4\alpha)(1 + 5 \alpha)}{(1 +  
\alpha)^2(2 + \alpha)(1+ 2 \alpha)} $                  \\
      $Z_{[41^2]}$ &$-12\alpha^2(\alpha-1)$ &  $12\alpha^3$
       & $ \frac{10\alpha^2(1 + 3\alpha)(1 + 4\alpha)(1 + 5\alpha)}{(1 +  
\alpha)^2(1 + 2 \alpha)}  $                      \\
      $Z_{[321]}$  & $-\alpha(2\alpha^2-13\alpha+2)$&   
$4\alpha^2(\alpha-1)$
        & $\frac{720 \alpha^3 (1 + \alpha)(1 + 3 \alpha)
        (1 + 4 \alpha)(1+ 5 \alpha)}{(2 + \alpha)^2(1 + 2 \alpha)(3 +  
2\alpha)(2 + 3\alpha)} $ \\
      $Z_{[31^3]}$ &$12\alpha(\alpha-1)$ &  $-12\alpha^2$
       &   $\frac{10 \alpha^3(1 + 2 \alpha)(1 + 3\alpha)(1 +  
4\alpha)(1+5\alpha)}{(1+\alpha)^2
         (2+\alpha)^2}$                         \\
      $Z_{[2^3]}$ & $12\alpha(\alpha-1)$ &    
$2\alpha(\alpha-2)(\alpha-1)$
      &  $\frac{30 \alpha^4(1 + 3 \alpha)(1 + 4\alpha)(1+ 5\alpha)}{(1 +  
\alpha)^2(2+\alpha)^2(3+\alpha)}$                          \\
      $Z_{[2^21^2]}$ &$4\alpha(\alpha-5)$ & $-6\alpha(\alpha-1)$
       &  $\frac{90\alpha^4(1 + 2\alpha)(1 + 3 \alpha)(1 + 4\alpha)(1 +  
5\alpha)}{(1 + \alpha)^2
       (3 + \alpha)(4 + \alpha)(3 + 2 \alpha)}$                      \\
      $Z_{[21^4]}$    &$-24(\alpha-1)$ & $24\alpha$
       & $\frac{15\alpha^4(1 + 2\alpha)(1 + 3\alpha)(1+4\alpha)(1 +  
5\alpha)}{(2 + \alpha)^2(3 + \alpha)(5 +\alpha)}$                       
\\
      $Z_{[1^6]}  $ & 144 &   $-120$       &  $\frac{\alpha^5(1+ 2  
\alpha)(1 + 3\alpha)(1 + 4\alpha)
      (1+5\alpha)}{(2 + \alpha)(3 + \alpha)(4 + \alpha)(5 + \alpha)}$     
                   \\
\hline
\end{tabular}

\vskip 2mm
\begin{tabular}{|   l|| l| }
\hline
                 &$Z_p(\rm I)$  \\
\hline
      $Z_{[6]}$   & $    k(k+\alpha)(k+2 \alpha)(k+ 3\alpha)(k+  
4\alpha)(k+ 5\alpha)         $ \\
      $Z_{[5,1]}$  &$  k(k+\alpha)(k+2\alpha)(k+3\alpha)(k+4\alpha)(k-1)  
$ \\
      $Z_{[42]}$  &$     k(k+\alpha)(k+2\alpha)(k+3\alpha)(k-1)  
(k+\alpha-1)      $ \\
      $Z_{[3^2]}$ &$   k(k+\alpha)(k+2\alpha)(k-1) (k+\alpha-1) (k+  
2\alpha-1)                $ \\
      $Z_{[41^2]}$ &$   k(k+\alpha)(k+2\alpha)(k+3\alpha)(k-1) (k-2)      
                          $   \\
      $Z_{[321]}$  & $  k(k+\alpha)(k+2\alpha)(k-1) (k+\alpha-1) (k-2)    
                 $\\
      $Z_{[31^3]}$ &$   k(k+\alpha)(k+2\alpha)(k-1)  (k-2)  (k-3)         
                           $ \\
      $Z_{[2^3]}$ & $  k(k+\alpha)(k-1)(k+\alpha-1)(k-2)(k+\alpha-2)      
$      \\
      $Z_{[2^21^2]}$ &$  k(k+\alpha)(k-1)(k+\alpha-1)(k-2)(k-3)           
                                                             $          
\\
      $Z_{[21^4]}$    &$  k(k+\alpha)(k-1)(k-2)(k-3)(k-4)                 
   $        \\
      $Z_{[1^6]}  $ &  $  k(k-1)(k-2)(k-3)(k-4)(k-5)                     
$       \\
\hline
\end{tabular}
\vskip 8mm

\vskip 5mm
{\bf Appendix B: The transformation of the paired products of
  symmetric functions $s_n(\tilde x)$ (power sum) to  $\tau$-polynomials}
\vskip 5mm

We consider the transformation of the paired products of the
classical symmetric function $s_n(\tilde x)$, and $s_n(\tilde \lambda)$  
to the $\tau$-polynomials,
where
$\tilde x_i = x_i -
\frac{1}{k}\sum x_j$, and $\tau_{ij}=(x_i - x_j)(\lambda_i -  
\lambda_j)$, and $s_n(\tilde x)
= \sum \tilde x_i^n$.

At oprder two, $s_2(\tilde x)s_2(\tilde \lambda) $ is expressed
in terms of $\tau_{ij}$ by
\be\label{s2bis}
s_2(\tilde x)s_2(\tilde \lambda)= \sigma_{\rm II}\cdot [ \tau_{ij}^2] +  
\sigma_{\Lambda}\cdot
   [\tau_{12}\tau_{13}] + \sigma_{\rm I,\rm I}\cdot [\tau_{12}\tau_{34}]
\ee
where the coefficients $\sigma$ are functions of $k$.
We apply the differential operators $D_{k,l}^{i,j}=
\frac{\partial^4}{\partial x_i \partial x_j \partial
\lambda_k \partial \lambda_l}$ on  both sides of the above equation.
We obtain
\be
D_{2,2}^{1,1}[\tau_{12}^2] = 4, \hskip 3mm
D_{2,2}^{1,1}[\tau_{12}\tau_{13}]= D_{2,2}^{1,1}[\tau_{12}\tau_{34}]=0
\ee
We use the notation $[\tau_{12}^2]$ for $\sum_{i<j} \tau_{ij}^2$,  
$[\tau_{12}\tau_{13}]$
for $\sum_{i<j<k} \tau_{i,j}\tau_{i,k}$,etc.
For the symmetric function $s_2(\tilde x)s_2(\tilde \lambda)$, we have
\be
D_{2,2}^{1,1}(s_2(\tilde x)s_2(\tilde \lambda))= \frac{4(k-1)^2}{k^2}
\ee
Therefore, we obtain the coefficient of $[\tau_{12}^2]$ as $\sigma_{\rm  
II}=\frac{(k-1)^2}{k^2}$.
Similarly, we apply $D_{2,3}^{1,1}$ and $D_{2,4}^{1,3}$, and then we  
find
the coefficients of $[\tau_{12}\tau_{13}]$ and $[\tau_{12}\tau_{34}]$,
  \be
D_{2,3}^{1,1}[\tau_{12}^2]=0,\,D_{2,3}^{1,1}[\tau_{12}\tau_{13}]=2,\,
D_{2,3}^{1,1}[\tau_{12}^2]=0,\,
D_{2,3}^{1,1}(s_2(\tilde x)s_2(\tilde \lambda))=-\frac{4(k-1)}{k^2}
\ee
These equations  give the coefficient of $[\tau_{12}\tau_{13}]$ as  
$\sigma_{\Lambda}=
  -\frac{2(k-1)}{k^2}$.
\be
D_{2,4}^{1,3}[\tau_{12}^2] = 0,\,  
D_{2,4}^{1,3}[\tau_{12}\tau_{13}]=0,\,D_{2,4}^{1,3}[\tau_{12}
\tau_{3,4}] =2,\, D_{3,4}^{1,3}(s_2(\tilde x)s_2(\tilde \lambda))=  
\frac{4}{k^2}
\ee
These equations give the coefficient of $[\tau_{12}\tau_{34}]$ as  
$\sigma_{\rm I,\rm I}=
\frac{2}{k^2}$. These results give (\ref{s2}).

We have examined all possible differential operators of order two for  
(\ref{s2bis}) ; they are
$D_{22}^{11},D_{12}^{11},D_{12}^{12},D_{11}^{11},D_{23}^{11},D_{13}^{12} 
,
D_{34}^{12}$.
These operators confirm the equation (\ref{s2bis}) with the  
coefficients determined hereabove. `
We have thus established  the identity  (\ref{s2bis}).

At order three, we transform $s_3(\tilde x)s_3(\tilde \lambda)$ to   
$\tau$-polynomials.
\ba\label{ss3}
s_3(\tilde X)s_3(\tilde \Lambda) &=& \sigma_{\rm III}\cdot[ \tau_{12}^3  
]+
\sigma_{\underline{\angle}}\cdot
[\tau_{12}^2\tau_{13}] +  
\sigma_{\bigtriangleup}\cdot[\tau_{12}\tau_{13}\tau_{23}]
\nonumber\\
&+&\sigma_{\rm Y} \cdot[\tau_{12}\tau_{13}\tau_{14}] + \sigma_{\rm  
N}\cdot [\tau_{12}\tau_{13}\tau_{34}]
+ \sigma_{\rm II,\rm I}\cdot[\tau_{12}^2\tau_{34}]\nonumber\\
&+&\sigma_{\Lambda,\rm I}\cdot[\tau_{12}\tau_{13}\tau_{45}]+  
\sigma_{\rm I,\rm I,\rm I}
\cdot[\tau_{12}\tau_{34}
\tau_{56}]
\ea
By applying $D_{2,2,2}^{1,1,1}= \frac{\partial^6}{\partial x_1^3  
\partial \lambda_2^3}$ on
both sides of (\ref{ss3}), we find two non-vanishing contributions :
\be
D_{2,2,2}^{1,1,1} [\tau_{12}^3]=-36, \,D_{2,2,2}^{1,1,1} (s_3(\tilde  
x)s_3(\tilde \lambda)) = 36
\frac{(k-1)^2(k-2)^2}{k^4}
\ee
 From this result, we obtain $\sigma_{\rm III} = -  
\frac{(k-1)^2(k-2)^2}{k^4}$.
 From $D_{2,2,3}^{1,1,1}$, we obtain
\be
D_{2,2,3}^{1,1,1}[\tau_{12}^2\tau_{13}] = -12,\, D_{2,2,3}^{1,1,1}  
(s_3(\tilde x)s_3(\tilde \lambda))
= - 36 \frac{(k-1)(k-2)^2}{k^4}
\ee
which gives $\sigma_{\underline{\angle}}= \frac{3(k-1)(k-2)^2}{k^4}$.
By the differentiation $D_{2,3,4}^{1,1,1}$, we obtain
\be
D_{2,3,4}^{1,1,1} [\tau_{12}\tau_{13}\tau_{14}] = -6,\,  
D_{2,3,4}^{1,1,1} (s_3(\tilde x)s_3(\tilde \lambda))
= 72 \frac{(k-1)(k-2)}{k^4}
\ee
which reads to $\sigma_{\rm Y} = - \frac{12}{k^4}(k-1)(k-2)$.
For the differentiation $D_{2,3,5}^{1,1,4}$, noting that  
$[\tau_{12}\tau_{13}\tau_{45}]$ includes
a sum of the relevant terms  
$\tau_{12}\tau_{13}\tau_{45}+\tau_{12}\tau_{15}\tau_{34}+\tau_{13}\tau_{ 
15}
\tau_{24}$, we obtain
\be
D_{2,3,5}^{1,1,4} [\tau_{12}\tau_{13}\tau_{45}] = -6,\,  
D_{2,3,5}^{1,1,4}(s_3(\tilde x)s_3(\tilde \lambda))
= - 72 \frac{k-2}{k^4}
\ee
which reads  $\sigma_{\Lambda,\rm I} = 12 \frac{k-2}{k^4}$.
By the differentiation $D_{2,4,6}^{1,3,5}$, we obtain
\be
D_{2,4,6}^{1,3,5} [\tau_{12}\tau_{34}\tau_{56}] = - 6,\,  
D_{2,4,6}^{1,3,5}
(s_3(\tilde x)s_3(\tilde \lambda)) = 144 \frac{1}{k^4}
\ee
which reads $ \sigma_{\rm I,\rm I,\rm I} = - \frac{24}{k^4}$.
Thus we determine 5 coefficients of $\sigma$ as the function of $k$,  
uniquely.
These coefficients are represented in (\ref{s3}).
For other 3 coefficients, $\sigma_{\bigtriangleup},\sigma_{\rm  
N},\sigma_{\rm II,\rm I}$,
we need other differential operators, which give coupled equations.

By the differentiation $D_{2,2,4}^{1,1,3}$ of (\ref{ss3}), two terms of  
$\tau$ are non-vanishing.
\ba
&&D_{2,2,4}^{1,1,3}\{ \sigma_{\rm N}\cdot[\tau_{12}\tau_{13}\tau_{24}]  
+ \sigma_{\rm II,\rm I}\cdot
[\tau_{12}^2 \tau_{34}]\}= -4 \sigma_{\rm N} - 4 \sigma_{\rm II,\rm  
I}\nonumber\\
&&D_{2,2,4}^{1,1,3}(s_3(\tilde x)s_3(\tilde \lambda)) = 36  
\frac{(k-2)^2}{k^4}
\ea
which reads
\be\label{cubic1}
  \sigma_{\rm N} +  \sigma_{\rm II,\rm I} = - 9 \frac{(k-2)^2}{k^4}.
\ee
   By the differentiation $D_{3,3,2}^{1,1,2}$, we obtain from (\ref{ss3})
\be
8 \sigma_{\underline{\angle}} - 4 \sigma_{\bigtriangleup} +  
4(k-3)\sigma_{\rm II,\rm I}=
36 \frac{(k-2)^2}{k^4}
\ee
Using the value of $\sigma_{\underline{\angle}}$, we have from above  
equation,
\be\label{cubic2}
   - \sigma_{\bigtriangleup} + (k-3)\sigma_{\rm II,\rm I} = \frac{(15 -  
6 k) (k-2)^2}{k^4}.
\ee

By the differentiation $D_{2,2,3}^{1,1,2}$, we obtain
\be
\sigma_{\underline{\angle}} + \sigma_{\bigtriangleup} + (k-3)  
\sigma_{\rm N} = 9 \frac{(k-2)^2}{k^4}
\ee
Using the obtained value of $\sigma_{\underline{\angle}}$, we have from  
above equation,
\be\label{cubic3}
\sigma_{\bigtriangleup} + (k-3)\sigma_{\rm N} =  
\frac{3(k-2)^2(4-k)}{k^4}.
\ee

The three relations of (\ref{cubic1}),(\ref{cubic2}) and (\ref{cubic3})  
are not linearly independent.
Indeed if we shift $\sigma_{\rm II,\rm I} \rightarrow \sigma_{\rm  
II,\rm I} + \alpha$,
$\sigma_{\rm N} \rightarrow \sigma_{\rm N} - \alpha$ and  
$\sigma_{\bigtriangleup}
\rightarrow \sigma_{\bigtriangleup} + \alpha (k-3)$, these relations  
are unchanged. It means that
there is a cubic identity,
\be
I_3 =[\rm II,\rm I] - [\rm N] + (k-3) [\bigtriangleup] = 0.
\ee
The proof of this identity is discussed in the appendix C.
Therefore, the coefficients $\sigma_{\rm II,\rm I},\sigma_{\rm  
N},\sigma_{\bigtriangleup}$ have ambiguities
up to a parameter $\alpha$, which is an arbitrary constant.
One can add a term  $\alpha I_3$, i.e.
zero, to the expression of $s_3(\tilde x)s_3(\tilde \lambda)$.

This may be used to write simple expressions for these coefficients,  
for instance,
\be
\sigma_{\rm II,\rm I} = - 6\frac{(k-2)^2}{k^4}, \hskip 5mm
\sigma_{\rm N} = - 3 \frac{(k-2)^2}{k^4},\hskip 5mm
\sigma_{\bigtriangleup}= 3\frac{(k-2)^2}{k^4}
\ee

We have examined   the transformation of (\ref{ss3})
by all possible differential operators $D_{l,m,n}^{i,j,k}$
on (\ref{ss3}). These  operators are,  
$D_{111}^{111},D_{112}^{111},D_{122}^{111},D_{223}^{112},
D_{123}^{111},D_{223}^{111}$,
$D_{223}^{113},D_{123}^{123},D_{123}^{112},D_{223}^{112},D_{234}^{111},D 
_{334}^{112},
D_{234}^{112},D_{124}^{123},D_{134}^{112}$,
$D_{145}^{123},D_{456}^{123},D_{124}^{123},D_{345}^{112}$.
This means that we have proved (\ref{ss3}) with explicit coefficients.

In  fourth order, the products of  symmetric functions $(s_2(\tilde  
x)s_2(\tilde \lambda))^2$,\\
$s_2(\tilde x)^2 s_4(\tilde \lambda) + s_4(\tilde x)s_2(\tilde  
\lambda)^2$, and $s_4(\tilde x)s_4(
\tilde \lambda)$ appear in the zonal polynomial expansions.
Since $s_2(\tilde x)s_2(\tilde \lambda)$ is transformed as  
(\ref{s2bis}),
  $(s_2(\tilde x)s_2(\tilde \lambda))^2$ is transformed as the square of  
(\ref{s2bis}).

We first consider the transformation of $s_4(\tilde x)s_4(\tilde  
\lambda)$. It is written as a sum
of 23 possible terms,
\ba\label{s4}
s_4(\tilde x)s_4(\tilde \lambda) &=&
\sigma_{[+]}[\tau_{12}\tau_{13}\tau_{14}\tau_{15}] +
\sigma_{[{\angle  
\angle}]}[\tau_{12}\tau_{13}\tau_{34}\tau_{35}]\nonumber\\
&+&\sigma_{[{\unrhd}]}[\tau_{12}\tau_{13}\tau_{14}\tau_{23}] +
\sigma_{[\Box]} [\tau_{12}\tau_{13}\tau_{24}\tau_{34}] \nonumber\\
&+&\sigma_{[\rm M]} [\tau_{12}\tau_{13}\tau_{34}\tau_{45}] +
\sigma_{[\rm N,\rm I]} [\tau_{12}\tau_{13}\tau_{34}\tau_{56}]\nonumber\\
&+&\sigma_{[\rm Y,\rm I]} [\tau_{12}\tau_{13}\tau_{14}\tau_{56}] +
\sigma_{[\bigtriangleup,\rm I]}  
[\tau_{12}\tau_{13}\tau_{23}\tau_{45}]\nonumber\\
&+&\sigma_{[\wedge,\wedge]} [\tau_{12}\tau_{13}\tau_{45}\tau_{46}] +
\sigma_{[\wedge,\rm I,\rm I]}  
[\tau_{12}\tau_{13}\tau_{45}\tau_{67}]\nonumber\\
&+&
\sigma_{[\rm I,\rm I,\rm I,\rm I]}  
[\tau_{12}\tau_{34}\tau_{56}\tau_{78}] +
\sigma_{[\hskip 1mm{\underline{\amalg}}\hskip 1mm]}  
[\tau_{12}^2\tau_{13}\tau_{24}]\nonumber\\
&+&
\sigma_{[\sqsupseteq]}  [\tau_{12}^2\tau_{13}\tau_{34}] +
\sigma_{[\hskip 1mm{\underline{\bigtriangleup}}\hskip 1mm]}  
[\tau_{12}^2\tau_{13}\tau_{23}]\nonumber\\
&+&\sigma_{[\ll]} [\tau_{12}^2\tau_{13}^2] +
\sigma_{[{\models}]} [\tau_{12}^2\tau_{13}\tau_{14}]\nonumber\\
&+& \sigma_{[\hskip 1mm{\underline{\angle},\rm I}]}  
[\tau_{12}^2\tau_{13}\tau_{45}] +
\sigma_{[\wedge,\rm II]} [\tau_{12}\tau_{13}\tau_{45}^2]\nonumber\\
&+&\sigma_{[\rm II,\rm I, \rm I]} [\tau_{12}^2\tau_{34}\tau_{56}] +
\sigma_{[\rm II,\rm II]} [\tau_{12}^2\tau_{34}^2]\nonumber\\
&+&\sigma_{[\hskip 1mm{\underline{\underline{\angle}}}\hskip 1mm]}
[\tau_{12}^3\tau_{13}] +
\sigma_{[\rm I \rm I \rm I,\rm I]} [\tau_{12}^3\tau_{34}]\nonumber\\
&+&\sigma_{[\hskip 1mm \rm IIII \hskip 1mm]}[\tau_{12}^4]
\ea

A systematic way to determine these coefficients $\sigma$ consists in  
classifying
the terms by the number  $l$ of the points in the graph.

For $l=8$, we make use of the differential operator $D_{5678}^{1234}$,  
and the graph is $[\rm I,\rm I,\rm I,\rm I]$.
Then, we find uniquely that
\be
\sigma_{\rm I,\rm I,\rm I,\rm I} = \frac{216}{k^6}
\ee
For l=7, we have two differential operators, $D_{4567}^{1123}$ and  
$D_{1567}^{1234}$, which
act on the possible two graphs $[\rm I,\rm I,\rm I,\rm I]$ and  
$[\Lambda,\rm I,\rm I]$ (other
graphs have  less than  7 points and do not contribute).
 From $D_{4567}^{1123}$, we obtain
\be
\sigma_{\Lambda,\rm I,\rm I} = -72 \frac{(k-3)}{k^6}
\ee
For l=6, in addition to the above two graphs, we have 4 graphs,
$[\rm N,\rm I]$,$[\rm Y, \rm I]$,\\
$[\Lambda,\Lambda]$,and $[\rm II,\rm  
I,\rm I]$.
We have 6 different differential operators,  
$D_{3456}^{1112}$,$D_{1456}^{1123}$,
$D_{3456}^{1122}$,$D_{2256}^{1134}$,
$D_{1256}^{1234}$,and $D_{2456}^{1123}$. From $D_{3456}^{1112}$, we  
find uniquely,
\be
  \sigma_{[\rm Y,\rm I]} = 72 \frac{(k^2 - 3 k + 3)}{k^6}
\ee
 From $D_{3456}^{1122}$, we find
\be
\sigma_{[\Lambda,\Lambda]} = - 72 \frac{(2k-3)}{k^6}.
\ee
 From $D_{1456}^{1123}$, we have
\be
-24 \sigma_{\rm II,\rm I,\rm I} + 12\sigma_{\rm Y,\rm I} - 48  
\sigma_{\rm N,\rm I} - 12 (k-6)
\sigma_{\Lambda,\rm I,\rm I} = - 24\cdot24\cdot 3\frac{(k-3)}{k^6}
\ee
which reads to
\be\label{1123}
\sigma_{\rm II,\rm I,\rm I} + 2 \sigma_{\rm N,\rm I} = 36 \frac{( 2 k^2  
- 10 k + 15)}{k^6}
\ee
 From $D_{2256}^{1134}$, we obtain the same relation.

 From $D_{1256}^{1234}$, we obtain
\be
8 \sigma_{\rm II,\rm I,\rm I} + 14 (k-6) \sigma_{\Lambda,\rm I,\rm I} +  
2 (k-6)(k-7)
\sigma_{\rm I,\rm I,\rm I,\rm I} + 4 \sigma_{\Lambda,\Lambda}+ 16  
\sigma_{\rm N,\rm I}
= 24\cdot 24\cdot 9 \frac{1}{k^6}
\ee
which gives again,
\be\label{1256}
\sigma_{\rm II,\rm I,\rm I} + 2 \sigma_{\rm N,\rm I} = 36 \frac{(2 k^2  
- 10 k + 15)}{k^6}
\ee
Thus we are unable to find  definite valuesfor  $\sigma_{\rm II,\rm  
I,\rm I}$ and $\sigma_{\rm N,\rm I}$.
The reason of this indefiniteness is  the existence of an identity,  
which is an extension
of the cubic identity.
In the appendix C, we have derived the following identity  
(\ref{quarticI}) :
\be
[\rm II,\rm I,\rm I] - [\rm N,\rm I] + (k-5)[\bigtriangleup, \rm I] = 0.
\ee
This identity gives  a freedom in the choice of their respective values.

At next order $l=5$ (five points), in addition to these 6 graphs, we  
have to consider six new
graphs, $[\rm X]$,$[\angle\angle]$,$[\rm M]$,$[\bigtriangleup,\rm  
I]$,$[\underline{\angle},\rm I]$,
$[\Lambda,\rm II]$.

For the differential operators $D_{ijkl}^{mnst}$, we have the following  
different kinds
for $l=5$,
\ba\label{12D}
&&D_{2345}^{1111},\hskip 3mm D_{2345}^{1112},\hskip 3mm  
D_{2345}^{1122},\hskip 3mm D_{2345}^{1123},
\hskip 3mm D_{3345}^{1112},\hskip 3mm D_{3345}^{1122}
\nonumber\\
&&D_{1235}^{1234},\hskip 3mm D_{1145}^{1123}, \hskip 3mm  
D_{1245}^{1123},\hskip 3mm D_{2235}^{1124},\hskip 3mm D_{2235}^{1134}
,\hskip 3mm D_{1345}^{1112}\nonumber\\
\ea
where we note that  
$D_{1445}^{1123}=D_{1225}^{1134}=D_{2235}^{1124}$,$D_{2345}^{1223}=D_{13 
45}^{1123}=D_{1245}^{1123}$.
(the equal sign means the equivalence for the operator in this problem).

We have uniquely from $D_{2345}^{1111}$,
\be\label{X}
    \sigma_{\rm X} = - 72 \frac{(k-1)(k^2 - 3 k + 3)}{k^6}
\ee
We next obtain the coupled equations.
 From $D_{3345}^{1122}$, we have
\be\label{M}
   \sigma_{\rm M} + \sigma_{\Lambda,\rm II} = 36 \frac{(k-3)(2 k -  
3)}{k^6}
\ee
 From $D_{2345}^{1112}$, we obtain
\be\label{XY}
\sigma_{\rm X} + (k-5)\sigma_{\rm Y,\rm I} = -288 \frac{(k^2 - 3 k +  
3)}{k^6}
\ee
 From $D_{3345}^{1112}$, we obtain
\be\label{LL}
\sigma_{\angle\angle}+ 2 \sigma_{\underline{\angle},\rm I} = -48  
\frac{(k-3)(k^2 - 3 k + 3)}{k^6}
\ee
 From $D_{1345}^{1112}$, we have
\be
\sigma_{\rm X} - 6 \sigma_{\angle\angle} - 3 (k-5) \sigma_{\rm Y,\rm I}
  - 12 \sigma_{\underline{\angle},\rm I} = 288 \frac{(k^2 - 3 k +  
3)}{k^6}
\ee
which reduces to (\ref{LL}).

 From $D_{2345}^{1122}$, we have
\be
2 \sigma_{\Lambda,\rm II} + 2 \sigma_{\rm M} + (k-5)  
\sigma_{\Lambda,\Lambda} =
144 \frac{(2k-3)}{k^6}
\ee
This reduces to (\ref{M}).

 From $D_{2235}^{1124}$, we get
\ba\label{2235}
&&\sigma_{\angle\angle}- 6 \sigma_{\rm M} - 2 (k-5) \sigma_{\rm N,\rm  
I} - 2 \sigma_{\bigtriangleup,\rm I}
-(k-5)\sigma_{\Lambda,\Lambda}\nonumber\\
&&- 2 \sigma_{\underline{\angle},\rm I} - 2 \sigma_{\Lambda,\rm II}
= 144 \frac{(k-3)^2}{k^6}
\ea
 From $D_{2235}^{1134}$, we have
\be\label{2235b}
- \sigma_{\rm M} +2 \sigma_{\angle\angle} + (k-5)\sigma_{\rm N,\rm I} -  
\sigma_{\bigtriangleup,\rm I}
+ 2 \sigma_{\underline{\angle},\rm I} +
(k-5) \sigma_{\rm II,\rm I,\rm I} + \sigma_{\Lambda,\rm II}
=
-144 \frac{(k-3)^2}{k^6}
\ee
 From $D_{1145}^{1123}$, we have
\ba\label{1145}
&&\sigma_{\rm X} - 8 \sigma_{\angle\angle} + 8 \sigma_{\rm M} + 8  
(k-5)\sigma_{\rm N,\rm I}
  - 4 \sigma_{\rm Y,\rm I}
+ 4 \sigma_{\bigtriangleup,\rm I}\nonumber\\
&&+ (k-5)(k-6)\sigma_{\Lambda,\rm I,\rm I} - 8  
\sigma_{\underline{\angle},\rm I} +
2 (k-5) \sigma_{\rm II,\rm I,\rm I}= 144 \frac{(k-3)^2}{k^6}
\ea

These three equations coincide when we put the known values of the  
coefficients.

The last differentiation in (\ref{12D}),
$D_{1235}^{1234}$, gives
\ba
&&\sigma_{\rm X} - 12 \sigma_{\angle\angle} - 12 \sigma_{\rm M} - 24  
(k-5)\sigma_{\rm N,\rm I}
- 4 (k-5) \sigma_{\rm Y,\rm I}\nonumber\\
&&- 9 (k-5) \sigma_{\Lambda,\Lambda} - 12 (k-5)(k-6)  
\sigma_{\Lambda,\rm I,\rm I} -
(k-5)(k-6)(k-7) \sigma_{\rm I,\rm I,\rm I,\rm I}\nonumber\\
&& - 12 \sigma_{\Lambda,\rm II} - 12 (k-5)
\sigma_{\rm II,\rm I,\rm I} - 24 \sigma_{\underline{\angle},\rm I}=  
5184\frac{1}{k^6}
\ea
However, this leads to known identities rather than to a new relation.

Several coefficients of the terms, which have less than 4 points,
  are
determined uniquely by the  differential operators. We list them here,
\ba
\sigma_{[\rm IIII]}&=& \frac{1}{k^6}(k-1)^2(k^2- 3 k + 3)^2,\hskip 5mm  
(D_{2,2,2,2}^{1,1,1,1})\nonumber\\
\sigma_{[\models]} &=& 12 \frac{(k-1)(k-3)(k^2-3 k + 3)}{k^6},\hskip  
5mm (D_{2,2,3,4}^{1,1,1,1})\nonumber\\
\sigma_{[\ll]} &=& 6 \frac{(k-1)(2 k - 3)(k^2 - 3 k + 3)}{k^6},\hskip  
5mm(D_{2,2,3,3}^{1,1,1,1})\nonumber\\
\sigma_{[\hskip 1mm\underline{\underline{\angle}}\hskip 1mm]}
&=& - 4 \frac{(k-1)(k^2 - 3 k + 3)^2}{k^6},\hskip 5mm  
(D_{2,2,2,3}^{1,1,1,1})\nonumber\\
\sigma_{\sqsupseteq} &=&
-12 \frac{(2k-3)(k^2- 3 k + 3)}{k^6}, \hskip 5mm (D_{2244}^{1113}).
\ea

We have coupled equations by other differential operators.
We list them with the differential operator $D_{\lambda}^{x}$, which  
are used.
\be\label{255}
\sigma_{[\rm III,\rm I]}  +  \sigma_{[\hskip  
1mm{\underline{\amalg}}\hskip 1mm]}  = 16 \frac{(k^2 - 3 k + 3)^2}{
k^6}
,\hskip 5mm (D_{2,2,2,4}^{1,1,1,3})
\ee
\be\label{256}
- 2 \sigma_{[\underline{\underline{\angle}}]} - (k-3)\sigma_{[\rm  
III,\rm I]} + \sigma_{[\underline{\bigtriangleup}]}
= 16 \frac{(k^2 - 3 k + 3)^2}{k^6}\hskip 5mm (D_{2,2,2,3}^{1,1,1,3}).
\ee
\ba\label{257}
&&3\sigma_{\sqsupseteq} +2\sigma_{[\rm II,\rm II]} + \sigma_{[\Box]}  
+(k-4)\sigma_{\rm M} + (k-4)
\sigma_{\Lambda,\rm II}=
- 72 \frac{(k-3)(2 k - 3)}{k^6},\nonumber\\
&&\hskip 85mm
(D_{2,2,3,4}^{1,1,3,3}).
\ea
\be\label{258}
2\sigma_{[\rm II,\rm II]} + \sigma_{[\Box]} = 36 \frac{(2 k -  
3)^2}{k^6},
\hskip 5mm
(D_{2,2,4,4}^{1,1,3,3}).
\ee
\ba\label{262}
&&- 2 (k-3) \sigma_{\unrhd}+6 \sigma_{\ll} - 6  
\sigma_{\underline{\bigtriangleup}}
+ (k-3)(k-4) \sigma_{\Lambda,\rm II} + 2(k-3)\sigma_{\rm II,\rm II}  
\nonumber\\
&&+ 4 (k-3) \sigma_{\sqsupseteq}  = 72 \frac{(2 k-3)^2}{k^6},
\hskip 5mm (D_{2,2,3,3}^{1,1,3,3}).
\ea
\be\label{263}
(k-4) \sigma_{\underline{\angle},\rm I}+\sigma_{\models} -  
\sigma_{\unrhd}+
\sigma_{\underline{\amalg}} = 48
\frac{(k-3)(k^2-3 k + 3)}{k^6},\hskip 4mm (D_{2,4,3,3}^{1,1,1,2}).
\ee
\ba\label{266}
   &&- 2 (k-4) \sigma_{\angle\angle} + 4 (k-4) \sigma_{\rm M} +  
(k-4)(k-5) \sigma_{
\Lambda,\Lambda}+ 2 (k-4)\sigma_{\Lambda,\rm II}\nonumber\\
&&+ 2\sigma_{\Box} -4 \sigma_{\sqsupseteq} - 8  
\sigma_{\underline{\amalg}} + 4 \sigma_{\models} = 144 \frac{(k-3)(2 k  
- 3)}{k^6},
\hskip 5mm (D_{2,2,3,4}^{1,1,2,2}).
\ea
\ba\label{267}
   && 2(k-4) \sigma_{\angle\angle} + 2 \sigma_{\Box}+ 6(k-4) \sigma_{\rm  
M} + 8 \sigma_{\rm II,\rm II}
+ 16 \sigma_{\sqsupseteq}+ 8 (k-4)\sigma_{\Lambda,\rm II}\nonumber\\
&& + 2(k-4)(k-5)\sigma_{\Lambda,\Lambda}+ 8 \sigma_{\underline{\amalg}}
  - 4 \sigma_{\models} = -432 \frac{(2 k - 3)}{k^6},\nonumber\\
&&\hskip 85mm (D_{1,2,3,4}^{1,1,2,2})
\ea
\ba\label{292}
&& 2 \sigma_{\unrhd}+ (k-4) \sigma_{\angle\angle}- 2  
\sigma_{\underline{\amalg}} + 2 \sigma_{\models}
= 48 \frac{(k-3)(k^2 - 3 k + 3)}{k^6},\hskip 3mm  
(D_{2234}^{1112})\nonumber\\
\ea
\ba\label{293}
&&(k-3)(k-4)\sigma_{\angle\angle} + 2 (k-3) \sigma_{\unrhd} +  
4(k-3)\sigma_{\underline{\amalg}} +
6 \sigma_{\underline{\bigtriangleup}}\nonumber\\
&& + 24 \sigma_{\underline{\underline{\angle}}} + 4 \sigma_{\ll}+  
2(k-3)\sigma_{\models}+ 2 (k-3)
\sigma_{\sqsupseteq}\nonumber\\
&&= 48 \frac{(2 k - 3)(k^2 - 3 k + 3)}{k^6} - 48  
\frac{1}{k-2}\sigma_{\rm IIII}, \hskip 5mm
(D_{1122}^{1112})
\ea

These coupled equations are equivalent to the following equations,
\ba\label{296}
   &&2 \sigma_{\rm II,\rm II} + \sigma_{\Box} = 36 \frac{(2 k  
-3)^2}{k^6},\nonumber\\
&&\sigma_{\rm III,\rm I} + \sigma_{\underline{\amalg}} = 16 \frac{(k^2  
- 3 k + 3)^2}{k^6},
\nonumber\\
&& - (k-3) \sigma_{\rm III,\rm I} + \sigma_{\underline{\bigtriangleup}}  
=
-8 \frac{(k-3)(k^2 - 3 k + 3)^2}{
k^6},\nonumber\\
&& \sigma_{\underline{\bigtriangleup}} + (k-3)  
\sigma_{\underline{\amalg}} =
  8 \frac{1}{k^6}(k-3)(k^2 - 3 k + 3)^2.\nonumber\\
\ea

 From (\ref{293}), we obtain by the use of (\ref{292}),
\be
  (k-3)\sigma_{\underline{\amalg}} +  \sigma_{\underline{\bigtriangleup}}
= 8 \frac{(k-3)(k^2 - 3 k + 3)^2}{k^6}
\ee
which is same as (\ref{296}).

We have the following equation by considering $D_{1233}^{1122}$,
\be
\frac{6}{k-3}\sigma_{\underline{\bigtriangleup}}+
\sigma_{\Box} + 2 \sigma_{\unrhd} + (k-4) \sigma_{\rm M} + 2  
\sigma_{\sqsupseteq}
= 72 \frac{( 2 k -3)}{k^6}
\ee

We find that many coefficients are not determined uniquely.
We have 23 coefficients and 10 coefficients are determined uniquely.
13 coefficients are not determined uniquely. There are 9 linear  
independent relations
among these undetermined coefficients.
These 9 linear independent relations are
(\ref{255}),(\ref{256}),(\ref{262}),(\ref{263}),(\ref{258}),(\ref{1256}) 
,(\ref{M}),(\ref{LL}),
(\ref{2235b}).
We find there are 4 quartic identities in appendix C,  
(\ref{quarticI})$\sim$(\ref{quarticIV}).
Therefore, we are able to choose freely 4 coefficients.

For instance, if we choose $\sigma_{\rm M}= -36 (k-3)^2$, we have
\ba\label{con}
&&\sigma_{\rm M} = - 36 (k-3)^2\nonumber\\
&& \sigma_{\Lambda,\rm II}= 108 (k-2)(k-3)\nonumber\\
&& \sigma_{\angle\angle} = -24 (k-3)(k^2 - 3 k + 3)\nonumber\\
&& \sigma_{\bigtriangleup,\rm I} = -18 (k-1)(2 k-3)\nonumber\\
&&\sigma_{\underline{\angle},\rm I} = -12 (k-3)(k^2 - 3 k + 3)
\ea

\vskip 5mm
{\bf Appendix C: The cubic and quartic identities}
\vskip 5mm

There is one cubic identity in the case  $k=4$ for the variables  
$\tau_{ij}=(x_i - x_j)(\lambda_i
- \lambda_j)$ as shown in \cite{BHc},
\be\label{identity1}
I_3=[\tau_{12}^2 \tau_{34}] - [\tau_{12}\tau_{13}\tau_{24}] +  
[\tau_{12}\tau_{13}\tau_{23}] = 0.
\ee
which may be conveniently depicted graphically by
\be
  I_3 =[\rm II,\rm I] - [\rm N] + [\bigtriangleup].
\ee

For $k > 4$, this cubic identity is modified by a factor $(k-3)$ as
\be\label{identity2}
I_3 =[\tau_{12}^2 \tau_{34}] - [\tau_{12}\tau_{13}\tau_{24}] + (k-3)  
[\tau_{12}\tau_{13}\tau_{23}] = 0.
\ee

We apply the differentiations $D_{abc}^{klm}$, where the indices take  
one of values among (1,2,3,...,k).
By the symmetry, it is enough to consider the values (1,2,3,4). We find  
all possible such
differentials satisfy this identity,
\be
  D_{abc}^{klm} I_3 = 0.
\ee
This is a proof that we have an identity of (\ref{identity2}).

The geometric proof is also possible. In the case k=4, we consider the  
coincidence of the
variable $x_4\rightarrow x_1$ and $\lambda_4 \rightarrow \lambda_1$.
In this limit, we obtain three type 3-point graphs.
 From $[\rm II,\rm I]$, we obtain $[\underline{\angle}]$, and this  
$[\underline{\angle}]$ is
exactly cancelled by the same graphs generated
  from $[\rm N]$ in this coincidence limit.
The graphs $[\rm N]$ also generate the triangle graphs  
$[\bigtriangleup]$, which
are cancelled by the existing triangle graphs of the last term of the  
identity $I_3$.
Thus we prove that $I_3 = 0$ for $k=4$. For $k>4$, we can prove this  
identity by the
inductive method, namely for $k=5$, when we take the limit  
$x_5,\lambda_5 \rightarrow
x_1,\lambda_1$, the graphs reduces to the graphs of $k=4$ by the  
cancellations.

Since we have this identity, the
coefficients $\sigma_{\rm II,\rm I},\sigma_{\rm  
N},\sigma_{\bigtriangleup}$ are not
uniquely determined as we have seen in the appendix B.

In the fourth order (4 line graphs), the identity $I_3=0$ at k=4 is  
generalized by the multiplication
of $[\tau_{12}]= \tau_{12}+ \tau_{13} + \cdots + \tau_{3,4}$. By  
writing all types of graphs, we find
easily  the following identity of k=4,
\ba\label{identity4}
I_{4} &=& I_3 \times [\tau_{12}] \nonumber\\
&=& [\rm III,\rm I] + 2[\rm II,\rm II] - [\unrhd] + [\hskip  
2mm\underline{\bigtriangleup}\hskip 2mm]
  - 4 [\Box]
- [\hskip 2mm \underline{\amalg}\hskip 2mm]=0
\ea

This identity of $I_4=0$ at k=4 is verified directly by the  
differentiations
$D_{abcd}^{klmn}$ (a,b,c,d,k,l,m,n
are 1,2,3 or 4).

For the general k $(k>4)$, we use the cubic identity of  
(\ref{identity2}).
  Multipling $[\tau_{12}]$
on each terms of the cubic identity,
we get (the indices are now from 1 to k),
\be
[\rm II, \rm I] [\tau_{12}] =
2 [\rm II,\rm I,\rm I] + [\rm III,\rm I] + 2[\rm II,\rm II] +  
[\underline{
\angle},\rm I] + [\sqsupseteq] + 2[\Lambda,\rm II]
\ee
\be
-[\rm N][\tau_{12}] = - 2[\rm M] - [\rm N,\rm I] - 2 [\unrhd] -  
[\underline{\amalg}]
-[\sqsupseteq]
- 4 [\Box] - 2 [\angle\angle]
\ee
\be
(k-3)[\bigtriangleup][\tau_{12}] = (k-3)[\bigtriangleup,\rm I] +  
(k-3)[\hskip 2mm \underline{\bigtriangleup}
\hskip 2mm] + (k-3) [ \unrhd ]
\ee
Adding these three equations, we obtain an identity.
\ba\label{quarticidentity}
&&2[\rm II,\rm I,\rm I] - [\rm N,\rm I] +
(k-3)[\bigtriangleup,\rm I] + [\rm III,\rm I] + 2[\rm II,\rm II]
  + [\underline{\angle},\rm I] - 2[\rm M]\nonumber\\
&& + 2[\rm II,\La] +(k-5) [\unrhd] - [\hskip 2mm \underline{\amalg}
\hskip 2mm]
- 4[\Box] - 2 [\angle\angle] + (k-3)[\hskip 2mm  
\underline{\bigtriangleup} \hskip 2mm] = 0
\nonumber\\
\ea
This long identity is devided into 4 sub-identities.

By the analogy of the cubic identity, we extract from  
(\ref{quarticidentity}) an identity,
\be\label{quarticI}
2[\rm II,\rm I,\rm I] - [\rm N,\rm I] + (k-5)[\bigtriangleup,\rm I] = 0
\ee
This identity is the cubic identity plus a separate  line. The factor  
$(k-5)$ means that when $k=5$
all terms are vanishing. ($[\bigtriangleup,\rm I]$ is nonvanishing,  
therefore it should have
a factor $(k-5)$ as a coefficient, since other two terms do not exist  
for k=5.)

This simple (fundamental) identity at fourth order is proved by  all  
possible differentiations
$D_{\lambda_j,...}^{x_i,...}$.
For instance, with the differentials of $D_{1123}^{1123}$, we find
$D_{1123}^{1123}[\rm II,\rm I,\rm I] = 4 (k-3)(k-4)(k-4)$,  
$D_{1123}^{1123}[\rm N,\rm I]
= 16 (k-3)(k-4)(k-5)$ and  $D_{1123}^{1123}[\bigtriangleup,\rm I] = 8  
(k-3)(k-4)$. These values
satisfy (\ref{quarticI}). All differentiations which appeared in  
appendix B satisfy this identity.

The second identity in (\ref{quarticidentity}) is a generalization of  
(\ref{identity4}), obtained by
adding a factor $(k-3)$ for $[\hskip 2mm \underline{\bigtriangleup}  
\hskip 2mm]$,
\be\label{quarticII}
  [\rm III,\rm I] + 2[\rm II,\rm II] - [\unrhd] + (k-3)[\hskip  
2mm\underline{\bigtriangleup}\hskip 2mm]
  - 4 [\Box]
- [\hskip 2mm \underline{\amalg}\hskip 2mm]=0
\ee
This identity is verified by the differentiations of $D_{abcd}^{klmn}$,
where the indices take values from one to four. Indeed in appendix B we  
have evaluated
various differentiations, and we have found the contributions of every  
graph which enters
  in this second identity.
We have checked all possible differentiations.

As  third identity in (\ref{quarticidentity}), we find
\be\label{quarticIII}
2[\La, \rm II]  - 2[\rm M]+ 4 [\bigtriangleup, \rm I] + 2(k-4)[\Box] -
(k-4) [\rm II,\rm II] =0
\ee
This identity is also proved by the consideration of all possible  
differentiations.

Subtracting these three identities from (\ref{quarticidentity}), we  
obtain
the fourth identity,
\be\label{quarticIV}
  [\underline{\angle},\rm I] - 2 [ \angle\angle] - 2 [  
\bigtriangleup,\rm I]
+ (k-4) [\hskip 2mm
\unrhd\hskip 2mm]
  - 2(k-4)[\Box] + (k-4) [\rm II,\rm II]\nonumber\\
= 0
\ee

We have checked this identity by the evaluation of all differentiations
$D_{abcd}^{klmn}$ in (\ref{12D}).

Further, we check them by the differentiations $D_{abcd}^{klmn}$,
(a,b,c,d,k,l,m,n = 1,2,3,4), which appeard in
the calculations in appendix B. They are
$D_{2234}^{1112}$,$D_{1122}^{1112}$,\\
$D_{1234}^{1122}$,$D_{2234}^{1122}$,$D_{2433}^{1112}$,
$D_{2233}^{1133}$,
$D_{2244}^{1133}$,
$D_{2234}^{1133}$,$D_{2223}^{1113}$,$D_{2224}^{1113}$.

In the appendix B, we were left with 13 undetermined coefficients  
$C_{[graph]}$ . We have
9 linear independent equations for them here. Since 13 - 9 = 4,
given that one has  4 quartic identities
(\ref{quarticI}) $\sim$ (\ref{quarticIV})
  all the  coefficients may be determined at the expense  of introducing  
4 arbitrary parameters.

In the case $\beta=4$, there are no double or multiple line graphs.  
Therefore, we have no
cubic or higher order identities, and all coefficients of the  $\tau$  
expansion are uniquely determined
without ambiguities. In the next appendix D, we discuss the case of  
$\beta=4$ in a more systematic way.

\vskip 5mm

{\bf Appendix D : Characterization of the $\tau$ expansion by specific  
differentiations}

As  discussed in appendix B, the coefficients of the $\tau$ expressions  
for the products of
  symmetric
functions $s_4(\tilde x)s_4(\tilde \lambda)$ are determined by focusing  
on
the monomials $x_{i_1}^{p_1}
x_{i_2}^{p_2} \cdots \lambda_{j_1}^{q_1}\lambda_{j_2}^{q_2}\cdots$.  
Such terms may be selected
  with the help of the differentiations $D_{j_1 j_2 \cdots}^{i_1 i_2  
\cdots}$. Many terms
are thereby uniquely determined
since they are characterized by the existence of  particular  
combinations
  of $x_i$ and $\lambda_j$.
Since each $\tau$ term has a specific   topological structure
  when  represented graphically, one may
  characterize such $\tau$ terms by the differentiations
$D_{j_1 j_2 \cdots}^{i_1 i_2 \cdots}$. For instance, the term  
$[\tau_{12}^2] = \sum_{i<j}\tau_{ij}^2$
is uniquely characterized by $x_1^2\lambda_2^2$, which is equivalent to  
$D_{22}^{11}$. Other
types of
$\tau$-terms, such as  $[\tau_{12}\tau_{13}]$,$[\tau_{12}\tau_{34}]$,  
have no such $x_1^2\lambda_2^2$
factor. Therefore $x_1^2 \lambda_2^2$  is a unique factor which selects  
$[\tau_{12}]$.

   Indeed we have used such properties for the determination of  
$\sigma$, which is a coefficient
  of the  $\tau$
expression for $s_2(\tilde x)s_2(\tilde \lambda)$. We study such
correspondences between the unique
differentiations  and the $\tau$ terms (graphs), because we can
apply these correspondences directly to determine  the coefficients $C$  
for the HIZ integrals.

This one-to-one correspondence between $\tau$ terms and   
differentiations $D$ may be used as follows.
We write  the function $f$ in (\ref{eq2}) as a sum,
\be\label{ff}
f= 1 + f_{(1)} + f_{(2)} + f_{(3)} + \cdots
\ee
where $f_{(l)}$ is the l-th order term of the $\tau$-expansion.
The function $f$ is given by
\be\label{expandf}
f = e^{-\frac{1}{k}\sum_{i<j}  
\tau_{ij}}\sum_{m=0}^{\infty}\frac{1}{m!}\frac{1}{\prod_{q=0}^{m-1}
( 1 + q \alpha)}\sum_{p}\chi_p \frac{Z_p(\tilde x)Z_p(\tilde  
\lambda)}{Z_p(\rm I)}
\ee
which is expanded as
\ba
f_{(1)} &=&  - \frac{1}{k}\sum_{i<j} \tau_{ij}\nonumber\\
f_{(2)} &=&  \frac{1}{2 k^2}(\sum_{ij} \tau_{ij})^2
+ \frac{\alpha}{2 (k+ \alpha)(k-1)}s_2(\tilde x)s_2(\tilde \lambda)
\ea

\vskip 20mm
\begin{picture}(150,100)
\put(30,140){\bf Table D-1: Characteristic differentials of $\tau$  
terms and $f$}
\put(1,120){\line(2,0){390}}
\put(20,100){$[\rm I] \hskip 20mm D_{2}^{1}[\rm I] = -1,
\hskip 20mm D_2^{1}( f_{(1)} )= \frac{1}{k}$}
\put(1,100){l=1}
\put(1,85){\line(2,0){390}}
\put(1,70){l=2}
\put(20,70){$[\Lambda] \hskip 20mm \rm D_{23}^{11}[\Lambda]= 2,
  \hskip 20mm D_{23}^{11}(f_{(2)})= \frac{2}{
k (k+ \alpha)}$}
\put(20,40){$[\rm II] \hskip 20mm D_{22}^{11}[\rm II] = 4,\hskip 20mm  
D_{22}^{11}(f_{(2)}) =
\frac{2(1 + \alpha)}{k(k+ \alpha)}$}
\put(20,10){$[\rm I,\rm I] \hskip 17 mm D_{34}^{12}[\rm I,\rm I] = 2,  
\hskip 15mm D_{34}^{12}(f_{(2)})
= \frac{2}{k(k+ \alpha)}( 1+ \frac{\alpha}{k-1})$}
\put(1,-20){l=3}
\put(1,-5){\line(2,0){390}}
\put(20,-20){$[\rm I,\rm I,\rm I] \hskip 14mm D_{456}^{123}[\rm I,\rm  
I,\rm I]=-6,
\hskip 5mm D_{456}^{123}(f_{(3)}) = \frac{6}{k(k+\alpha)(k+2\alpha)}[1 +
\frac{3\alpha}{k-1}+\frac{2\alpha^2}{(k-1)(k-2)}]$}
\put(20,-50){$[\Lambda,\rm I] \hskip 17 mm D_{345}^{112}[\La,\rm  
I]=-6,\hskip 5mm D_{345}^{112}
(f_{(3)}) = \frac{6}{k(k+\alpha)(k+2\alpha)}( 1 + \frac{2\alpha}{k-1})$}
\put(20,-80){$[\rm Y] \hskip 20mm D_{234}^{111}[\rm Y]=-6, \hskip 10mm  
D_{234}^{111}(f_{(3)}) =
\frac{6 }{k (k+ \alpha)(k+ 2 \alpha)}$}
\put(20,-110){$[\rm III] \hskip 17mm D_{222}^{111}[\rm III]=-36,\hskip  
8mm
D_{222}^{111}(f_{(3)}) = \frac{6(1 + \alpha)(1 + 2  
\alpha)}{k(k+\alpha)(k + 2 \alpha)}$}
\put(20,-140){$[\hskip 1mm \underline{\angle}\hskip 1mm] \hskip 17mm
\rm D_{223}^{111}[\hskip 1mm \underline{\angle}\hskip 1mm]=-12,\hskip  
5mm
D_{223}^{111}(f_{(3)})= \frac{6(1 + \alpha)}{k(k+\alpha)(k+ 2 \alpha)}  
$}
\put(20,-170){$[\rm II,\rm I],[\rm N] \hskip 10mm D_{224}^{113}(C_{\rm  
II,\rm I}[\rm II,\rm I]
+ C_{\rm N}[\rm N]) = -4 C_{\rm II,\rm I} - 4 C_{\rm N},$}
\put(150,-200){$D_{224}^{113}(f_{(3)})= \frac{2  
(\alpha+3)}{k(k+\alpha)(k+ 2 \alpha)}
+ \frac{4 \alpha(\alpha+2)}{k(k+\alpha)(k+2\alpha)(k-1)}$}
\put(20,-230){$[\bigtriangleup],[\rm II,\rm I],[\hskip 1mm  
\underline{\angle} \hskip 1mm]
\hskip 5mm D_{332}^{112}(C_{\bigtriangleup}[\bigtriangleup] +
C_{\rm II,\rm I}[\rm II,\rm I] + C_{ \underline{\angle}}[\hskip 1mm  
\underline{\angle}\hskip 1mm])
$}
\put(120,-250){$=-4 C_{\bigtriangleup} +4(k-3) C_{\rm II,\rm I} +  
8C_{\underline{\angle}},
  $      }
\put(150,-270){$ D_{332}^{112}(f_{(3)}) = \frac{-2(1 + \alpha) k^2 - 4  
(\alpha^2 - 2) k + ( 8\alpha^2 +
10 \alpha - 6)}{k(k+ \alpha )(k + 2 \alpha)( k-1)}$}
\put(1,-290){\line(2,0){390}}
\end{picture}
\vskip 4mm

\newpage

When the $\tau$-terms (graphs) are selected by a single differentiation  
$D$,
  the coefficients
of the corresponding term in the $\tau$ expansion are  given by
\be\label{ratio}
C_{[graph]} = \frac{D(f_{(l)})}{D[graph]}
\ee
For instance, from the above table, we obtain
\be
C_{\La} = \frac{1}{k(k+\alpha)}
\ee
The values of the coefficients $C_{[graph]}$ coincide with the values  
in Table B, when we put
$\alpha = 2/(2 - \beta)$.

The third order term $f_{(3)}$ in (\ref{ff}) follows from the Jack  
polynomial expansion,
\ba
f_{(3)} &=& - \frac{1}{6 k^3}(\sum_{i<j} \tau_{ij})^3 - \frac{\alpha}{2  
k (k+ \alpha)(k-1)}(
\sum_{i<j}\tau_{ij})s_2(\tilde x)s_2(\tilde \lambda)\nonumber\\
&+& \frac{\alpha^2 k}{3 ( k+ \alpha)(k + 2 \alpha)(k-1)(k-2)}s_3  
(\tilde x)s_3(\tilde \lambda)
\ea
The coefficients of $[\rm II,\rm I],[\rm N]$ have arbitrariness due to  
the cubic identity,
which has been discussed in the appendix C. We take here the reasonable  
constraint that
the double line graphs have $\frac{1 + \alpha}{2}$ as  overall factor.
This means that $C_{\rm II,\rm I}$ is proportional to  
$\frac{1+\alpha}{2}$, and
it is vanishing for $\alpha=-1 (\beta=4)$. For the single line graphs,  
their coefficients
are simply $-\frac{1}{k^3}$ in the large k limit, for the third order  
terms.
With these constraints, we have from the table D-1,
\be\label{CII,I}
C_{\rm II,\rm I} = - \frac{1+\alpha}{2k(k+ \alpha)(k + 2 \alpha)}( 1 +  
\frac{2 \alpha}{k-1})
\ee
\be
C_{\rm N} = - \frac{1}{k(k+ \alpha)(k+ 2\alpha)}( 1 +  
\frac{\alpha}{k-1})
\ee
which coincide with the values in Table B by the substitution of  
$\alpha = \frac{2}{2 - \beta}$.

By the differentiation $D_{332}^{112}$, we have from Table D-1,
\be
- 4 C_{\bigtriangleup} = - 4 (k-3)C_{\rm II,\rm I} - 8  
C_{\underline{\angle},\rm I} +
D_{332}^{112}(f_{(3)})
\ee
Using the value of $C_{\rm II,\rm I}$ in (\ref{CII,I}), we obtain
\be
C_{\bigtriangleup} = - \frac{1}{k(k+\alpha)(k+ 2 \alpha)}( 1 -  
\frac{\alpha^2}{k-1})
\ee
which coincides with the value in Table B.

For the fourth order, we have from the expression of Table in the  
appendix A,
\ba
f_{(4)} &=& \frac{1}{24 k^4} (\sum_{i<j} \tau_{ij})^4 + \frac{\alpha}{4  
k^2 (k+ \alpha)(k-1)}
(\sum_{i<j} \tau_{ij})^2 s_2(\tilde x)s_2(\tilde \lambda)\nonumber\\
&-& \frac{\alpha^2}{3(k+ \alpha)(k+ 2 \alpha)(k-1)(k-2)}(\sum_{i<j}  
\tau_{ij})
s_3(\tilde x)s_3(\tilde \lambda)\nonumber\\
&+& \frac{\alpha^2(18\alpha^2 + 18 \alpha k - 18 \alpha^2 k + 6 k^2 -  
18 \alpha k^2 + 6 \alpha^2 k^2 - 5 k^3
+5 \alpha k^3 + k^4)}{8 k (k+ \alpha)(k+ 2 \alpha)(k+ 3 \alpha) (k+  
\alpha -1)(k-1)(k-2)(k-3)}
\nonumber\\
&\times& (s_2(\tilde x)s_2(\tilde \lambda))^2\nonumber\\
&+& \frac{\alpha^3 k (k^2 + \alpha k - k + \alpha)}{4 (k+ \alpha)(k+ 2  
\alpha)
(k + 3 \alpha) (k+ \alpha-1)(k-1)(k-2)(k-3)} s_4(\tilde x)s_4(\tilde  
\lambda)\nonumber\\
&-& \frac{\alpha^3 ( 2k^2 + 3 \alpha k - 3 k - 3 \alpha)}{4(k+  
\alpha)(k + 2 \alpha) (k + 3 \alpha)
(k+ \alpha-1)(k-1)(k-2)(k-3)}\nonumber\\
&\times&
[s_2(\tilde x)^2s_4(\tilde \lambda) + s_4(\tilde x)(s_2(\tilde  
\lambda))^2]
\ea

If we apply $D_{2345}^{1111}$ on $f_{(4)}$, we obtain
\be
D_{2345}^{1111}(f_{(4)}) = \frac{24}{k(k+ \alpha)(k+ 2\alpha)(k+  
3\alpha)}
\ee
This $D_{2345}^{1111}$ is a characteristic differentiation of the  
$\tau$ term $[\rm X]$, namely
it is the unique differentiation for $[\rm X]$ term. Since we have  
$D_{2345}^{1111}[\rm X] =
24$, one obtains
\be
C_{\rm X} = \frac{1}{k(k+ \alpha)(k+ 2 \alpha)(k+ 3\alpha)}
\ee
which coincides with the value given in Table B, if we put $\alpha=  
\frac{2}{2 - \beta}$.
Other differentiations of $\tau$ terms and $f_{(4)}$ are represented in  
the following table
D-2.

\newpage
\vskip 30mm
\begin{picture}(150,100)
\put(30,140){\bf Table D-2: Characteristic differentiations of $\tau$  
terms and $f_{(4)}$}
\put(1,120){\line(2,0){390}}
\put(1,100){l=4}
\put(20,100){$ [\rm X] \hskip 5mm D_{2345}^{1111}[\rm X] = 24, \hskip  
5mm
D_{2345}^{1111}(f_{(4)}) = \frac{24}{k(k+ \alpha)(k+ 2\alpha)(k+  
3\alpha)}$}
\put(20,70){$[\rm Y,\rm I] \hskip 5mm D_{3456}^{1112}[\rm Y,\rm I] =  
24, \hskip 5mm
D_{3456}^{1112}(f_{(4)})=  
\frac{24(k+3\alpha-1)}{k(k-1)(k+\alpha)(k+2\alpha)(k+3\alpha)}$}
\put(20,40){$ [ \Lambda,\Lambda] \hskip 5mm  
D_{3456}^{1122}[\Lambda,\Lambda] = 24,
\hskip 5mm D_{3456}^{1112}(f_{(4)})=  
\frac{24(k+2\alpha-1)(k+3\alpha-1)}{k
(k+\alpha)(k+2\alpha)(k+3\alpha)(k+\alpha-1)(k-1)}$}
\put(20,10){$ [ \Lambda,\rm I,\rm I]  \hskip 5mm D_{4567}^{1123}[  
\Lambda,\rm I,\rm I] = 24,$}
\put(60,-7){$
D_{4567}^{1123}(f_{(4)})= \frac{24}{k(k+\alpha)(k+2 \alpha)(k+  
3\alpha)}(1 - \frac{2\alpha}{k+\alpha-1}
+ \frac{7\alpha}{k-1} + \frac{6\alpha^2}{(k-1)(k-2)}-  
\frac{2\alpha^2}{(k-2)(k+\alpha-1)})$}
\put(20,-25){$ [\rm I,\rm I,\rm I,\rm I] \hskip 5mm D_{5678}^{1234}[\rm  
I,\rm I,\rm I,\rm I]
= 24,$}
\put(60,-45){$D_{5678}^{1234}(f_{(4)})=  
\frac{24}{k(k+\alpha)(k+2\alpha)(k+ 3\alpha)}(1 +
\frac{6\alpha}{k+\alpha-1} + \frac{9\alpha^2}{(k-1)(k+\alpha-1)}
+\frac{8\alpha^2}{(k-2)(k+\alpha-1)}$}
\put(80,-65){$+\frac{25\alpha^3}{(k-1)(k-3)(k+\alpha-1)}
-\frac{8\alpha^3}{(k-2)(k-3)(k+\alpha-1)}+\frac{6\alpha^4}{(k-1)(k-2)
(k-3)(k+\alpha-1)})$}
\put(20,-95){$[\sqsupseteq] \hskip 5mm D_{2244}^{1113}[\sqsupseteq] =48,
\hskip 5mm D_{2244}^{1113}(f_{(4)})=  
\frac{24(1+\alpha)}{k(k+\alpha)(k+2\alpha)(k+3\alpha)}
( 1+ \frac{2\alpha}{k-1})$}
\put(20,-125){$[\ll] \hskip 5mm D_{2233}^{1111}[\ll]=96,
\hskip 5mm D_{2233}^{1111}(f_{(4)})=\frac{24 (1 +  
\alpha)^2}{k(k+\alpha)(k+2 \alpha)(k+ 3 \alpha)}$}
\put(20,-155){$[\models] \hskip 5mm D_{2234}^{1111}[\models] = 48,
\hskip 5mm D_{2234}^{1111}(f_{(4)}) = \frac{24(1 +  
\alpha)}{k(k+\alpha)(k+2\alpha)(k + 3\alpha)}$}
\put(20,-185){$[\hskip 1mm \underline{\underline{\angle}}\hskip 1mm]  
\hskip 5mm
D_{2223}^{1111}[\hskip 1mm \underline{\underline{\angle}}\hskip 1mm]=  
144,
\hskip 5mm D_{2223}^{1111}(f_{(4)})=
\frac{24(1 + \alpha)(1+ 2 \alpha)}{k(k+\alpha)(k+2\alpha)(k + 3  
\alpha)}$}
\put(20,-215){$[\rm IIII] \hskip 5mm D_{2222}^{1111}[\rm IIII] = 576,
\hskip 5mm D_{2222}^{1111}(f_{(4)}) =
\frac{24(1 + \alpha)(1+2 \alpha)(1+3\alpha)}{k(k+\alpha)(k+ 2  
\alpha)(k+ 3 \alpha)}$}
\put(1,-245){\line(2,0){390}}
\end{picture}
\vskip 100mm

\newpage
\vskip 30mm
\begin{picture}(150,100)
\put(30,140){\bf Table D-3: Characteristic differentiations of $\tau$  
terms and $f_{(4)}$}
\put(1,130){\line(2,0){420}}
\put(20,110){$ D_{3345}^{1112}(C_{[\angle\angle]}[\angle\angle]+
C_{[\underline{\angle},\rm I]}[\underline{\angle},\rm  
I])=12C_{[\angle\angle]}+
24 C_{[\underline{\angle},\rm I]}$}
\put(40,80){$ D_{3345}^{1112}(f_{(4)})=  
\frac{12}{k(k+\alpha)(k+2\alpha)(k+3\alpha)}[(1+\alpha)(
1 + \frac{3\alpha}{k-1}) + (1+ \frac{2\alpha}{k-1})]    $}
\put(20,50){$ D_{2345}^{1122}(
C_{[\Lambda,\rm II]}[\Lambda,\rm II]+
C_{[\rm M]}[\rm M]+C_{[\Lambda,\Lambda]}[\Lambda,\Lambda]) $}
\put(90,30){$
= -24C_{[\Lambda,\rm II]}-24
C_{[\rm M]}- 12(k-5)C_{[\Lambda,\Lambda]}$}
\put(40,0){$ D_{2345}^{1122}(f_{(4)})= - \frac{12(k+ 2 \alpha -1)(k^2 +  
4 \alpha k - 3 k + 3 \alpha^2 - 9
\alpha + 2)}{k(k-1)(k+ \alpha-1)(k+ \alpha)(k+ 2 \alpha)(k + 3 \alpha)}  
     $}
\put(20,-30){$D_{2224}^{1113}(C_{[\rm III,\rm I]}[\rm III,\rm I] +  
C_{[\underline{\amalg}]}[\amalg])=
36 C_{[\rm III,\rm I]} + 36 C_{[\underline{\amalg}]} $}
\put(40,-60){$D_{2224}^{1113}(F_{(4)})= \frac{12(1 + \alpha)(\alpha k +  
2 k + 3\alpha^2 + 2 \alpha -2)}
{k(k+\alpha)(k+ 2\alpha)(k+ 3 \alpha)(k-1)}  $}
\put(20,-90){$D_{2244}^{1133}(C_{[\rm II,\rm II]}[\rm II,\rm II] +  
C_{[\Box]}[\Box]) =
32 C_{[\rm II,\rm II]} + 16 C_{[\Box]}$}
\put(30,-120){$D_{2244}^{1133}(f_{(4)})=   
\frac{8(1+a)^2}{k(k+\alpha)(k+2\alpha)(k+3\alpha)}(
1+  
\frac{4\alpha}{k-1}+\frac{2\alpha^2(2+\alpha)}{(1+\alpha)(k 
-1)(k+\alpha-1)})$}
\put(60,-140){$+\frac{16}{k(k+\alpha)(k+2\alpha)(k+3\alpha)}(1 +  
\frac{2\alpha}{k-1})$}
\put(20, 
-170){$D_{2223}^{1113}(C_{[\underline{\underline{\angle}}]}[\underline{
\underline{\angle}}]
+C_{[\rm III,\rm I]}[\rm III,\rm I] +  
C_{[\underline{\bigtriangleup}]}[\underline{\bigtriangleup}])=
-72 C_{\underline{\underline{\angle}}}-36 (k-3)C_{\rm III,\rm I} + 36  
C_{\underline{\bigtriangleup}}$}
\put(40,-200){$D_{2223}^{1113}(f_{(4)})=  -\frac{6(1+\alpha)(2\alpha  
k^2 + k^2 + 6 \alpha^2 k - \alpha k
- 5 k - 12 \alpha^2 - 7\alpha +  
4)}{k(k+\alpha)(k+2\alpha)(k+3\alpha)(k-1)} $}
\put(20,-230){$D_{1456}^{1123}(C_{[\rm II,\rm I,\rm I]}[\rm II,\rm  
I,\rm I] + C_{[\rm Y,\rm I]}[\rm Y,\rm I]
+ C_{[\rm N,\rm I]}[\rm N,\rm I] + C_{[\Lambda,\rm I,\rm  
I]}[\Lambda,\rm I,\rm I])$}
\put(40,-260){$=  -24 C_{[\rm II,\rm I,\rm I]}+12 C_{[\rm Y,\rm I]} -  
48 C_{[\rm N,\rm I]}-12(k-6)
C_{[\Lambda,\rm I,\rm I]} $}
\put(30,-290){$D_{1456}^{1123}(f_{(4)})=  - 12  
\frac{1+\alpha}{(k+\alpha)(k+2 \alpha)(k+ 3 \alpha)}
[ 1 - \frac{2\alpha}{k+\alpha-1} + \frac{7 \alpha}{k-1} +  
\frac{6\alpha^2}{(k-1)(k-2)}
-\frac{2\alpha^2}{(k-2)(k+\alpha-1)}]$}
\put(50,-320){$ -\frac{48}{k(k+\alpha)(k+\alpha)(k+2\alpha)(k+3  
\alpha)}[1 + \frac{8\alpha}{k-1}
-\frac{4\alpha}{k-2} -\frac{\alpha^2}{(k+\alpha-1)(k-2)}+  
\frac{(1+\alpha)(4\alpha k + 3 \alpha^2 - 4\alpha)}{
(k-1)(k-2)(k+\alpha-1)}]$}
\put(20,-350){$D_{2234}^{1112}(C_{[\unrhd]}[\unrhd]+  
C_{[\angle\angle]}[\angle\angle]+
C_{[\underline{\amalg}]}[\underline{\amalg}]+ C_{[\models]}[\models])$}
\put(40,-380){$ = -24 C_{[\unrhd]}-12 (k-4)C_{[\angle\angle]}+ 24  
C_{[\underline{\amalg}]}
-24 C_{[\models]}  $}
\put(40,-410){$ D_{2234}^{1112}(f_{(4)})= -\frac{12(k^2 + 2\alpha k - 3  
k - 3\alpha^2 - 7 \alpha + 2)}{
k(k+\alpha)(k+ 2 \alpha)(k+ 3\alpha)(k-1)} $}
\put(20,-440){$D_{2235}^{1124}(C_{[\angle\angle]}[\angle\angle]  
+C_{[\rm M]}[\rm M]+C_{[\rm N,\rm I]}
[\rm N,\rm I] +C_{[\bigtriangleup,\rm I]}[\bigtriangleup,\rm I]+  
C_{[\Lambda,\Lambda]}[\Lambda,\Lambda]$}
\put(30,-460){$ +C_{[\underline{\angle},\rm I]}
[\underline{\angle},
\rm I]+ C_{[\Lambda,\rm II]}[\Lambda,\rm II])$}
\put(30,-480){$ = 4 C_{[\angle\angle]} -24C_{[\rm M]}- 8(k-5)C_{[\rm  
N,\rm I]}
  - 8 C_{[\bigtriangleup,\rm I]}
  - 4 (k-5) C_{[\Lambda,\Lambda]} - 8 C_{[\underline{\angle},\rm I]}
- 8 C_{[\Lambda,\rm II]}$}
\put(40,-500){$D_{2235}^{1124}(f_{(4)})$}
\put(20,-520){$= -\frac{4(12-78 \alpha+ 120 \alpha^2 - 36 \alpha^3 -  
6\alpha^4 -36
k + 151 \alpha k - 120\alpha^2 k + 7\alpha^3 k +39 k^2-90 \alpha k^2 +  
27 \alpha^2 k^2 - 18 k^3 + 17 \alpha k^3
+ 3 k^4)}{k(k-1)(k-2)(k+\alpha-1)(k+ \alpha)(k+ 2 \alpha)(k+ 3\alpha)}$}
\put(1,-537){\line(2,0){420}}
\end{picture}
\vskip 30mm
\newpage
The coefficients in Table D-3 have ambiguities due to the quartic  
identities discussed in Appendix C.
For the coefficients $C_{[\underline{\angle},\rm  
I]},C_{[\angle\angle]}$, we have
\be
   24 C_{[\underline{\angle},\rm I]} + 12 C_{[\angle\angle]} =
\frac{12}{k(k+\alpha)(k+2\alpha)(k+3\alpha)}[(1+\alpha)(
1 + \frac{3\alpha}{k-1}) + (1+ \frac{2\alpha}{k-1})]
\ee
If we assume that $C_{[\underline{\angle},\rm I]}$ has an overall  
factor $(1 + \alpha)$,
which implies that it vanishes for $\beta=4$, we uniquely determine  
these two coefficients as
\be
  C_{[\underline{\angle},\rm I]}= \frac{1+\alpha}{2 k (k+ \alpha)(k+  
2\alpha)(k + 3 \alpha)}(1+
\frac{3\alpha}{k-1})
\ee
and
\be
C_{\angle\angle} =  \frac{1}{ k (k+ \alpha)(k+ 2\alpha)(k + 3  
\alpha)}(1+
\frac{2\alpha}{k-1}).
\ee
These values coincide with Table B, and for $\beta=4$, they are  
consistent with Table A.

Subtracting the value of $-12(k-5)C_{\Lambda,\Lambda}$, which was
obtained in Table D-2, from $D_{2345}^{1122}(f_{(4)})$, we obtain
\ba
C_{\Lambda,\rm II} + C_{\rm M} &=&
\frac{1+\alpha}{2 k (k+ \alpha)(k+ 2\alpha)(k+ 3\alpha)}( 1+  
\frac{\alpha}{k+\alpha-1})
(1 + \frac{3 \alpha}{k-1})\nonumber\\
&+& \frac{1}{k (k+ \alpha)(k+ 2\alpha)(k+ 3\alpha)}(1  
+\frac{\alpha}{k+\alpha-1})(
1 + \frac{2\alpha}{k-1})\nonumber\\
\ea
Normalizing again  the double bond with the factor $(1 + \alpha)/2$  
which vanishes for $\beta=4$,
  we obtain
\be
C_{\Lambda,\rm II} =
\frac{1+\alpha}{2 k (k+ \alpha)(k+ 2\alpha)(k+ 3\alpha)}( 1+  
\frac{\alpha}{k+\alpha-1})
(1 + \frac{3 \alpha}{k-1})
\ee
and
\be
C_{\rm M} =\frac{1}{k (k+ \alpha)(k+ 2\alpha)(k+ 3\alpha)}(1  
+\frac{\alpha}{k+\alpha-1})(
1 + \frac{2\alpha}{k-1})
\ee

We  write $D_{2224}^{1113}(f_{(4)})$ as
\ba
D_{2224}^{1113}(f_{(4)})&=& \frac{6(1+\alpha)(1 + 2  
\alpha)}{k(k+\alpha)(k+ 2 \alpha)(k+ 3\alpha)}
( 1+ \frac{3\alpha}{k-1})\nonumber\\
&+& \frac{18(1+\alpha)}{k(k+\alpha)(k+ 2 \alpha)(k+ 3 \alpha)}( 1+  
\frac{\alpha}{k-1})
\ea
Then we find that
\be
C_{\rm III,\rm I} =\frac{(1+\alpha)(1 + 2 \alpha)}{6k(k+\alpha)(k+ 2  
\alpha)(k+ 3\alpha)}
( 1+ \frac{3\alpha}{k-1})
\ee
and
\be
C_{\underline{\amalg}}=\frac{(1+\alpha)}{2k(k+\alpha)(k+ 2 \alpha)(k+ 3  
\alpha)}( 1+ \frac{\alpha}{k-1})
\ee
  We have from Table D-3,
\be
C_{\rm II,\rm II} = \frac{(1+\alpha)^2}{4  
k(k+\alpha)(k+2\alpha)(k+3\alpha)}(
1+  
\frac{4\alpha}{k-1}+\frac{2\alpha^2(2+\alpha)}{(1+\alpha)(k 
-1)(k+\alpha-1)})
\ee
and
\be
C_{\Box} = \frac{1}{k(k+\alpha)(k+2\alpha)(k+3\alpha)}(1 +  
\frac{2\alpha}{k-1})
\ee
For the differentiation $D_{2223}^{1113}$, we use obtained values of  
$C_{\underline{\underline{\angle}}}$,
$C_{\rm III,\rm I}$, then we find uniquely,
\be
C_{\underline{\bigtriangleup}}= \frac{1+\alpha}{2 k  
(k+\alpha)(k+2\alpha)(k+3\alpha)}(1 - \frac{2\alpha^2}{
k-1})
\ee
  For $D_{1456}^{1123}$, we subtract the two known expressions of  
$C_{\rm Y,\rm I}$ and $C_{\Lambda,\rm I,\rm I}$,
and obtain the following two coefficients,
\ba
C_{\rm II,\rm I,\rm I}&=&  \frac{1+\alpha}{(2 k+\alpha)(k+2 \alpha)(k+  
3 \alpha)}
[ 1 - \frac{2\alpha}{k+\alpha-1} + \frac{7 \alpha}{k-1} +  
\frac{6\alpha^2}{(k-1)(k-2)}\nonumber\\
&&-\frac{2\alpha^2}{(k-2)(k+\alpha-1)}]
\ea
\ba
C_{\rm N,\rm I} &=& \frac{1}{k(k+\alpha)(k+2\alpha)(k+3 \alpha)}[1 +  
\frac{8\alpha}{k-1}
-\frac{4\alpha}{k-2} \nonumber\\
&&-\frac{\alpha^2}{(k+\alpha-1)(k-2)}+
\frac{(1+\alpha)(4\alpha k + 3 \alpha^2 - 4\alpha)}{
(k-1)(k-2)(k+\alpha-1)}]
\ea
   We have chosen the overall factor of $C_{\rm II,\rm I,\rm I}$ as  
$\frac{1+\alpha}{2}$.
The factor $[ 1 - \frac{2\alpha}{k+\alpha-1} + \frac{7 \alpha}{k-1} +  
\frac{6\alpha^2}{(k-1)(k-2)}
-\frac{2\alpha^2}{(k-2)(k+\alpha-1)}]$ in $C_{\rm II,\rm I,\rm I}$, is  
the same as for
$C_{\Lambda,\rm I,\rm I}$.

For $D_{2234}^{1112}$, we obtain after the subtraction of the known  
three terms,
\be
C_{\unrhd} = \frac{1}{k(k+\alpha)(k+ 2\alpha)(k+ 3\alpha)}( 1 -  
\frac{\alpha(\alpha-1)}{k-1})
\ee
This expression becomes $\frac{(k-3)}{k(k-1)^2(k-2)}$ for $\alpha=-1$,  
and it is consistent with the
value given in Table A.
For $D_{2235}^{1124}$, using the previously determined 6 coefficients,  
one obtains
$C_{\bigtriangleup,\rm I}$ as
\ba
C_{\bigtriangleup,\rm I} &=& \frac{1}{k(k+\alpha)(k+2 \alpha)(k+  
3\alpha)}[1- \frac{\alpha(\alpha-3)}{
k+\alpha-1}
-\frac{\alpha^2 (4 \alpha -3)}{(k-2)(k+\alpha-1)}\nonumber\\
&&-\frac{\alpha^2( 3\alpha^2 - 2\alpha + 3)}{
(k-1)(k-2)(k+\alpha-1)}]
\ea
This expression reduces to  $\frac{(k-3)^2}{k(k-1)^2(k-2)^3}$ for  
$\alpha=-1$, and this agrees
  with the
result given  in Table A.

   We have thus obtained explicitely all the coefficeints up to  fourth  
order.
They satisfy a remarkable property.
When the parameter $\alpha$ vanishes
$$ \sum_{m=0}^{\infty}\frac{1}{m!}\frac{1}{\prod_{q=0}^{m-1}
( 1 + q \alpha)}\sum_{p}\chi_p \frac{Z_p(\tilde x)Z_p(\tilde  
\lambda)}{Z_p(\rm I)}$$ becomes
unity, and $f$ in (\ref{expandf}) is simply given by
\be
f=e^{-\frac{1}{k}\sum_{i<j}\tau_{ij}}.
\ee
Then the coefficients are given by
\be
C= (-1)^l \frac{g}{ k^l}
\ee
where$l$ is the order of the  $\tau$ term, and $C$ the corresponding  
coefficient ;
$g$ is a degeneracy factorfor the multiple lines of the graphs. For the  
double line, $g=1/2!$,
and for the triple line, $g=1/3!$ etc.

The pole terms with denominators $(k-1)$,$(k-2)$,$(k+\alpha-1)$,...,  
which come from $\frac{1}{Z_p(\rm I}$
in (\ref{expandf})
disappear in the expression of the coefficients in the  limit  
$\alpha=0$.
This leads to the fact that the pole terms $\frac{1}{k-1}$,  
$\frac{1}{k-2}$,$\frac{1}{k+\alpha-1}$,...,
always appear with a factor proportional to $\alpha$ or to a power of  
$\alpha$. Then,
the coefficients $C$ may be  factorized as
\be\label{largek}
C= (-1)^l \frac{g}{k (k+\alpha)(k+ 2\alpha) \cdots (k+ (l-1)\alpha)}(  
1+ O(\alpha))
\ee
where $g$ is a degeneracy factor, and for the single multiple line  
graph case, it is given by
\be
g = \frac{\prod_{m=1}^{q-1} ( 1 + m \alpha)}{q!}
\ee
$q$ is the number of multiple lines. When there are $n$ mutiple lines,
($q_1$,$q_2$,...,$q_n$) \\
in a graph,
the degeneracy factor $g$ becomes
\be
g= \prod_{i=1}^n \frac{\prod_{m=1}^{q_i-1} ( 1 + m \alpha)}{q_i!}
\ee
For instance, in the case of $C_{\rm II,\rm II}$, the degeneracy factor  
$g$ is
$\displaystyle{\frac{(1+\alpha)^2}{2!2!}}$.

The correction of order $\alpha$ in (\ref{largek}) is
order of $\frac{1}{k}$ in the large k limit. Therefore, in the large k  
limit, all the coefficients
of the $\tau$-expansion  are given by (\ref{largek}).

The second remark is about the correction terms to (\ref{largek}).  
Common corrections appear
when graphs have common structures.
For instance, star graphs have no corrections. As star graphs, we have  
up to the fourth order,
$C_{\rm I}$, $C_{\Lambda}$,$C_{\rm II}$, $C_{\rm Y}$,$C_{\rm  
III}$,$C_{\underline{\angle}}$,
$C_{\rm  
X}$,$C_{\ll}$,$C_{\models}$,$C_{\underline{\underline{\angle}}}$,$C_{\rm 
  IIII}$.
They are given exactly by the leading term of (\ref{largek}).

For the graphs  with one separate line, the correction terms are agaion  
all the same
  if the remaining part
is a star graph.
For instance, we find a common factor for $C_{\Lambda,\rm I}$ and  
$C_{\rm II,I}$, which is
$\displaystyle{(1 + \frac{2 \alpha}{k-1})}$.
For $C_{\rm Y,\rm I}$,$C_{\underline{\angle},\rm I}$,$C_{\rm III,\rm  
I}$, the common  is
$\displaystyle{(1+ \frac{3\alpha}{k-1})}$.

\newpage

%%%%%%%%%%%%%%%%%%%%%%%%%%%%%%%%%%%%%%%%%%%%%%%%%%%%%%%%%%%%%%%

%%%%%%%%%%%%%%%%%%%%%%%%%%%%%%%%%%%%%%%%%%%%%%%%%%%%%%%%%%%%%%%%%%%%%%%% 
%%%

\end{document}